\documentclass[8pt,review]{article}
\usepackage{amsmath,amssymb,amsfonts,latexsym}
\usepackage{color}
\usepackage{graphicx}
\usepackage{subfigure}
\usepackage{amssymb,latexsym,amsmath,amsfonts}
\usepackage{graphicx}
\usepackage{subfigure}
\usepackage{color}
\usepackage[english]{babel}
\usepackage{authblk}
\usepackage{subfigure}
\usepackage[utf8]{inputenc}
\usepackage{pdfpages}

\catcode `\@=11 \@addtoreset{equation}{section}

\catcode `\@=12

\newcommand{\be}{\begin{equation}}
\newcommand{\en}{\end{equation}}
\newcommand{\bea}{\begin{eqnarray}}
\newcommand{\ena}{\end{eqnarray}}
\newcommand{\beano}{\begin{eqnarray*}}
\newcommand{\enano}{\end{eqnarray*}}
\newcommand{\bee}{\begin{enumerate}}
\newcommand{\ene}{\end{enumerate}}
\newcommand{\R}{\mathcal{R}}

\newcommand{\K}{\mathcal{K}}

\newcommand{\Sc}{\mathcal{S}}

\newcommand{\C}{\mathcal{C}}
\newcommand{\Pc}{\mathcal{P}}

\title{A ~Phenomenological Operator Description \\of Dynamics of Crowds: Escape Strategies}
\author{F. Bagarello$^{1}$, F. Gargano$^2$, F. Oliveri$^{3}$,
       \\
      \tiny{ 
       $^1$DEIM -  Universit\`a di Palermo\\ Viale delle Scienze, I--90128  Palermo, Italy.\\
       $^2$National Research Council (CNR), Institute for Coastal Marine Environment (IAMC),\\
             Via L. Vaccara 61, 91026, Mazara del Vallo (TP), Italy.\\
       $^3$Dipartimento di Matematica e Informatica, Universit\`a di Messina\\
Viale F. Stagno d'Alcontres 31, I--98166 Messina, Italy.\\
       }\small{\textit{Email addresses:}\\
       fabio.bagarello@unipa.it (F.B.),
       francesco.gargano@unipa.it (F.G),\\
       francesco.oliveri@unime.it (F.O)}}
\begin{document}
\date{}
\maketitle

%

\begin{abstract}
We adopt an operatorial method, based on creation, annihilation and number operators, to describe one or two populations mutually interacting
and moving in a two--dimensional region. In particular, we discuss how the two populations, contained in a certain two-dimensional
region with a non--trivial topology, react when some alarm occurs. We consider the cases of both low and high densities of the populations,
and discuss what is changing as the strength of the interaction increases. We also analyze what happens when the region has either a single exit or two ways out.
\end{abstract}


\section{Introduction and preliminaries}
\label{sec:introduction}

In a recent paper, Ref.~\cite{ff}, two of us (F.B. and F.O.)  used an operatorial approach, based on some well known features of quantum mechanical methods, to describe the interaction between two different populations constrained in a finite, closed, two-dimensional region. In particular, two different applications have been considered. In the first one, there are two populations located in different regions, a \emph{poor area} and a different, \emph{richer}, zone, and it is investigated how the two populations spread during the time evolution. In the second application, namely  a simplified view to a predator--prey system, the two populations interact adopting essentially the same mechanisms as before, with the major difference that, at $t=0$, they are located in the same area. Again, the main interest was in the time dispersion of the two species.

Here, adopting the same operatorial mechanism, we consider a different but somehow similar problem, for which
a quite different interpretation is needed. We have two different populations, $\Pc_a$ and $\Pc_b$, forced to stay together in a certain two-dimensional region $\R$ with
a non--trivial topology. What we  have in mind is essentially the following: $\Pc_a$ and $\Pc_b$ are two groups of,
say, young and aged people, staying in  a shop (or in some closed space), with some exits and some obstacles around
(like shelves, columns, \ldots). We want to explore what happens when some alarm starts to ring. How do the two populations react?
How fast do they leave the room? How  different the behaviors of the two populations are?
Is there any reasonable way to help $\Pc_a$ and $\Pc_b$ to move faster? We will consider a somehow fixed topology, and a fixed initial condition,
playing with the parameters of the model and with the escape strategies in order to find some sort of \emph{optimal path}, or, more generally,
some optimal escape strategy. Also, we will consider the case in which the two populations do not interact, which is reasonable where there is only a
small number of people inside the room at $t=0$, and the case in which some interaction between $\Pc_a$ and $\Pc_b$ is expected, \emph{i.e.}, for a sufficiently
large number of people.\\

This kind of problem has clear implications in concrete situations, and for this reason it has attracted, along the years,
the attention of several researchers who dealt with escape dynamics in many physical environments. Various methodological approaches to
simulate crowd evacuation can be found in the vast scientific literature developed in the last years: each of these approaches has pros and cons,
and clearly none of them can solve the problem  completely. In fact, some approaches work better than the others according to the physical effects
one wants to include in the model, to the level of the description (\emph{microscopic}  or \emph{macroscopic}), or to the computational effort.
Among these methods we can quote those based on lattice gas and cellular automata
(see, for example, Refs.~\cite{baglietto,varas,nagai,kirchener,yuan,guo})
which revealed really consistent in several contexts: they are interesting microscopic models which often can not properly simulate some typical \emph{high--pressure}
phenomena arising from the contact forces during the evacuation of a crowd. To account for these high--pressure phenomena,
as well as other collective behaviors of a crowd, approaches based on force--models can be used (see, for example, Refs.\cite{evers,charaibi,helbing}).
The force--based description is generally motivated by the observation that
the motion of pedestrians deviates in the presence of other
pedestrians, and this effect seems as induced by a force that must be included in the model to correctly simulate the pedestrian motion.
Nevertheless, some unrealistic behaviors may arise, especially for high densities:  overlapping, oscillations phenomena,
or backwards movement due to negative velocities.
To avoid or to control these phenomena,
the equations of motion must implement other procedures, \emph{e.g.}, collision detection
algorithms, having as a counterpart a substantial increasing of the complexity of the model.
Methods based on the interactions of individuals or collective agents are taken into account in the agent--based--methods (see  Refs.~\cite{dai,basset,marsili}):
these methods rely on the techniques of cognitive science and they are able
to capture or predict some \emph{emergent} phenomena arising during a crowd evacuation, even if there are some difficulties to model
a form of intelligence for each agent.
The macroscopic point of view is generally the framework in which the fluid--dynamic models are built: the idea here is to consider the motion of a
crowd like a fluid motion (see Ref.~\cite{helbing2}). This viewpoint is quite reasonable, but it is limited by the requested hypothesis
of high density of the crowd,
which is not always the condition one wants to consider.
Other recent methodological approaches are based on game theory
(see  Refs.~\cite{shi,heliovara}):
even if these methods take into account some wanted effect of strategic thinking that can characterize, for obvious reasons, the behavior of a crowd,
it is really difficult, especially with a large number of players, to find the
appropriate payoff matrix required for a game.
We refer the interested reader to some interesting review papers in which a more detailed analysis of the problem is performed (Refs.~\cite{bellomo,zheng,xiaoping}).

Our approach to the problem of an escaping crowd is really different and it is based on operatorial methods of quantum mechanics explained in details in
the next section. We only mention here that this kind of approach has revealed successful not only to describe the migration of a population, as previously said,
but also the dynamics of several other macroscopic models, such as  stock markets, love affairs or closed ecosystems (see Ref.~\cite{bagbook,ff_eco,baghav}).

The paper is organized as follows. In Section~\ref{sec:spreading},
we introduce the hamiltonian operator describing the dynamics
of the populations $\Pc_a$ and $\Pc_b$, determine the equations of motion and write down the solution.
We refer again to Ref.~\cite{bagbook} for the general ideas behind our settings. In Section~\ref{sec:numerical},
we apply our strategy to the derivation of the dynamics of the two populations, and we discuss in details the meaning of the parameters of our model.
Our conclusions are given in Section~\ref{sec:conclusions}.
Finally, the Appendix contains some snapshots of our video simulations which are available, for the interested readers, upon request to the authors.

\section{The dynamical model}
\label{sec:spreading}

Let us consider a 2D--region $\R$ in which, in
principle, the two populations $\Pc_a$ and $\Pc_b$ are distributed.  The (\emph{e.g.}, rectangular or square) region $\R$, is divided in $N$ cells
(see Figure~\ref{fig1}), labeled by $\alpha=(i,j)$, $i=1,\ldots,L_x$, $j=1,\ldots,L_y$; to simplify the notation, when needed, we refer to the cell $(1,1)$ as cell 1, the cell $(2,1)$ as cell 2, \ldots, the cell $(1,2)$ as the
cell $L_x+1$, \ldots and the cell $(L_x,L_y)$ as the cell $N$. In the rest of this paper, we will always assume that $L_x=L_y=L$. It should be stressed that, contrarily to what we have done in Ref.~\cite{ff}, here not all the cells can in principle be occupied, since there are obstacles in $\R$ where the populations can not go. Moreover, $\R$ is not a closed region as it was in Ref.~\cite{ff}, but, on the contrary, there are exits somewhere on the borders, exits which $\Pc_a$ and $\Pc_b$ want to reach, as fast as they can, to leave $\R$ under some emergency.

\begin{figure}
\begin{center}
\begin{picture}(400,240)
\put(40,240){\thicklines\line(1,0){320}}
\put(40,0){\thicklines\line(0,1){240}}
\put(40,0){\thicklines\line(1,0){320}}
\put(360,0){\thicklines\line(0,1){240}}
\put(40,0){\line(1,0){320}}
\put(40,40){\line(1,0){320}}
\put(40,80){\line(1,0){320}}
\put(40,120){\line(1,0){320}}
\put(40,160){\line(1,0){320}}
\put(40,200){\line(1,0){320}}
\put(80,0){\line(0,1){240}}
\put(120,0){\line(0,1){240}}
\put(160,0){\line(0,1){240}}
\put(200,0){\line(0,1){240}}
\put(240,0){\line(0,1){240}}
\put(280,0){\line(0,1){240}}
\put(320,0){\line(0,1){240}}
\put(360,0){\line(0,1){240}}
\put(60,20){\makebox(0,0){\footnotesize$(1,1)$}}
\put(100,20){\makebox(0,0){\footnotesize$(2,1)$}}
\put(140,20){\makebox(0,0){\footnotesize$\cdots$}}
\put(300,20){\makebox(0,0){\footnotesize$\cdots$}}
\put(340,20){\makebox(0,0){\footnotesize$(L_x,1)$}}
\put(60,60){\makebox(0,0){\footnotesize$(1,2)$}}
\put(100,60){\makebox(0,0){\footnotesize$(2,2)$}}
\put(140,60){\makebox(0,0){\footnotesize$\cdots$}}
\put(60,220){\makebox(0,0){\footnotesize$(1,L_y)$}}
\put(100,220){\makebox(0,0){\footnotesize$\cdots$}}
\put(300,220){\makebox(0,0){\footnotesize$\cdots$}}
\put(340,220){\makebox(0,0){\footnotesize$(L_x,L_y)$}}
\end{picture}
\end{center}
\vspace*{5mm}
\caption{\label{fig1} The two--dimensional lattice for the spatial model.}
\end{figure}
\vspace*{.1cm}

As widely discussed in Ref.~\cite{bagbook}, in our approach the dynamics of the system $\Sc$ under analysis is defined by means of a self--adjoint Hamiltonian operator which contains all the mechanisms we expect could take place in $\Sc$.  Following Ref.~\cite{ff}, we assume that in each cell $\alpha$ the two populations, whose related relevant operators are $a_\alpha$, $a_\alpha^\dagger$
and $\hat n^{(a)}_\alpha=a_\alpha^\dagger a_\alpha$ for what concerns $\Pc_a$, and $b_\alpha$, $b_\alpha^\dagger$ and $\hat
n^{(b)}_\alpha=b_\alpha^\dagger b_\alpha$ for $\Pc_b$, are described by the hamiltonian  \be H_\alpha=H_\alpha^0+\lambda_\alpha H_\alpha^I,\qquad
H_\alpha^0=\omega_\alpha^a a_\alpha^\dagger a_\alpha+\omega_\alpha^b b_\alpha^\dagger b_\alpha, \quad H_\alpha^I=a_\alpha^\dagger
b_\alpha+b_\alpha^\dagger a_\alpha. \label{31} \en
We refer to Ref.~\cite{ff} for more details on the meaning of $H_\alpha$. The operators involved in (\ref{31}) satisfy the following anticommutation rules: \be \{a_\alpha,a_\beta^\dagger\}=\{b_\alpha,b_\beta^\dagger\}=\delta_{\alpha,\beta}, \qquad \{a_\alpha^\sharp,b_\beta^\sharp\}=0.
\label{32} \en
As discussed in Ref.~\cite{ff},  it is natural to interpret the
mean values of the operators $\hat n^{(a)}_\alpha$ and $\hat n^{(b)}_\alpha$ as \emph{local density operators} (the local densities are in the
sense of mixtures; hence, we may sum up local densities relative to different cells) of the two populations in the cell $\alpha$: if the mean
value of, say, $\hat n^{(a)}_\alpha$, in the state of the system is equal to one, this means that the density of $\Pc_a$ in the cell $\alpha$
is very high. On the other hand, if the mean value of, say, $\hat n^{(b)}_\alpha$, in the state of the system is equal to zero,
we interpret this saying that, in cell $\alpha$, we can find only very few members of $\Pc_b$.
{Hence, the interaction hamiltonian $H_\alpha^I$ in (\ref{31}) can be easily understood: it describes the fact that once the density of, say, $\Pc_a$ increases in cell $\alpha$, the density of $\Pc_b$ in the same cell must decrease. This means that one species tends to exclude the other one}. Notice that $H_\alpha=H_\alpha^\dagger$,
since all the parameters, which are in general  assumed to be cell--depending (to allow
for the description of an anisotropic situation), are real (and positive) numbers.

 The full hamiltonian $H$ must consist of a sum of all the different $H_\alpha$ plus another contribution, $h$, responsible
for the diffusion of the populations all around the lattice: $H=\sum_\alpha H_\alpha+h$. A natural choice for $h$, which extends 
that in Ref.~\cite{ff}, is the following one: \be
h=\sum_{\alpha,\beta}\left\{p_{\alpha,\beta}^{(a)}\left(a_\alpha a_\beta^\dagger+a_\beta a_\alpha^\dagger\right)+p_{\alpha,\beta}^{(b)}\left(b_\alpha
b_\beta^\dagger+b_\beta b_\alpha^\dagger\right)\right\}, \label{33} \en where  $p_{\alpha,\beta}^{(a)}$ and $p_{\alpha,\beta}^{(b)}$ are real
quantities, and,{{ for instance, $a_\alpha a_\beta^\dagger$ describes movement of $\Pc_a$ from cell $\alpha$ to cell $\beta$}}\footnote{This is because the presence of $a_\alpha$ causes a lowering of the density of $\Pc_\alpha$ in the cell $\alpha$, density which increases in cell $\beta$ because of $a_\beta^\dagger$.}. The role of  $p_{\alpha,\beta}^{(a,b)}$ is more complex than that of the analogous quantities in Ref.~\cite{ff}, since they contain the following information: (i) they are equal to zero whenever the populations can not move from cell $\alpha$ to cell $\beta$; (ii) in general, we could have $p_{\alpha,\beta}^{(a)}\neq p_{\alpha,\beta}^{(b)}$, since the two populations may have different behaviors; hence, they describe some difference between $\Pc_a$ and $\Pc_b$, related, as we will see, to the different \emph{mobilities} of the two populations; (iii) in our approach they are used to \emph{suggest the populations} the fastest paths toward the exits. We will discuss this aspect in detail later on. As already said,  when $p_{\alpha,\beta}^{(a,b)}=0$, the two populations can not move from cell $\alpha$ to cell $\beta$. This happens, for instance, when $\beta$ is a cell in which an obstacle (or part of it) is located. In this case, in fact, neither $\Pc_a$ nor $\Pc_b$ can occupy that cell. Also, $p_{\alpha,\beta}^{(a,b)}=0$ whenever $\alpha$ and $\beta$ are not nearest neighbors.
Another difference with respect to Ref.~\cite{ff} is that we will allow here also for the \emph{diagonal} movements. Notice, finally, that we have assumed here that $p_{\alpha,\beta}^{(a,b)}=p_{\beta,\alpha}^{(a,b)}$. This is because, even if we are interested to introduce some anisotropy in the model, because of the structure of the Heisenberg equations of motion \cite{bagbook}, this would not be the most natural way to proceed, since the differential equations we would get in this way would depend on the sum $p_{\alpha,\beta}^{(a,b)}+p_{\beta,\alpha}^{(a,b)}$, and not  on just $p_{\alpha,\beta}^{(a,b)}$ or $p_{\beta,\alpha}^{(a,b)}$. Therefore, the resulting coefficients in the differential equations would be symmetrical in $\alpha$ and $\beta$ anyway, even if the $p_{\alpha,\beta}^{(a,b)}$ were not. Therefore, to simplify the treatment, we adopt the symmetric choice $p_{\alpha,\beta}^{(a,b)}=p_{\beta,\alpha}^{(a,b)}$ from the very beginning.

{Another interesting aspect of our general framework, and of the use of the anticommutation rules (\ref{32}) 
in particular, is that they automatically implement the impossibility of having too many elements of a single
population in a given cell. This is because ${a_\alpha^\dagger}^2={b_\alpha^\dagger}^2$.
This could be seen as a first attempt to implement a {\em no-collision rule}. 
However, in our model, collisions between $\Pc_a$ and $\Pc_b$ are possible, in principle.}

What we are mainly interested to is the time evolution of the densities of both $\Pc_a$ and $\Pc_b$ inside $\R$. This means that first we have to compute the time evolution $\hat n^{(a)}_\alpha(t)$ and $\hat n^{(b)}_\alpha(t)$ for each $\alpha=1,2,\ldots,N$, and then take their mean values on a vector state describing the initial status of $\Sc$, \emph{i.e.}, describing the densities, at $t=0$, of $\Pc_a$ and $\Pc_b$ in each cell of $\R$, see Refs.~\cite{ff,bagbook}. In order to deduce $\hat n^{(a)}_\alpha(t)$ and $\hat n^{(b)}_\alpha(t)$, it is convenient  first to look for the time evolution of both $a_\alpha$ and $b_\alpha$, by writing the Heisenberg differential equation $\dot a_\alpha(t)=i[H,a_\alpha(t)]$ and  $\dot b_\alpha(t)=i[H,b_\alpha(t)]$, which read as follows:
 \be
\left\{
    \begin{array}{ll}
\dot a_\alpha=-i\omega_\alpha^a a_\alpha -i\lambda_\alpha b_\alpha +2i\sum_{\beta=1}^{L^2} p_{\alpha,\beta}^{(a)}a_\beta,\\
\dot b_\alpha=-i\omega_\alpha^b b_\alpha -i\lambda_\alpha a_\alpha +2i\sum_{\beta=1}^{L^2} p_{\alpha,\beta}^{(b)}b_\beta.
\end{array}
        \right.
\label{34} \en

Recall that, since $p_{\alpha,\alpha}^{(a,b)}=0$, the sums in the right-hand sides of (\ref{34}) are really restricted to $\beta\neq\alpha$. System (\ref{34}) is linear, and it can be rewritten as
\be
\dot X_{L^2}=i \K_{L^2} X_{L^2},
\label{35}
\en
where $\K_{L^2}=2T_{L^2}-P_{L^2}$, $T_{L^2}$ and $P_{L^2}$ being two $L^2\times
L^2$ matrices defined as follows:
\[
T_{L^2}=\left(
          \begin{array}{cc}
            V_{L^2}^{(a)} & 0 \\
            0 & V_{L^2}^{(b)} \\
          \end{array}
        \right),
\qquad
P_{L^2}=\left(
          \begin{array}{cc}
            \Omega^{(a)} & \Lambda \\
            \Lambda & \Omega^{(b)} \\
          \end{array}
        \right).
\]
We have introduced here the following matrices: $\Omega^{(a)}=\hbox{diag}\{\omega_1^a,\omega_2^a,\ldots,\omega_{L^2}^a\}$,
$\Omega^{(b)}=\hbox{diag}\{\omega_1^b,\omega_2^b,\ldots,\omega_{L^2}^b\}$,
$\Lambda=\hbox{diag}\{\lambda_1,\lambda_2,\ldots,\lambda_{L^2}\}$, while $V_{L^2}^{(a)}$ is the $L^2\times L^2$ matrix with entries different from zero, and equal to $p_{\alpha,\beta}^{(a)}$ only for those matrix elements corresponding to the allowed movements in $\R$ (\emph{e.g.}, in the positions $(1,2)$, $(1,L+1)$, $(1,L+2)$, $(2,1)$, $(2,3)$, and so on). Similarly, the matrix $V_{L^2}^{(b)}$ has entries  zero or equal to $p_{\alpha,\beta}^{(b)}$ in the same positions as for $V_{L^2}^{(a)}$.
Finally, the transpose of the unknown vector, $X_{L^2}^T$, is defined as
$$
X_{L^2}^T=\left(
            \begin{array}{cccccccccccc}
              A_1(t) & A_2(t) & \cdots & \cdots & A_{L^2}(t) & B_1(t) & B_2(t) & \cdots & \cdots & B_{L^2}(t)  \\
            \end{array}
          \right),
$$
where $A_j(t)=a_j(t)e^{i\omega_j^at}$ and $B_j(t)=b_j(t)e^{i\omega_j^bt}$.

The solution of equation (\ref{35}) is
\[
X_{L^2}(t)=\exp\left(i\,\K_{L^2}t\right)X_{L^2}(0).
\]
Let us now call $f_{\alpha,\beta}(t)$ the generic entry of the matrix $\exp\left(i\,\K_{L^2}t\right)$, and let us assume that, at $t=0$, the system is described
by the vector $\varphi_{\mathbf{n}^a,\mathbf{n}^b}$, where ${\bf n}^a=(n_1^a,n_2^a,\ldots,n_{L^2}^a)$ and  ${\bf
n}^b=(n_1^b,n_2^b,\ldots,n_{L^2}^b)$\footnote{This is the way in which the initial condition of $\Sc$ are taken into account, Ref.~\cite{bagbook}.}. Hence, the mean values of the time evolution of the number operators in the cell $\alpha$, assuming these initial conditions, are
\be \label{35bis}
\begin{aligned}
&N_\alpha^a(t)=\left<\varphi_{\mathbf{n}^a,\mathbf{n}^b},a_\alpha^\dagger(t) a_\alpha(t)\varphi_{\mathbf{n}^a,\mathbf{n}^b}\right>=\left<\varphi_{\mathbf{n}^a,\mathbf{n}^b},A_\alpha^\dagger(t) A_\alpha(t)\varphi_{\mathbf{n}^a,\mathbf{n}^b}\right>,\\
&N_\alpha^b(t)=\left<\varphi_{\mathbf{n}^a,\mathbf{n}^b},b_\alpha^\dagger(t) b_\alpha(t)\varphi_{\mathbf{n}^a,\mathbf{n}^b}\right>=\left<\varphi_{\mathbf{n}^a,\mathbf{n}^b},B_\alpha^\dagger(t) B_\alpha(t)\varphi_{\mathbf{n}^a,\mathbf{n}^b}\right>,
\end{aligned}
\en
which can be written as \cite{ff} \be \label{36}
\begin{aligned}
&N_\alpha^a(t)=\sum_{\theta=1}^{L^2}|f_{\alpha,\theta}(t)|^2\,n_\theta^a+
\sum_{\theta=1}^{L^2}|f_{\alpha,L^2+\theta}(t)|^2\,n_\theta^b,\\
&N_\alpha^b(t)=\sum_{\theta=1}^{L^2}|f_{L^2+\alpha,\theta}(t)|^2\,n_\theta^a+
\sum_{\theta=1}^{L^2}|f_{L^2+\alpha,L^2+\theta}(t)|^2\,n_\theta^b.
\end{aligned}
\en

In Ref.~\cite{ff}, these formulas have been used to deduce the local
densities of the two populations $\Sc_1$ and $\Sc_2$ in three or two different regions,
respectively for migration or for the predator--prey system we have considered.
{ It is worth to stress 
that a main advantage of this approach lies in the fact that we
have derived an exact formulation for the densities of the populations in \eqref{36},
with obvious advantage in terms of computational effort. }

\section{Numerical simulations}
\label{sec:numerical}
\begin{figure}
\begin{center}
\includegraphics[width=8cm]{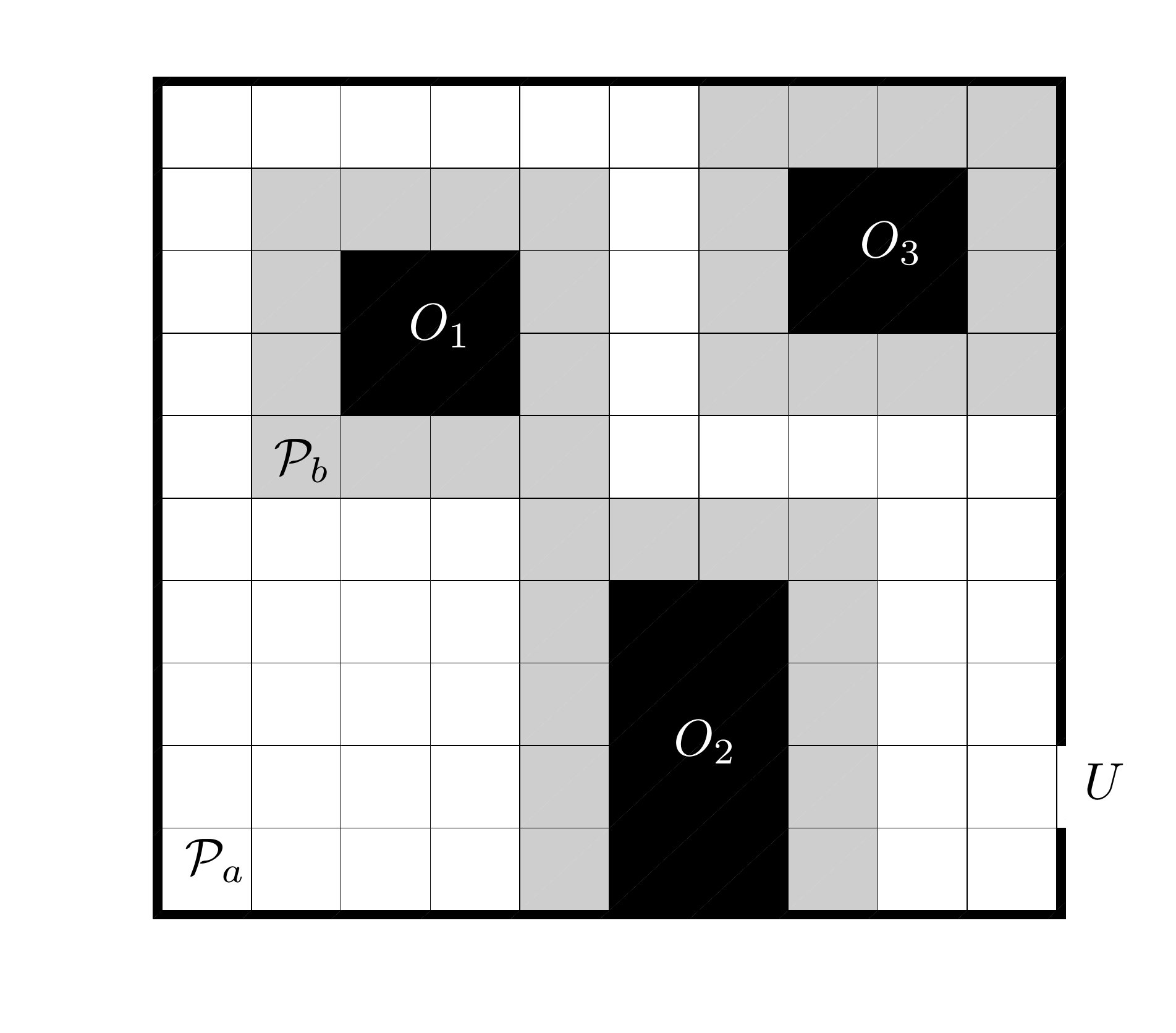}
\vspace*{-0.25cm}\caption{\textbf{Setting $S_2$}: the region $\R$ is a square of $L_x\cdot L_y=10\cdot10$ cells plus the exit cell $U$ is located at (11,2). At $t=0$ the populations $\Pc_a$ and $\Pc_b$
are located in the cells $(1,1)$ and $(2,6)$ respectively.
The black cells in $O_1,O_2,O_3$ represents the obstacles, surrounded by a region, $\partial O$, which might be slightly different from the rest of $\R$.
}
\label{setup_p2_u1_po}
\end{center}
\end{figure}

The differential equations in (\ref{35}), and the functions in (\ref{36}), will now be used considering some slightly different contexts with the aim of suggesting some \emph{optimal escape strategy}. However, before  considering the different configurations we are interested to, we need to introduce some efficient mechanism to describe the simple fact that \emph{people want to go out of the room} as fast as they can. Suppose that, at $t=0$,  the populations $\Pc_a$ and $\Pc_b$ are distributed in $\R$ as shown
in Figure \ref{setup_p2_u1_po}.

Since the hamiltonian $H$ commutes with the total density operator, $\hat N=\sum_{\alpha}\hat n_\alpha^{(a)}+\sum_{\alpha}\hat n_\alpha^{(b)}$, this implies that $\hat N(t)$, and its mean value as a consequence, remain constant in time. Of course, this would not be compatible with what we aim to describe: nothing can enter and nothing can leave $\R$, otherwise $\hat N(t)$ would change in time. For this reason, we have considered here two different strategies, adopting at the end the most effective one.

The first strategy refers to what has been recently proposed in Ref.~\cite{ff_eco}, \emph{i.e.}, an \emph{effective} mechanism to describe damping.
In fact, it has been shown that, adding a small negative imaginary part to a single parameter of the free hamiltonian produces such a damping. This is quite close to what is done in many models in quantum optics, to describe some decay.  Then, our idea here was to consider $\R$ as a subregion of a larger, closed, area $\R_{big}$, considering $\C:=\R_{big}\setminus\R$ as a sort of courtyard surrounding the room the people have to leave. This courtyard could only be reached through the exit $U$. In this case, the hamiltonian $H$ introduced previously refers not only to $\R$, but also to $\C$. In other words, the sums over $\alpha$ and $\beta$ is extended to all the lattice cells in $\R_{big}$. The main difference between $\R$ and $\C$ is that, while all the parameters of $H$ "inside" $\R$ are all real, those in $\C$ could be complex valued, with a small negative imaginary part. In this way, when some elements of $\Pc_a$ or $\Pc_b$ reach $\C$, they begin to \emph{effectively disappear} and, as a consequence, they can not return back to $\R$: the densities of the populations decrease simply because they have reached the courtyard and they are, of course, not willing to return back inside the room. Our numerical attempts show that, for this procedure to be efficient, $\C$ must be sufficiently large whereas the imaginary parts can be rather small.

The other strategy which we have considered, and which has proven to be much more efficient for our purposes, is the following one: we have first fixed a (small) time interval $\Delta T$ (not to be confused with the numerical time step!). We have computed $N_\alpha^a(t)$ and $N_\alpha^b(t)$ as in (\ref{36}) in each lattice cell $\alpha$. In particular, we have computed these functions in the exit cell(s). Let us call $N_U^a(t)$ and $N_U^b(t)$ these particular densities, assuming, for the moment, that there is only one exit, $U$.  During the computation we check the values of
$N_U^a(k\Delta T)$ and $N_U^b(k\Delta T)$ ($k$ positive integer). For the smallest value of $k$ when $N_U^a(k\Delta T)$ and/or $N_U^b(k\Delta T)$ do exceed  certain threshold values $N_{thr}^a$ and $N_{thr}^b$, we stop the computation and go back to the solution (\ref{36}), but considering \emph{new} initial conditions, \emph{i.e.}, those in which, at the \emph{new} initial time $t_0=k\Delta T$, there is no population $\Pc_a$ (if $N_U^a(k\Delta T)>N_{thr}^a$) and $\Pc_b$ (if $N_U^b(k\Delta T)>N_{thr}^b$) at all in $U$: those which have reached $U$ have just leaved $\R$, so that they do not longer contribute to $\Pc_a$ and/or $\Pc_b$! This choice, besides being natural, is also faster than the previous one since in that case the numerical computations involve all of $\R_{tot}$ and not only $\R$: a larger domain involves more degrees of freedom and, consequently, a larger dimension of the Hilbert space. The related matrix $\K_{L^2}$ in (\ref{35}) becomes much larger, and therefore the numerical computations slow down significantly. Moreover, some oscillations in the densities which are observed with the first approach, and which suggest that a small percentage of the populations reaching $U$ bounces back to $\R$, almost disappear completely using the second strategy. This is clearly more realistic, since we do not expect that people running away from $\R$ is willing to return back to the dangerous place! However, we should mention that, adopting this strategy, formula (\ref{36}) should be considered in a generalized sense, since not all the $n_\theta^{a,b}$ at time $k\Delta T$ are equal to zero or one, any longer. In other words, we use (\ref{36}) as our main equation, even if it can no longer be deduced from (\ref{35bis}), in principle.

Another preliminary remark, concerning our numerical procedure and its settings, is the following one: to fix the ideas we have always considered three obstacles in $\R$, always located in the same places. As for the two populations at $t=0$, we have considered two main situations. In the first one, they are located in just two different (but fixed) cells. In the second case, we have \emph{distributed} $\Pc_a$ and $\Pc_b$  quite a bit around $\R$. Our idea is that in the first case the interaction should not play any crucial role, while it becomes more and more relevant when the densities grow up.

\subsection{One population, one exit (setting $S_1$)}

\begin{figure}
\begin{center}
\includegraphics[width=8cm]{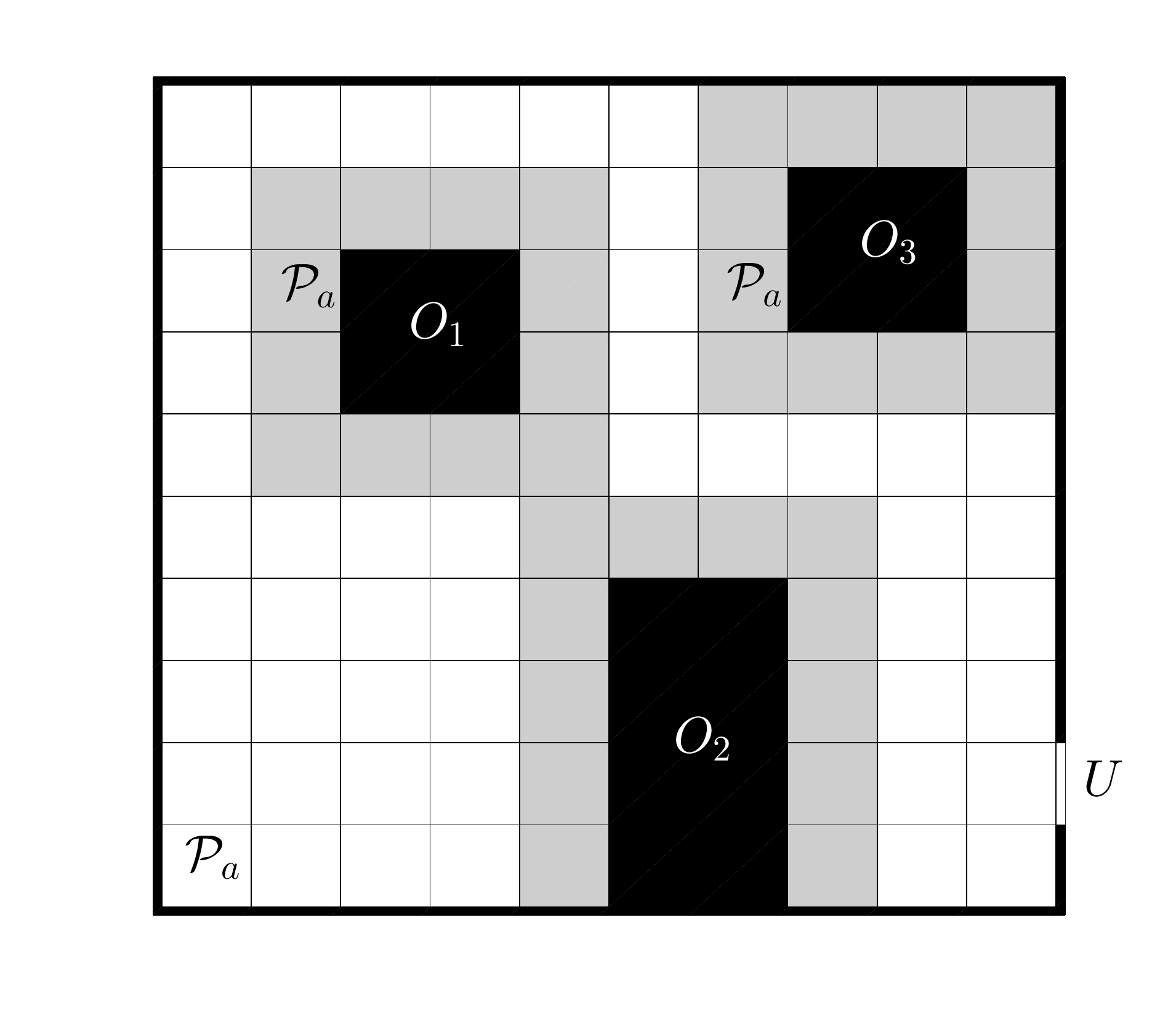}
\vspace*{-0.25cm}\caption{\textbf{Setting $S_1$}: at $t=0$ the population $\Pc_a$ is located in the cells $(1,1),(2,8)$ and $(7,8)$.
The black cells $O_1,O_2,O_3$ represents the obstacles, and the exit cell $U$ is located at (11,2).}
\label{setup_p1_u1_po}
\end{center}
\end{figure}

In order to propose an optimal escape strategy, our main goal is to define $p^{(a)}_{\alpha,\beta}$
 in a convenient way. It is clear, for example, that if part of the population is initially located in the cell (1,1)
 as in Figure \ref{setup_p1_u1_po}, a convenient escape path to reach the exit cell $U$
should go around the obstacle $O_2$, while
an escape path going around $O_1$ or $O_3$ seems not to be a good choice for obvious reasons.
Hence, we define a procedure to determine $p^{(a)}_{\alpha,\beta}$ by requiring that $\Pc_a$
takes the shortest path possible to reach the exit cell $U$.
To apply this procedure, we suppose that $\Pc_a$ is initially located in the cells
$\alpha_1,\alpha_2,...,\alpha_n$. Then we write
\begin{eqnarray}
p^{(a)}_{\alpha,\beta}=\rho_a \hat\delta_{\alpha,\beta}(\gamma^{(a)}_{\alpha}+\gamma^{(a)}_{\beta})/2,
\label{pab}
\end{eqnarray}
where $\rho_a$ is a positive real parameter, whose meaning will be discussed soon, $\hat\delta_{\alpha,\beta}$ is a symmetric tensor which is equal to 1 if the
population can move from $\alpha$ to $\beta$ and 0
otherwise,
and  $\gamma^{(a)}_{\alpha}$ is defined by
\begin{eqnarray}
\tilde\gamma^{(a)}_{\alpha}= \max_{j=1...n}\left[\left(\frac{g_j(\alpha)}{M_j}\right)^{\sigma_a}\right], \label{function_ya}\\
g_j(\alpha)=(f_d(\alpha,\alpha_j)+f_d(\alpha,U))^{-1} \quad \forall j=1...n,\label{function_gj}\\
M_j=\max_{\alpha \in \R}g_j(\alpha)  \quad \forall j=1...n.\label{function_Mj}
\end{eqnarray}
Then we put \begin{eqnarray}\gamma^{(a)}_{\alpha}=\frac{\tilde\gamma^{(a)}_{\alpha}}{\max_\alpha(\tilde\gamma^{(a)}_{\alpha})}. \label{function_ga}\end{eqnarray}
The function $f_d(\alpha,\beta)$ in \eqref{function_gj} is the well known $Dijkstra-function$, see Ref.~\cite{Dijkstra},
which returns the length of the minimal path among the paths going from the cell $\alpha$ to the cell $\beta$ under the constraints given by the definition of
 $\hat\delta_{\alpha,\beta}$ and the presence of the obstacles,  with the assumption that all cells have the same weight.
Therefore, $\gamma^{(a)}_{\alpha}$ takes the value 1 if $\alpha$
is a cell contained in a minimal path going from a given cell $\alpha_i$ to $U$, and it decreases to zero
 if $\alpha$ gets further from the minimal path;
the parameter $\sigma_a$ in \eqref{function_ya} tunes how rapidly $\gamma^{(a)}_{\alpha}$ decreases. More explicitly, if $\sigma_a\rightarrow \infty$ then  $\gamma^{(a)}_{\alpha}=1$
 if $\alpha$ is in the minimal path,
 and $\gamma^{(a)}_{\alpha}=0$ in all the other cells.
By  increasing or decreasing the value of the parameter $\rho_a$ in \eqref{function_ya}, we speed up or slow down $\Pc_a$.
Then, $p^{(a)}_{\alpha,\beta}$ can be different from 0 only if $\alpha$ and $\beta$ are
neighboring cells or if $\beta$ is not a  cell of some obstacles. As we have already commented above, we assume that $p^{(a)}_{\alpha,\beta}=p^{(a)}_{\beta,\alpha}$ holds.
The greatest values of $p^{(a)}_{\alpha,\beta}$ are obtained if $\alpha$ and $\beta$ are along a
minimal path going from the given cell $\alpha_i$ to $U$,
while $p^{(a)}_{\alpha,\beta}$ decrease to zero when the direction from cell $\alpha$ to cell $\beta$ is not along a minimal path.
 We stress that the construction of $p^{(a)}_{\alpha,\beta}$ through \eqref{pab}-\eqref{function_Mj} is not the only way to do this:
for example, one could define another metric through the function $g_j$ in \eqref{function_gj} or using an exponential decay
in \eqref{function_ya}. However, our choice seems to work very well for the escape strategy to impose to $\Pc_a$, as the numerical results show. Obviously, the coefficients $p^{(b)}_{\alpha,\beta}$ are defined in a similar way.

\begin{figure}
\begin{center}
\vspace*{-1cm}\includegraphics[width=8cm]{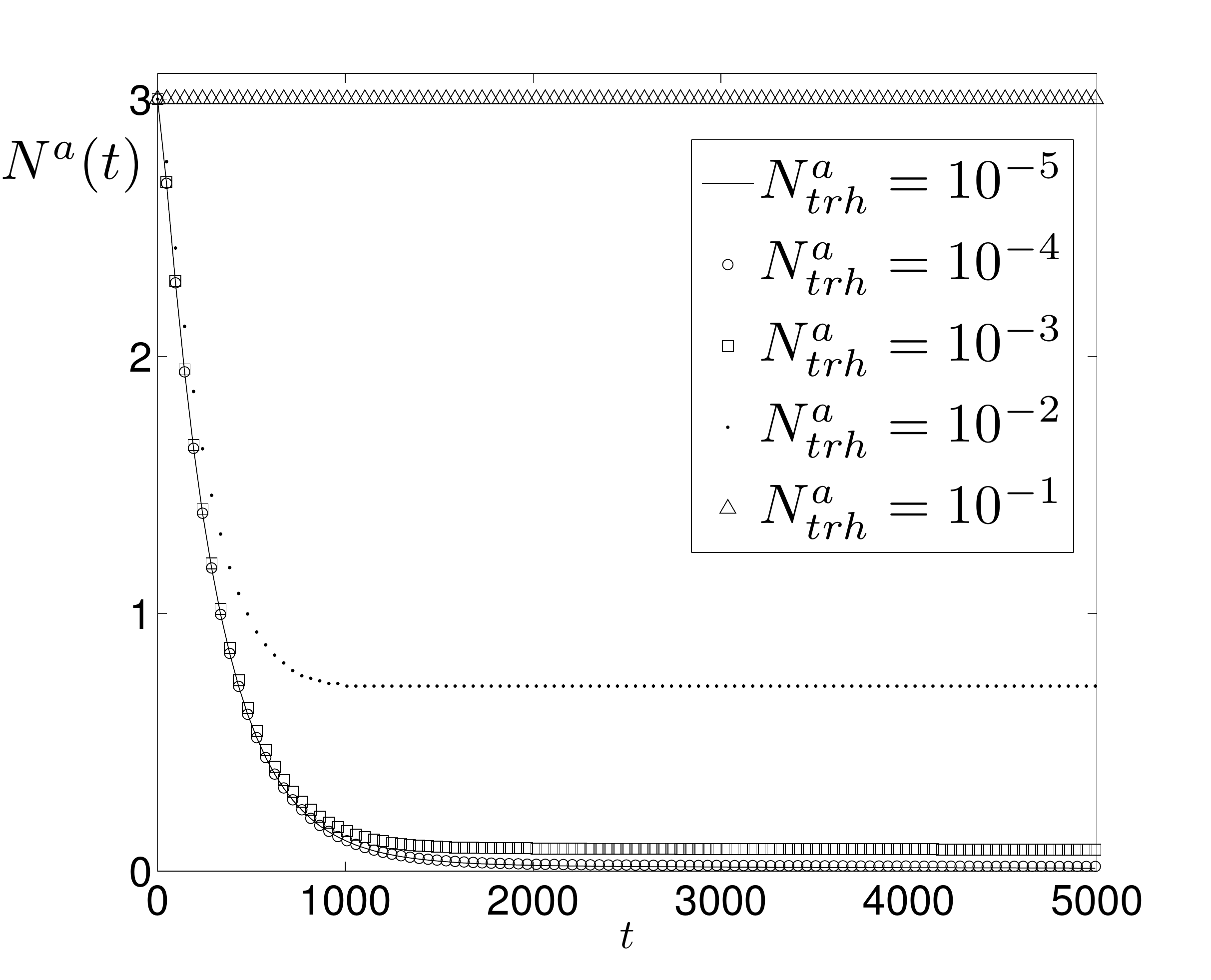}
\caption{\textbf{Setting $S_1$}: the total density $N^a(t)$ is plotted for $\Delta T=0.08$ and different values of $N^a_{trh}$.
For $N^a_{trh}=10^{-5}$ and $N^a_{trh}=10^{-4}$ the numerical results are almost indistinguishable, suggesting that
$N^a_{trh}=10^{-5}$ can be considered an optimal value of threshold for this particular setting.}
\label{p1_u1_vartrh}
\end{center}
\end{figure}

\begin{figure}
\begin{center}
\includegraphics[width=8cm]{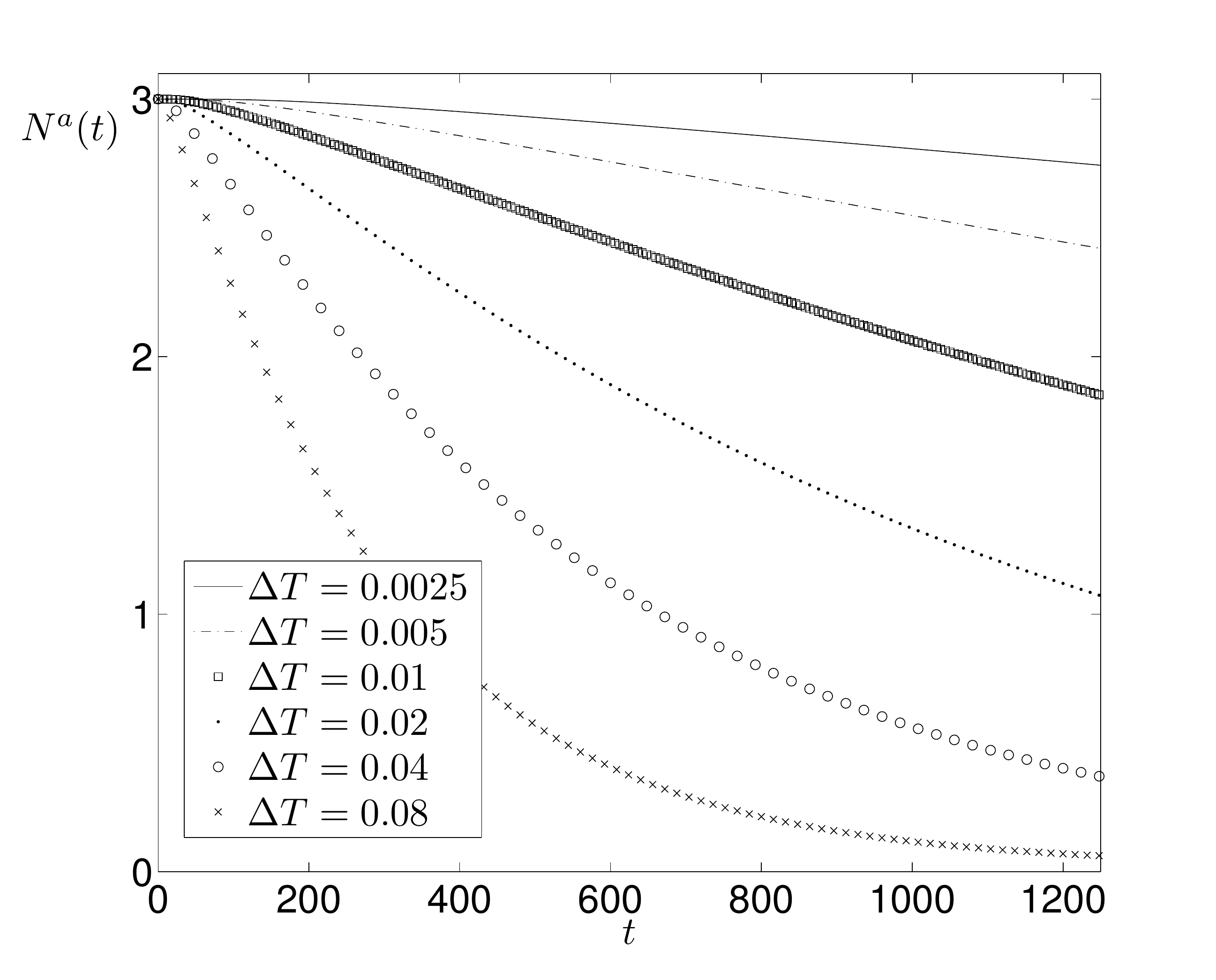}
\vspace*{-0.25cm}\caption{\textbf{Setting $S_1$}: the total density $N^a(t)$ is plotted for $N^a_{trh}=10^{-5}$ and different values of $\Delta T$.
As $\Delta T$ increases a larger amount of population accumulates in the exit cell and therefore at each interval $\Delta T$ we remove a larger amount of population: this explains the faster decay of $N^a(t)$. }
\label{p1_u1_vardt}
\end{center}
\end{figure}

\begin{figure}
\begin{center}
\includegraphics[width=8cm]{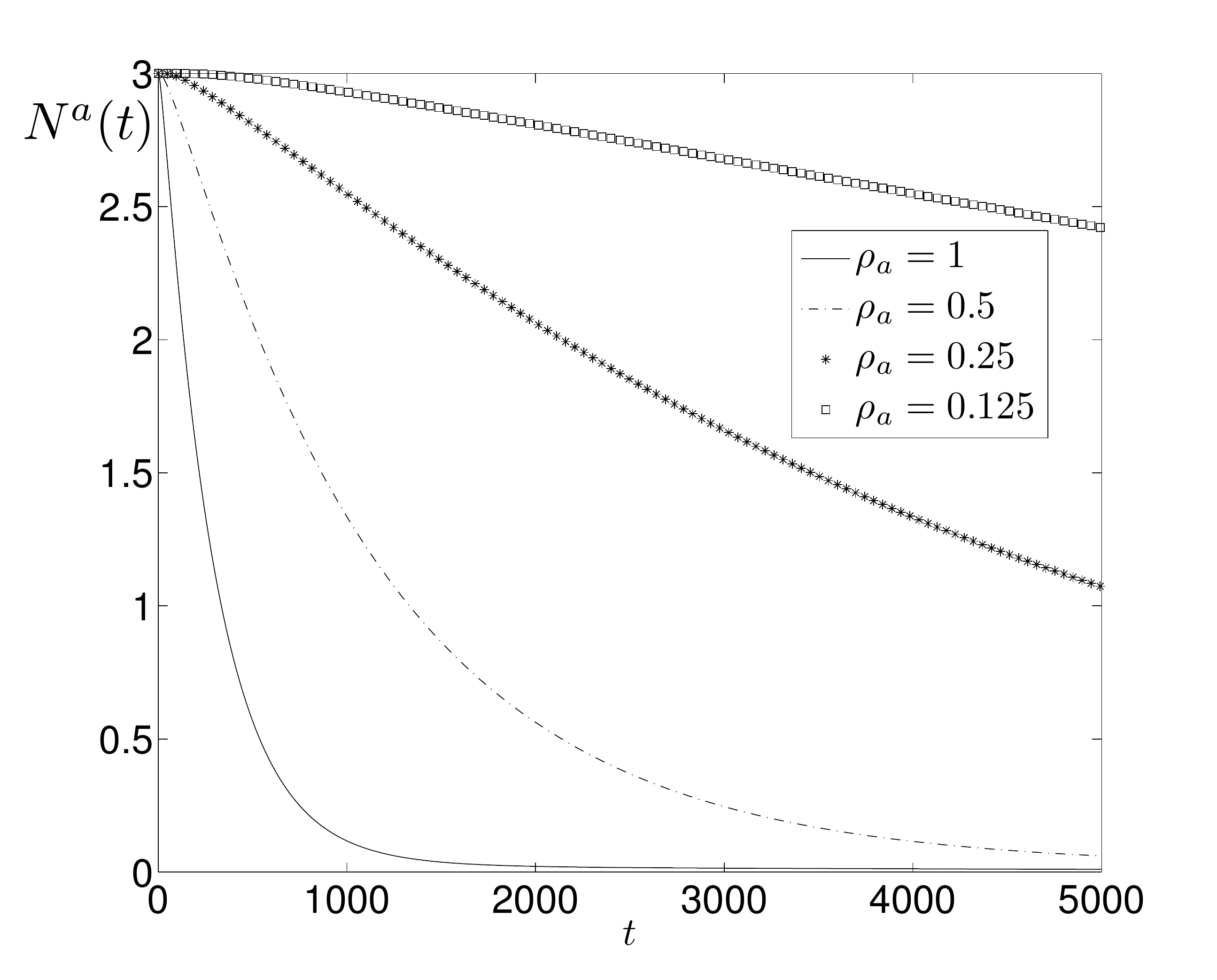}
\vspace*{-0.25cm}\caption{\textbf{Setting $S_1$}: the total density $N^a(t)$ of the population $\Pc_a$ is plotted for $\Delta T=0.08, N^a_{trh}=10^{-5},\omega^a_{\alpha}=1\quad \forall \alpha $,
 and different values of the parameter $\rho_a$.
By varying the parameter $\rho_a$ we tune the speed of the mobility of the population $\Pc_a$.
As expected, by decreasing $\rho_a$ we obtain a slowing-down effect for the population $\Pc_a$.}
\label{p1_u1_vargamma}
\end{center}
\end{figure}
\begin{figure}
\begin{center}
\includegraphics[width=8cm]{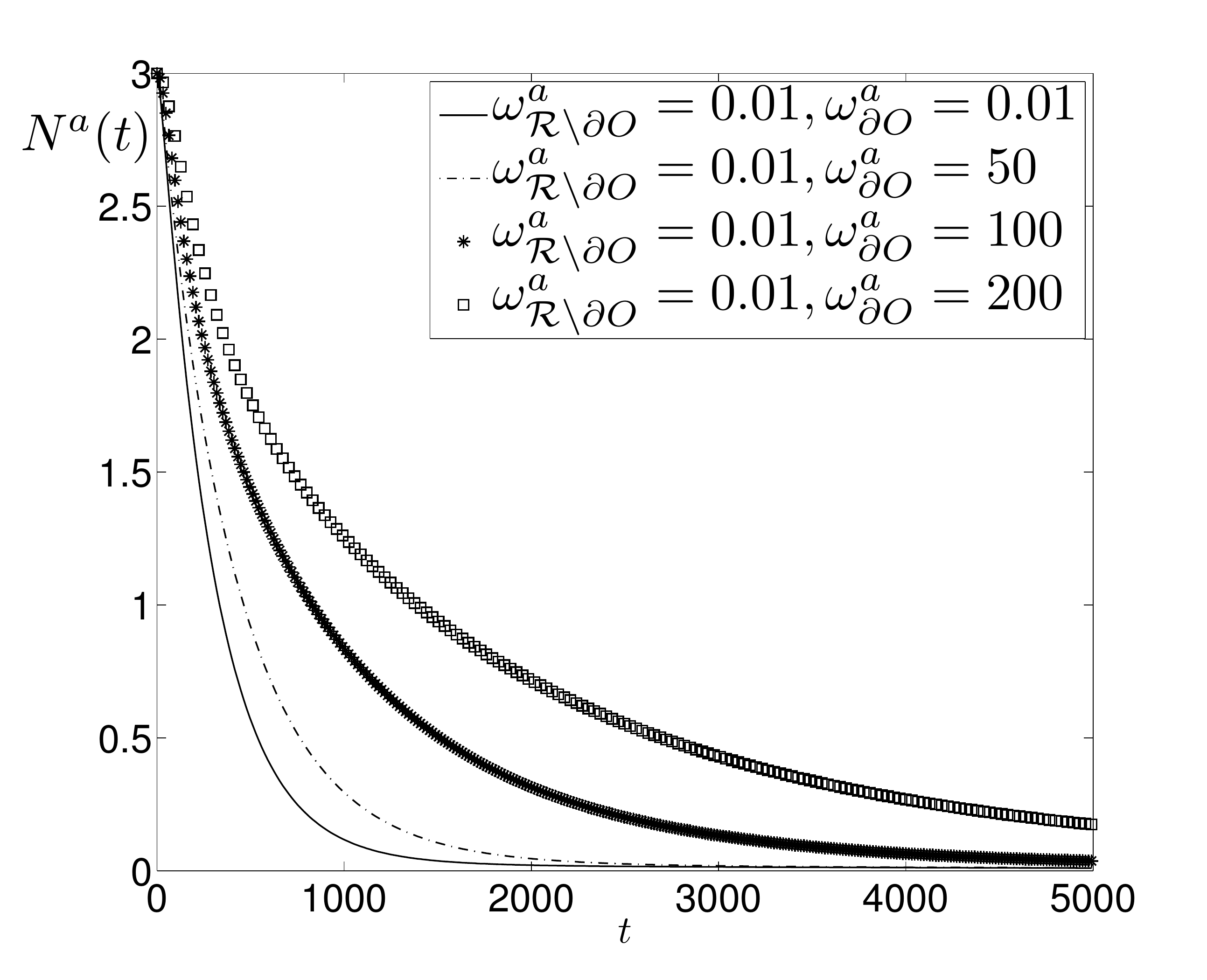}
\vspace*{-0.25cm}\caption{\textbf{Setting $S_1$}: the total density $N^a(t)$ of the population $\Pc_a$
is plotted for $\Delta T=0.08, N^a_{trh}=10^{-5},\rho_a=1$ and different values of the parameter $\omega^a_{\alpha}$.
 An increasing value of $\omega^a_{\partial O}$ means more staticity of the population $\Pc_a$ in $\partial O$ and, in fact, the population slows down.}
\label{p1_u1_varomega}
\end{center}
\end{figure}

\begin{figure}
\begin{center}
\includegraphics[width=8cm]{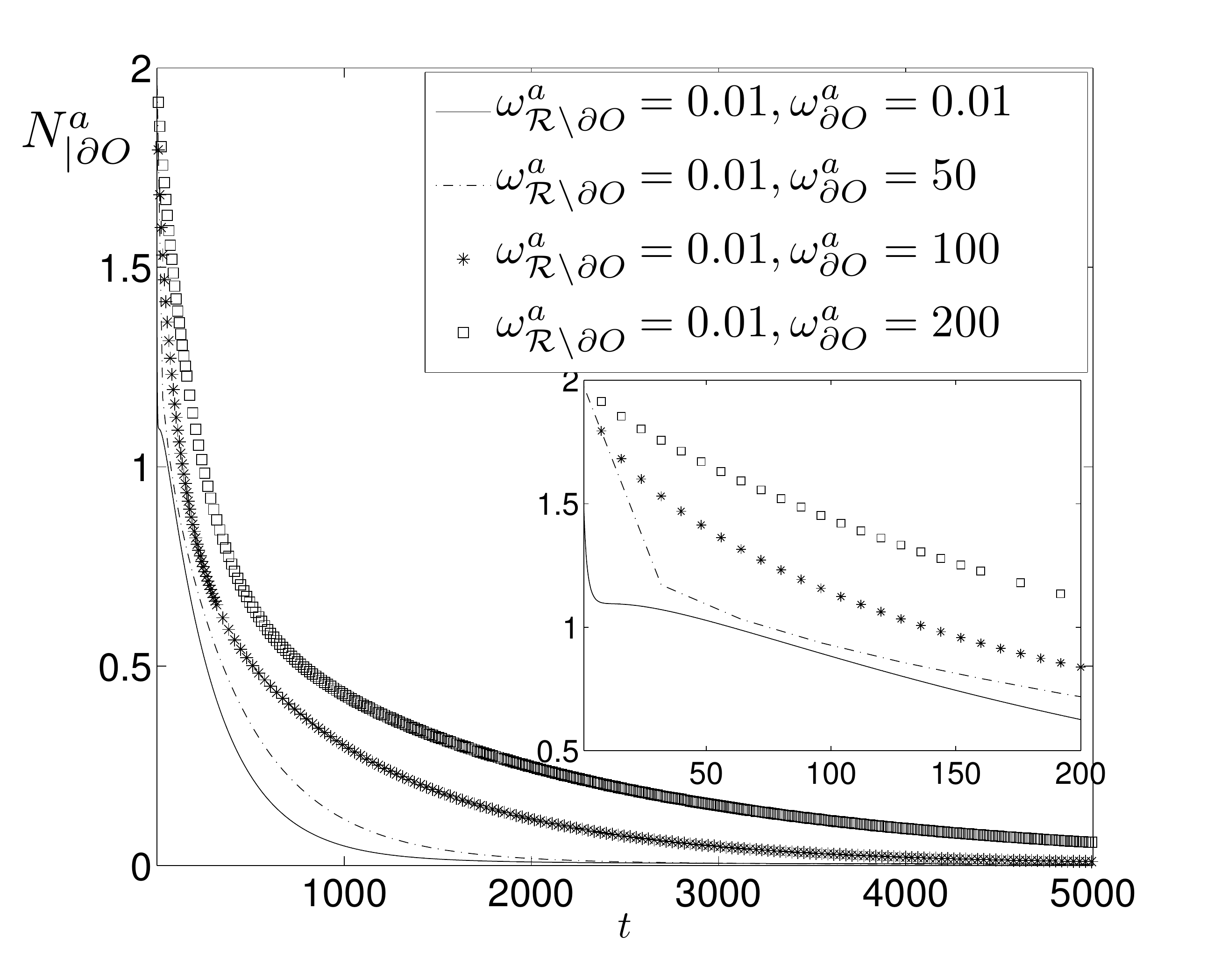}
\vspace*{-0.25cm}\caption{\textbf{Setting $S_1$}: the total density $N^a_{|\partial O}(t)$ of the population $\Pc_a$ evaluated in the region $\partial O$
is plotted for $\Delta T=0.08, N^a_{trh}=10^{-5},\rho_a=1$ and different values of the parameter $\omega^a_{\alpha}$.
 In the inset the density $N^a_{|\partial O}(t)$
is shown up to time $t=200$. An increasing value of $\omega^a_{\partial O}$ means more staticity of the population $\Pc_a$ in $\partial O$ and, in fact, the population slows down.}
\label{p1_u1_varomega_PO}
\end{center}
\end{figure}

A single population can be easily described by our Hamiltonian simply assuming that, at $t=0$, the vector ${\bf
n}^b=(n_1^b,n_2^b,\ldots,n_{L^2}^b)$ introduced before is all made by zeros. Of course, it is also natural, in this case (even if not strictly necessary), to fix $\lambda_\alpha=0$, $\forall\,\alpha$.
We assume here that $\Pc_a$ is originally
located as shown
in Figure \ref{setup_p1_u1_po}.

Our general procedure requires, first of all, two input ingredients: the threshold value $N^a_{trh}$ and  the time interval $\Delta T$ we have introduced above. To fix their values, we have
performed several simulations, some of them reported in Figures \ref{p1_u1_vartrh} and \ref{p1_u1_vardt}.

In Figure \ref{p1_u1_vartrh}
we plot the different densities of $\Pc_a$ inside $\R$ for different values of $N^a_{trh}$ with fixed $\Delta T=0.08$.
We see that the numerical results for $N^a_{trh}=10^{-5}$ and $N^a_{trh}=10^{-4}$ are almost indistinguishable. Hence, at least for the given value of $\Delta T$ and for this setting,
$N^a_{trh}=10^{-5}$ can be considered a good threshold value.
  It is interesting to observe that there is evidence that after a time $\tilde t$ (depending on $N^a_{trh}$), the temporal variation of the density $N^a(t)$
is very low ($dN^a(t)/dt\approx0$ for $t>\tilde t$). For example for $N^a_{trh}=10^{-1}$ in practice  $\tilde t\approx 0$ , while for $N^a_{trh}=10^{-2}$ we find
$\tilde t\approx 1000$.
Figure \ref{p1_u1_vardt} is deduced varying the value of $\Delta T$ while keeping fixed the optimal value of $N^a_{trh}$. We see that increasing $\Delta T$
 improves convergence  to zero of the density of $\Pc_a$. This can be easily understood: when $\Delta T$ increases,
a larger amount of population can accumulate in $U$ and, because of our exit strategy, this larger amount quite likely exceeds $N^a_{trh}$ and, therefore, it is removed from $\R$ after $\Delta T$.
 For this reason, we do not want $\Delta T$ to be too large, to prevent all the populations to disappear {in few \emph{time steps}}.
 On the other hand, we can not even take $\Delta T$ to be too small, since otherwise the numerical computations become very slow. After some tests, we have found a good compromise by fixing  $\Delta T=0.08$.

In Figure \ref{p1_u1_vargamma} we plot the densities of $\Pc_a$ for different values of the parameter $\rho_a$ in \eqref{function_ya}, for the above
choices of $N^a_{trh}$ and $\Delta T$. It is evident that lower values of $\rho_a$ do not help the exit from the room, and $N^a(t)$ does not
decrease as fast as it happens for larger values of $\rho_a$. For this reason, in agreement with what deduced in Ref.\cite{ff}, we
interpret $\rho_a$ as the \emph{mobility} of $\Pc_a$: the higher its value, the faster $\Pc_a$ moves.

In Refs.~\cite{ff,bagbook} the role of the parameters of $H_0$ are shown to be related to a sort of inertia of those degrees of freedom to which they refer.
For this reason, and to validate further this interpretation, we have considered here two different values of $\omega^a$, an higher one in the cells
in $\partial O$ and a smaller one in the rest of $\R$. The results are given in Figures \ref{p1_u1_varomega}-\ref{p1_u1_varomega_PO} and confirm our previous
interpretation. It could be worth stressing that (i) big differences are needed in order to observe different behaviors; (ii) if we just
increases the value of $\omega^a$ in all of $\R$, nothing particular changes: it is the gradient between two different regions which is
responsible for this effect. Similar conclusions have been deduced also in very different contexts, see Ref.~\cite{bagbook}.

We conclude that, to decrease the time needed by $\Pc_a$ to leave $\R$, it is convenient to adopt the following strategy: (i) clarify, from each possible cell in $\R$, which direction goes straight to the exit (this can be done, for instance, clearly indicating the exit); (ii) try to increase the mobility of $\Pc_a$ (for instance, removing all the unnecessary obstacles in $\R$) (iii) try to keep a certain homogeneity in the accessible part of $\R$. Notice that, while (i) and (ii) are quite expected results, (iii) is not evident a priori, and its implementation could help improving the exit strategies. Of course, changing the topology of the room and the initial distribution of $\Pc_a$ would change the numerical outputs, but not our main conclusions.

\subsection{Two populations, one exit}

In this case neither ${\bf n}^a$ nor ${\bf n}^b$ are $\bf 0$: in fact, \emph{at least} one cell of $\R$ is surely occupied by \emph{at least} one population.
 We will now consider separately two different situations: in the first one $\Pc_a$ and $\Pc_b$ do not
 mutually interact. This is compatible with the fact that the densities of the two populations are small in $\R$ already at $t=0$.
 In the second situation, relevant in the case of higher densities, $\Pc_a$ and $\Pc_b$ do interact, since elements of $\Pc_a$ and $\Pc_b$ are more likely to meet while they are trying to reach the exit.

\subsubsection{Without interaction  (Setting $S_2^{ld}$)}\label{sect_s2ld}

\begin{figure}
\begin{center}
\includegraphics[width=8cm]{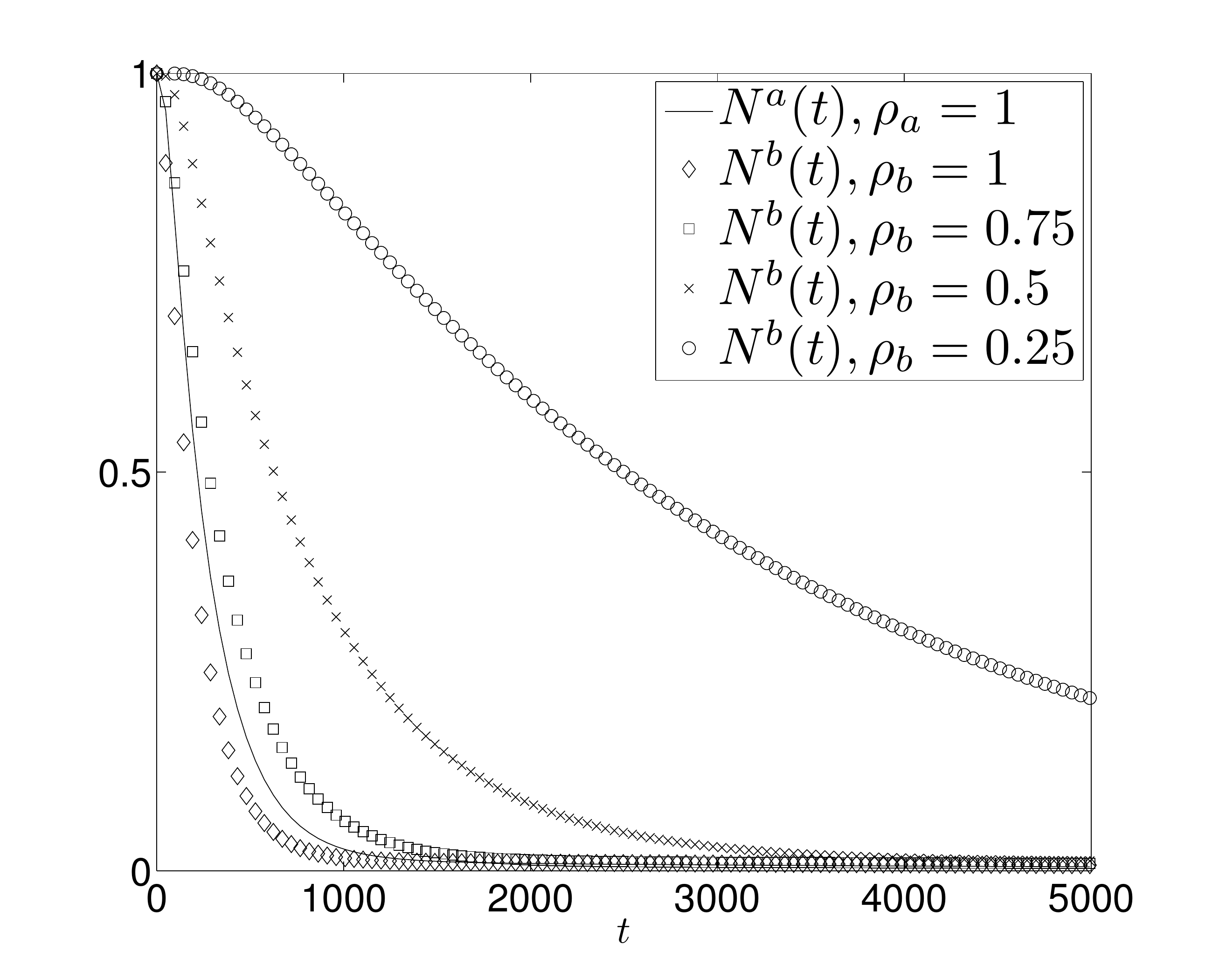}
\vspace*{0cm}\caption{\textbf{Setting $S_2^{ld}$}: the total densities $N^a(t),N^b(t)$ are shown for $\Delta T=0.08, N^a_{trh}=N^b_{trh}=10^{-5},
\omega^a_{\alpha}=\omega^b_{\alpha}=1,\, \forall \alpha , \rho_a=1$,
 and for different values of the parameter $\rho_b$.
As in the previous section, by varying the parameter $\rho_b$ we tune the speed of the population $\Pc_b$, making it faster or slower
than $\Pc_a$.}
\label{2p_1u_td_vargammab}
\end{center}
\end{figure}

We consider here the case in which the two populations $\Pc_a$ and $\Pc_b$ are originally
located as shown in Figure \ref{setup_p2_u1_po} and they have the same initial density, $N^a(0)=N^b(0)=1$. The suffix $ld$ in  $S_2^{ld}$ stands for \emph{low density}.
As the total densities are small, we neglect here the effect of the interaction between the two populations
($\lambda_{\alpha}=0$, $\forall \alpha \in \R$), since an interaction is more likely to occur
when both  densities are high, so that $\Pc_a$ and $\Pc_b$ can more likely meet somewhere in $\R$ during their motion. To determine the values of the coefficients $p^{(a)}_{\alpha,\beta}$ and $p^{(b)}_{\alpha,\beta}$ in the Hamiltonian we apply
the same procedure already outlined in the previous
section for both populations: of course, since the two populations have different
initial conditions then $p^{(a)}_{\alpha,\beta}$ and $p^{(b)}_{\alpha,\beta}$ are not identical. This is also a consequence of the different values of the mobilities we consider for $\Pc_a$ and $\Pc_b$. As in the previous section,
we fix $\Delta T=0.08$ and $N^a_{trh}=N^b_{trh}=10^{-5}$. We further take $\omega^a_{\alpha}=\omega^b_{\alpha}=1$, for all $\alpha$, $\rho_a=1$, and we consider the following values of $\rho_b$: $\rho_b=1,0.75,0.5,0.25$, in order to describe different mobilities for (the aged population) $\Pc_b$. \\
We should expect that if both the populations have the same $inertia$ and the same  $mobilities$\footnote{Some snapshots of the densities $N_{\alpha}^a(t)$, $N_{\alpha}^b(t)$ in $\R$ are shown in the Appendix, see Figures \ref{m21a}-\ref{m21b}.}  ($\omega^a_{\alpha}=\omega^b_{\alpha}$
 and $\rho_a=\rho_b$), then $N^b(t)$ should decay to zero faster than $N^a(t)$: in fact, the minimal path going from the cell occupied by $\Pc_b$ at $t=0$
to the exit cell $U$ has length 9, while for $\Pc_a$ the analogous path has length 10. In Figure \ref{2p_1u_td_vargammab} there is evidence of
this behavior, while, if we decrease $\rho_b$, we slow down $\Pc_b$, and this allows $N^a(t)$ to go to zero faster than $N^b(t)$: not surprisingly, the fact that a population leaves the room faster than the other is not only a matter of \emph{where the populations were originally located}, but also of \emph{how fast they can move}.

\begin{figure}

\subfigure[$N^a(t)$]{\vspace*{-0.75cm}\hspace*{-1.75cm}\includegraphics[width=8cm]{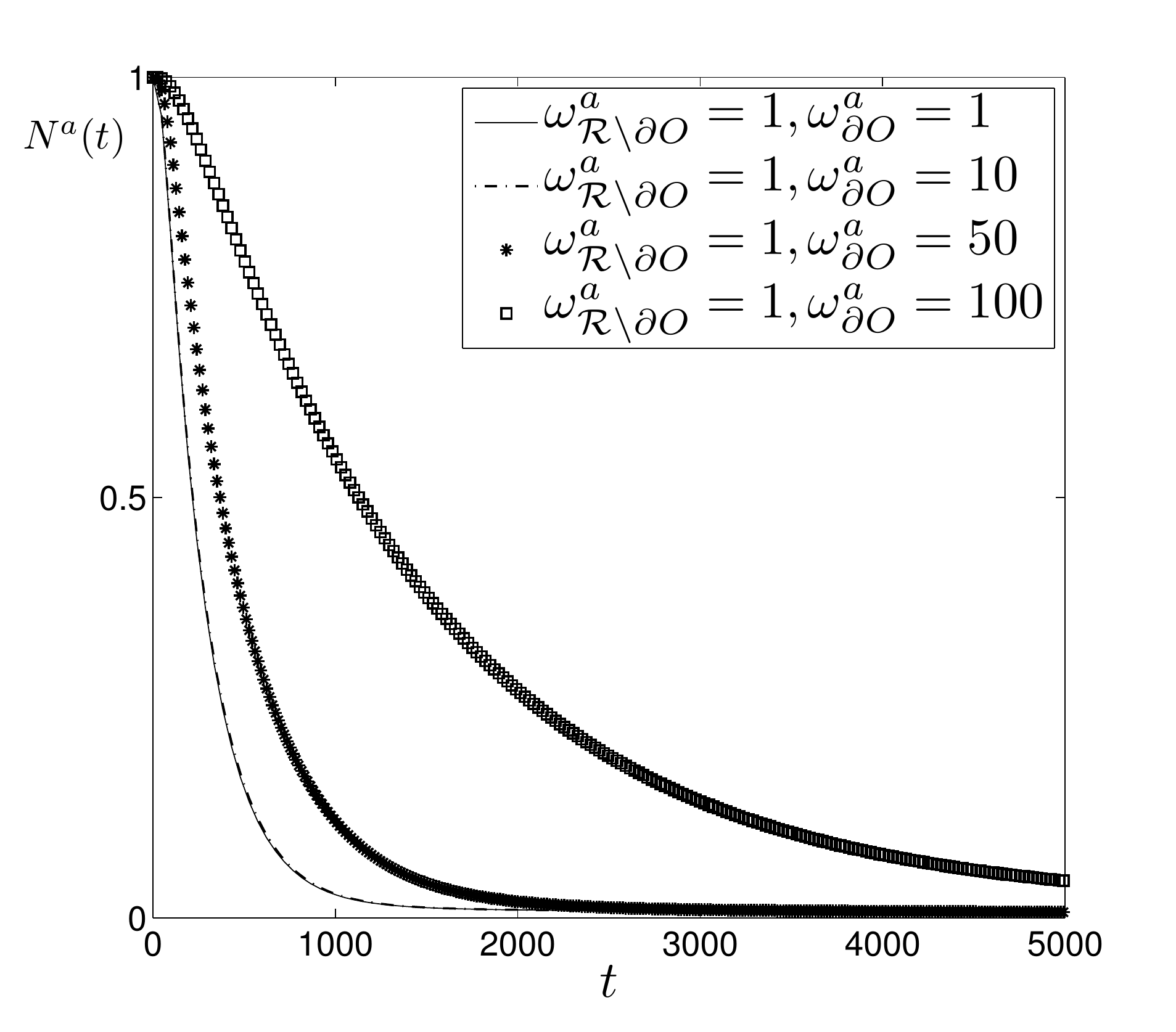}}
\subfigure[$N^b(t)$]{\vspace*{-0.75cm}\hspace*{-0.5cm}\includegraphics[width=8cm]{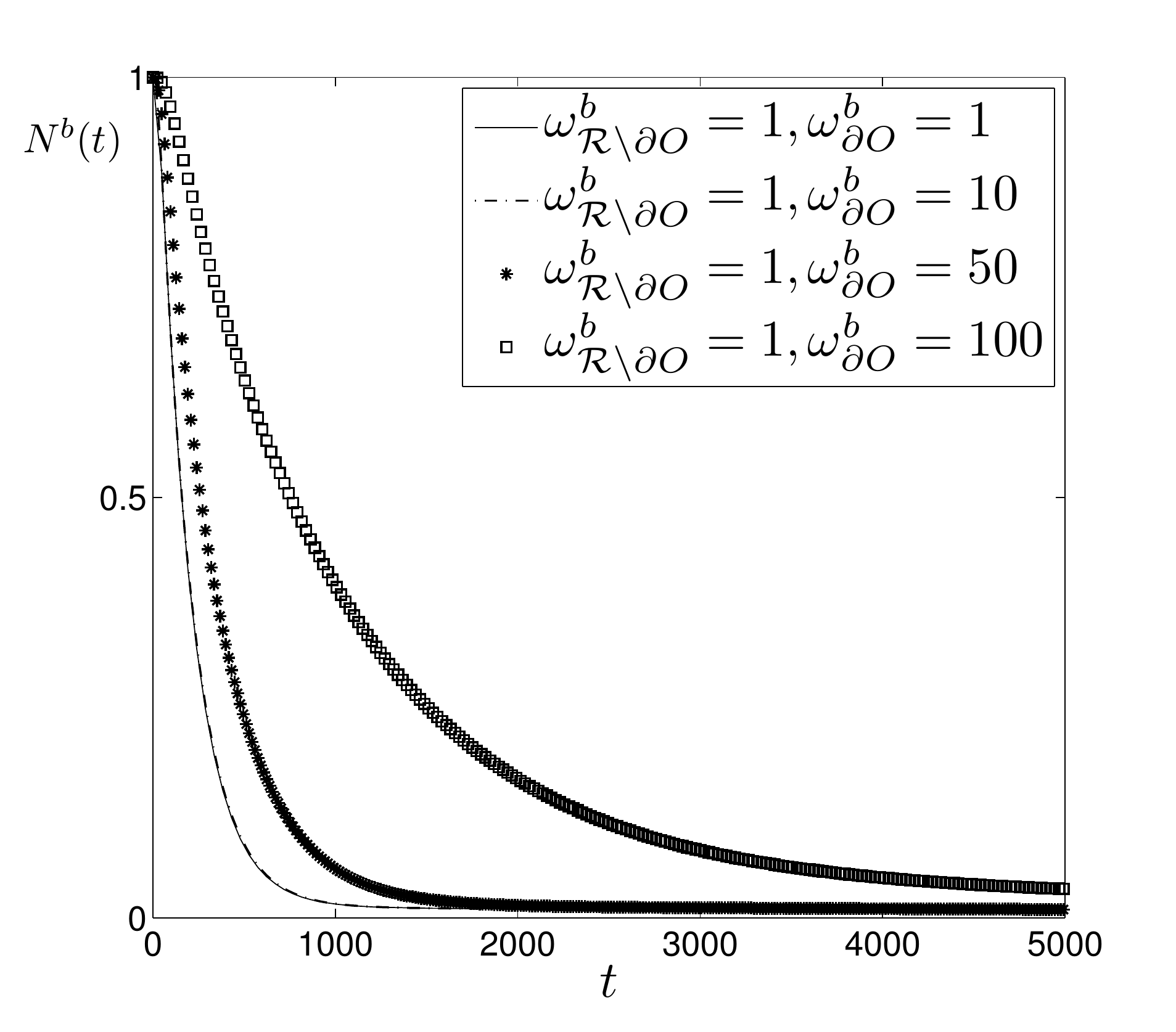}}
\vspace*{-0.5cm}\caption{\textbf{Setting $S_2^{ld}$}: the total densities $N^a(t),N^b(t)$ are shown for $\Delta T=0.08, N^a_{trh}=N^b_{trh}=10^{-5},
 \rho_a=\rho_b=1$ and different values of $\omega^a_{\alpha},\omega^b_{\alpha}$. A strong inhomogeneity within $\R$ has the effect to slow down both populations, and in particular increasing values
of $\omega^a,\omega^b$ in $\partial O$ means more staticity in $\partial O$ for both the populations.
For this kind of configuration the case $\omega^a_{\partial O}=\omega^b_{\partial O}=1$ and $\omega^a_{\partial O}=\omega^b_{\partial O}=10$ are almost indistinguishable, while
for $\omega^a_{\partial O}=\omega^b_{\partial O}=50,100$ a quite evident variation is visible with respect to the case $\omega^a_{\partial O}=\omega^b_{\partial O}=1$. }
\label{2p_1u_na_varomega}

\end{figure}

Also for this setting we consider the effect of a strong inhomogeneity of $\omega^{a}_\alpha$ and $\omega^{b}_\alpha$ in $\R$,
\emph{i.e.}, we consider the $\omega^{a,b}_\alpha$ inside $\partial O$  much bigger than outside. The results are shown in Figures \ref{2p_1u_na_varomega}(a)-\ref{2p_1u_na_varomega}(b),
where the densities $N^a(t),N^b(t)$ are plotted, respectively: as previously seen in the Setting $S_1$ (see Figures \ref{p1_u1_varomega}-\ref{p1_u1_varomega_PO}), increasing values
of $\omega^{a,b}_\alpha$ inside $\partial O$ corresponds to more staticity of the populations in the region $\partial O$, and, therefore, $N^a(t),N^b(t)$ decay slower. Again, a gradient
of the $\omega^{a,b}_\alpha$ between
two different regions is needed to obtain remarkable effects, and the suggestion is that if we want an optimal escape strategy we should avoid or minimize this gradient in $\R$.

\subsubsection{With interaction  (Setting $S_2^{hd}$)}
\begin{figure}
\begin{center}
\includegraphics[width=8cm]{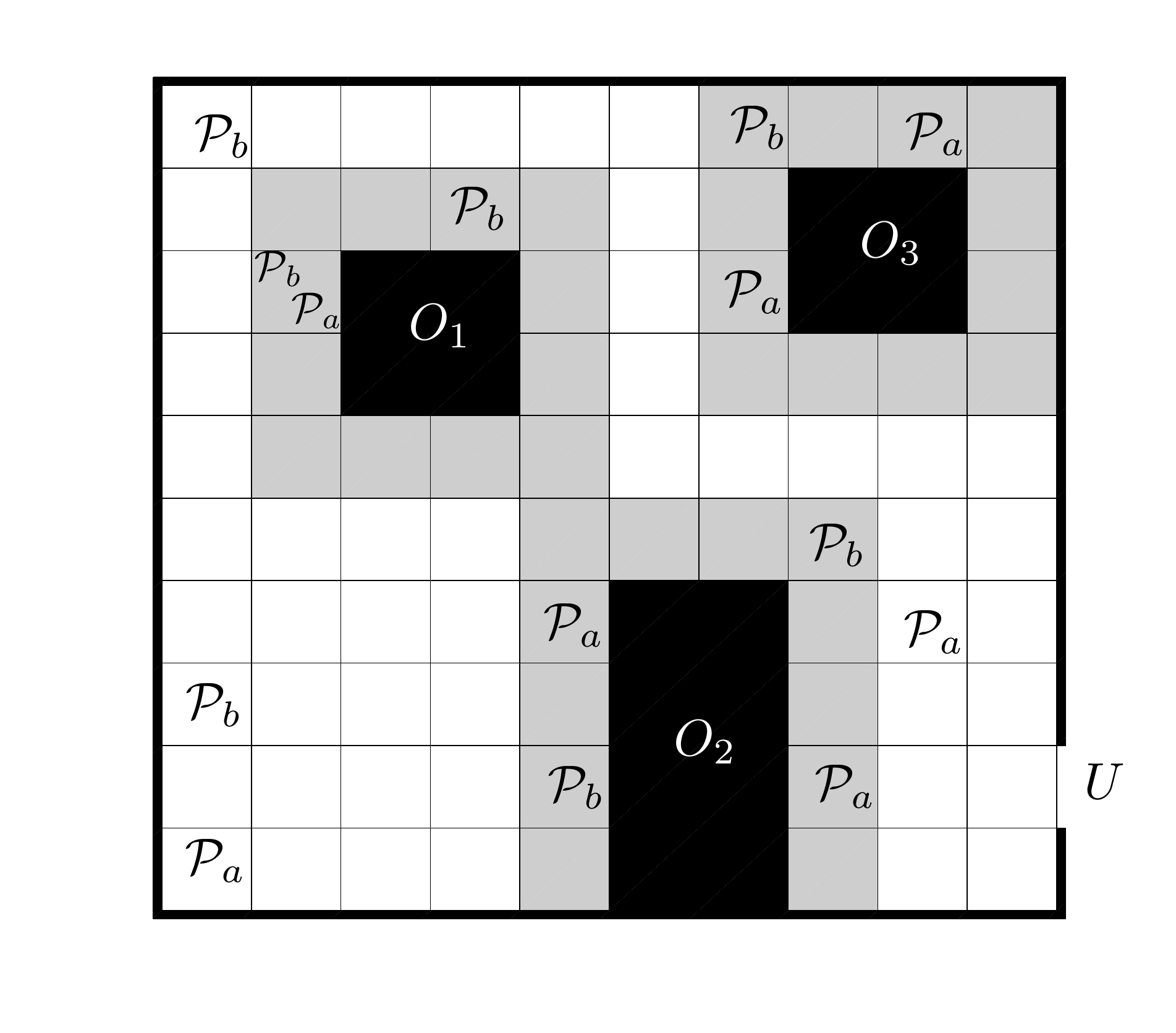}
\vspace*{-0.25cm}
\caption{ \textbf{Setting $S_2^{hd}$}: at $t=0$ the population $\Pc_a$ is located in the cells $(1,1),(8,2),(5,4),(9,4),(2,8),(7,8),(9,10)$, while
$\Pc_b$ is located in the cells $(5,2),(1,3),(8,5),(2,7),(2,8),(4,9), (1,10),(7,10)$. The
 black cells $O_1,O_2,O_3$ represents the obstacles, and the exit cell $U$ is located at (11,2).}
\label{setup_p2_u1_po_hd}
\end{center}
\end{figure}

\begin{figure}
\subfigure[$\lambda_{\alpha}=0.05$]{\hspace*{-1.75cm}\includegraphics[width=8cm]{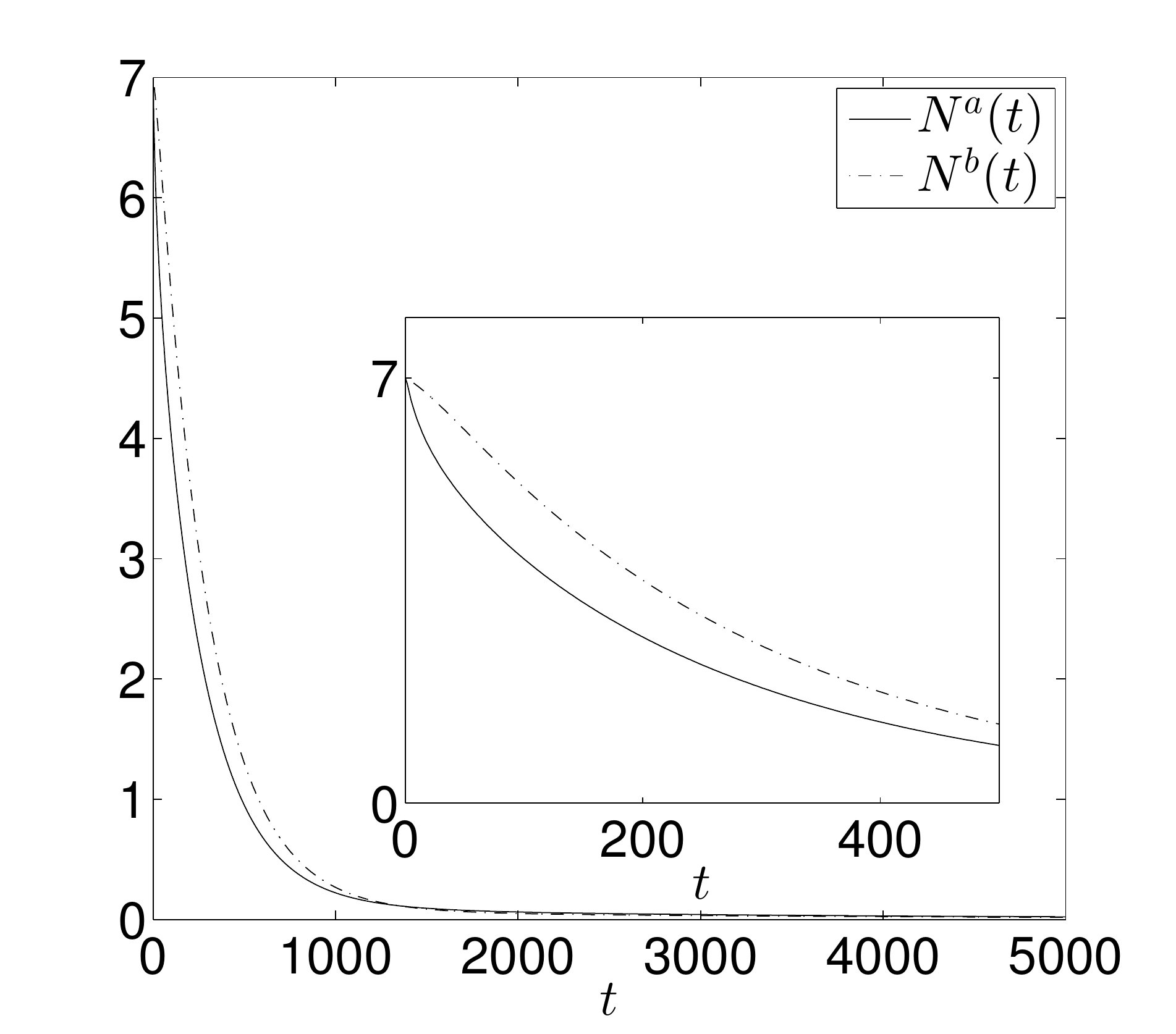} \label{2p_1u_td_hd_l005}}
\subfigure[$\lambda_{\alpha}=0.5$]{\hspace*{-0.5cm}\includegraphics[width=8cm]{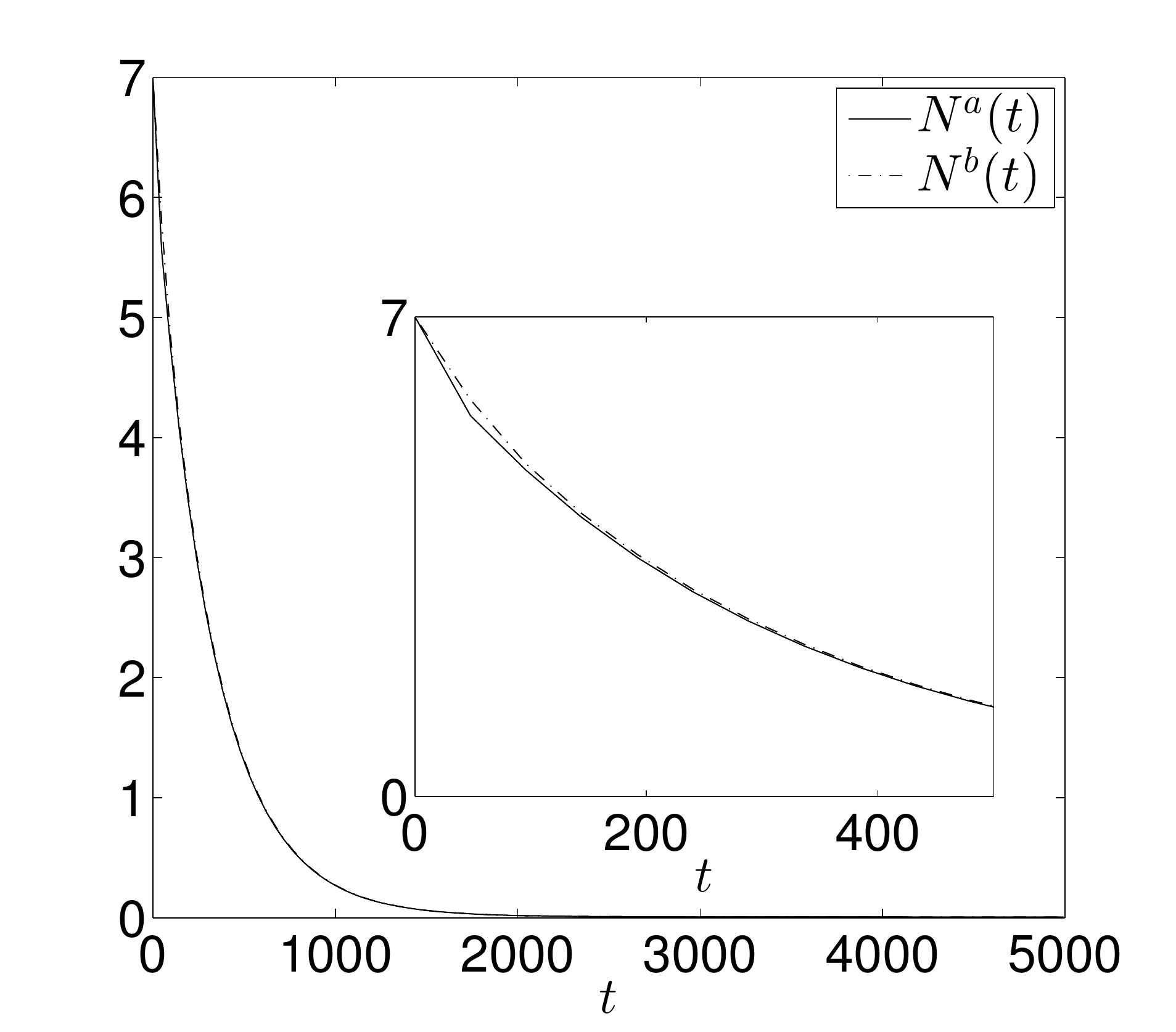}\label{2p_1u_td_hd_l05}}
\subfigure[$\lambda_{\alpha}=3$]{\vspace*{-0.5cm}\hspace*{-1.75cm}\includegraphics[width=8cm]{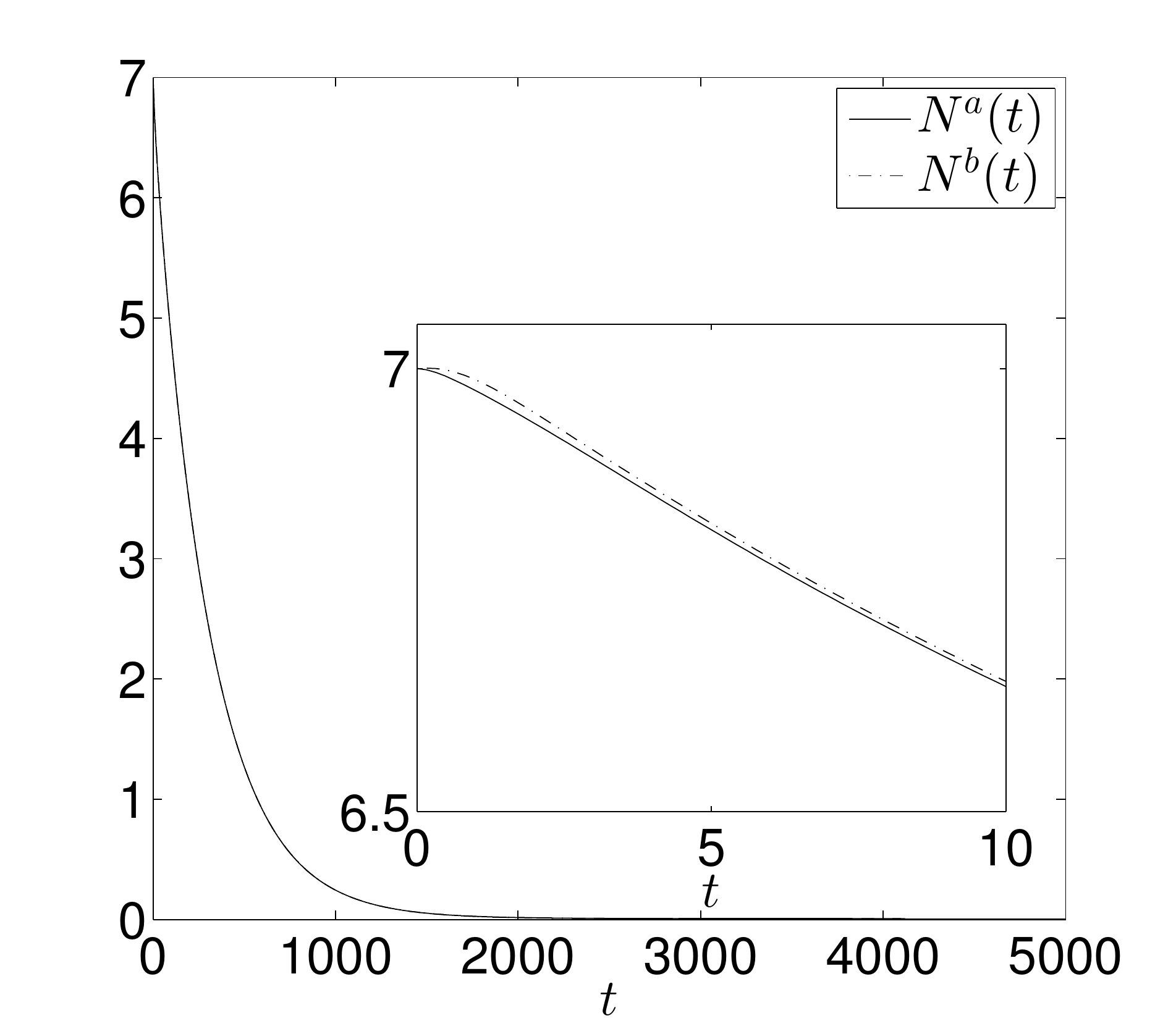} \label{2p_1u_td_hd_l3}}
\subfigure[$\lambda_{\alpha}=7$]{\hspace*{-0.5cm}\includegraphics[width=8cm]{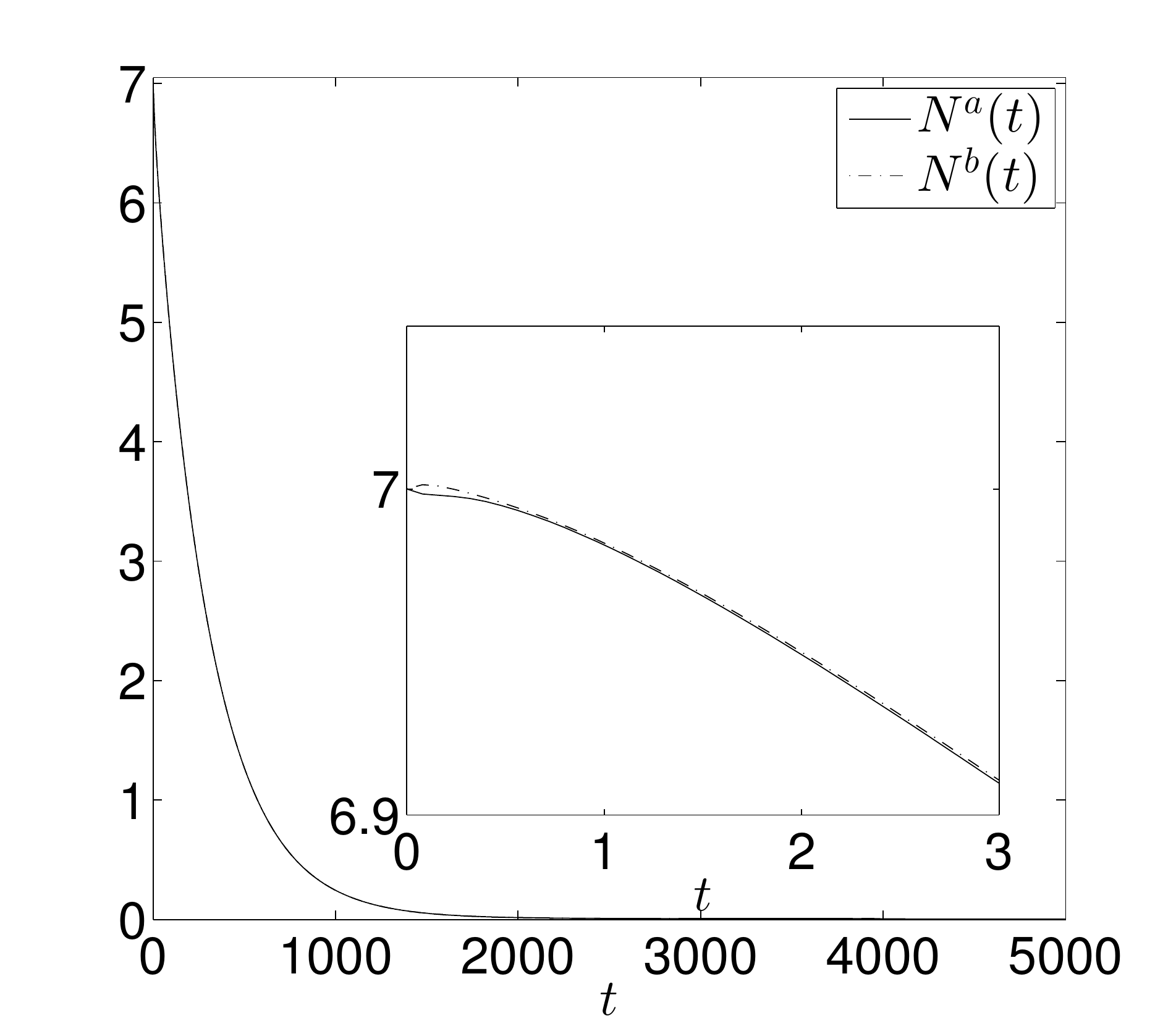} \label{2p_1u_td_hd_l7}}
\vspace*{-0.5cm}\caption{ \textbf{Setting $S_2^{hd}$}: the total densities $N^a(t),N^b(t)$ are shown for $\Delta T=0.08, N^a_{trh}=N^b_{trh}=10^{-5},
\omega^a_{\alpha}=\omega^b_{\alpha}=1\, \forall \alpha, \rho_a=\rho_b=1$,
 and different values of the parameter $\lambda_{\alpha}$. In this configuration the population $\Pc_a$ is \emph{globally} closer to the exit cell than $\Pc_b$
and, in fact, $N^a(t)$ decays faster than $N^b(t)$. However if we increase $\lambda_{\alpha}$ we obtain the effect to make $\Pc_a$ slower while $\Pc_b$ becomes faster
(see also Figures \ref{2p_1u_td_hd_na_ALL}- \ref{2p_1u_td_hd_nb_ALL}). }
\end{figure}

\begin{figure}
\subfigure[$N^a(t)$]{\hspace*{-1.5cm}\includegraphics[width=8cm]{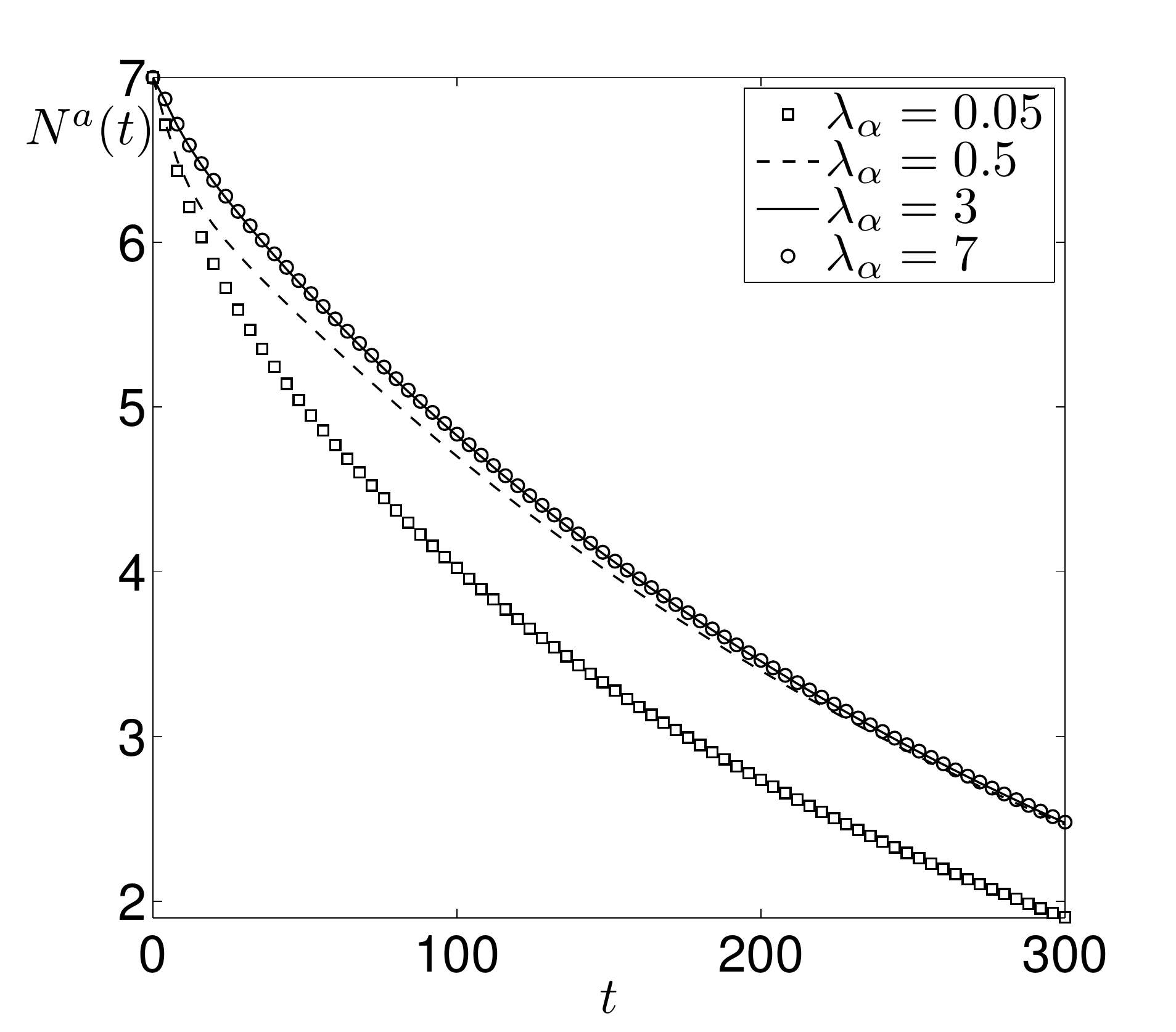} \label{2p_1u_td_hd_na_ALL}}
\subfigure[$N^b(t)$]{\hspace*{-0.5cm}\includegraphics[width=8cm]{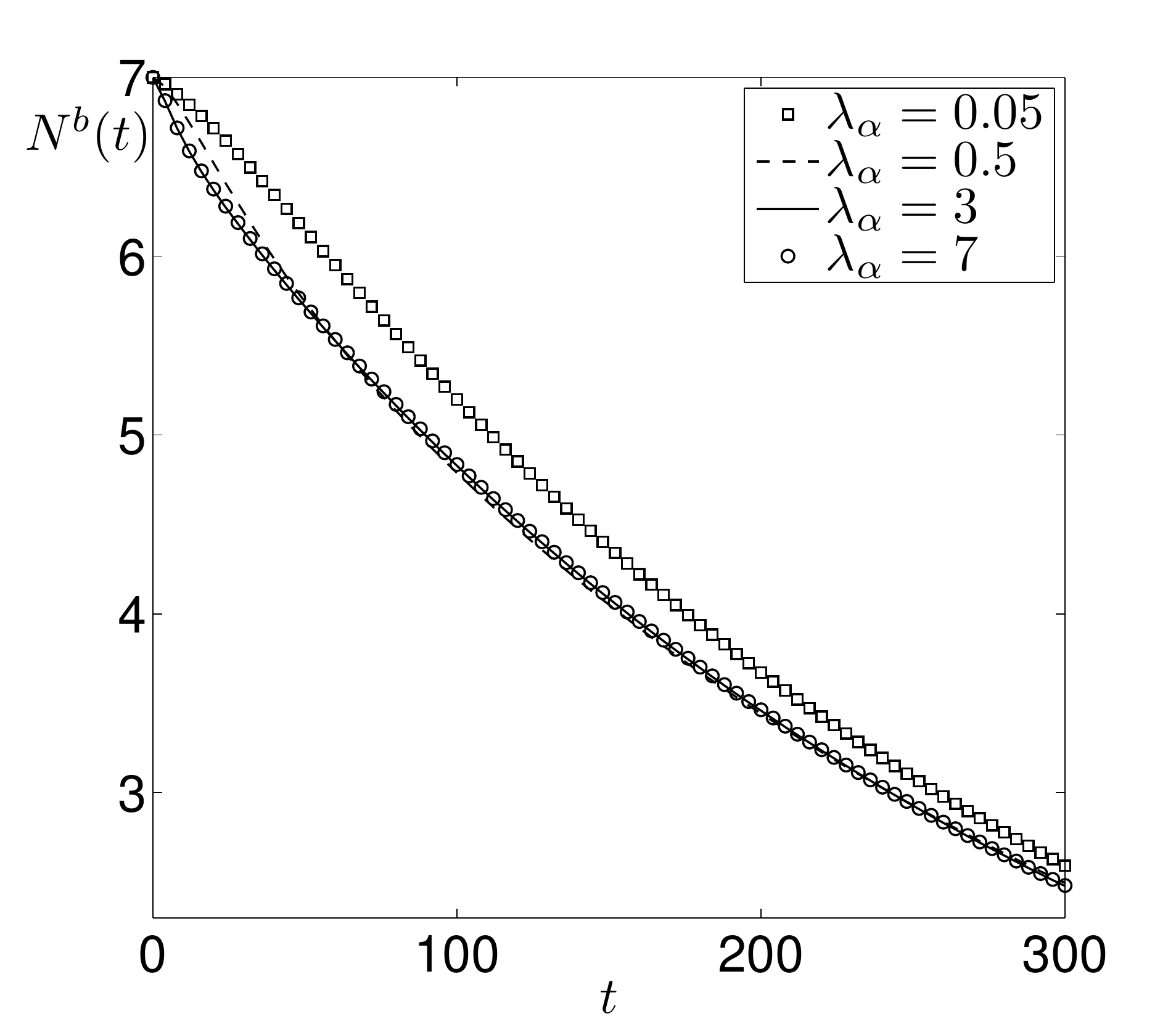}\label{2p_1u_td_hd_nb_ALL}}

\caption{ \textbf{Setting $S_2^{hd}$}: the total densities $N^a(t)$ (on the left) and $N^b(t)$ (on the right) are shown for $\Delta T=0.08, N^a_{trh}=N^b_{trh}=10^{-5},
\omega^a_{\alpha}=\omega^b_{\alpha}=1\quad \forall \alpha , \rho_a=\rho_b=1$,
 and different values of the parameter $\lambda_{\alpha}$. See also Figure \ref{2p_1u_td_hd_l005}-\ref{2p_1u_td_hd_l7}. }
\end{figure}

Suppose now that the populations $\Pc_a$ and $\Pc_b$ are originally
located as shown in Figure \ref{setup_p2_u1_po_hd}, so that they have the same initial density, $N^a(0)=N^b(0)=7$, significantly higher than in the previous situation. Here, the suffix $hd$ stands for \emph{high density}.
In this case the interaction between the two populations can quite reasonably occur, while the two populations try to reach the exit, and this has been considered by taking the \emph{local} interaction parameters $\lambda_\alpha$ different from zero.
The population $\Pc_a$ is, in this configuration, $globally$ closer to the exit cell $U$ than $\Pc_b$, as the sum of all the minimal paths
from the cells initially occupied by $\Pc_a$ to $U$ is 43, while that for $\Pc_b$ is 55.
We have performed several simulations by varying the $interaction$ parameter $\lambda_{\alpha}$, and the results are shown
in Figures \ref{2p_1u_td_hd_l005}-\ref{2p_1u_td_hd_l7} and \ref{2p_1u_td_hd_na_ALL}-\ref{2p_1u_td_hd_nb_ALL}.
These results shown that when the value of $\lambda_\alpha$ is small enough, for instance when $\lambda_{\alpha}=0.05$ for all $\alpha$,  the effect of the interaction is essentially negligible,
and in fact $N^a(t)$ decays faster than $N^b(t)$, as expected in view of our previous analysis. On the other hand, for larger values of the interaction parameters,  $\lambda_{\alpha}=7$ for all $\alpha$,
 the densities $N^a(t)$ and $N^b(t)$ decrease with almost the same speed\footnote{In fact, this already happens for $\lambda_\alpha=0.5$, in correspondence of which we observe almost negligible differences between $N^a(t)$ and $N^b(t)$.}. Roughly speaking, if  we increase $\lambda_{\alpha}$ we obtain the effect to slow $\Pc_a$ down,
 while $\Pc_b$ speeds up. This is well clarified by Figures \ref{2p_1u_td_hd_na_ALL}-\ref{2p_1u_td_hd_nb_ALL}. Therefore, the interaction between the populations, at least for these large values of $\lambda_{\alpha}$,
acts like an $equalizer$ between the populations $\Pc_b$  and $\Pc_a$. In all these plots we have taken $\rho_a=\rho_b=1$, since the role of the mobility is already understood and there is no need to analyze it further.


\subsection{Two populations, two exits }

We will now consider a more general situation, in which two different exits allow the populations to move away from $\R$ and neither ${\bf n}^a$ nor ${\bf n}^b$ are $\bf 0$.
As in the previous section we will consider separately the low-densities and high-densities cases.

Due to the presence of two exit cells, we will modify \eqref{function_ya}-\eqref{function_Mj}  in this way (the details are given for $\Pc_a$):
suppose that $\Pc_a$ is initially located in the cells
$\alpha_1,\alpha_2,...,\alpha_n$, and the exit cells are $U_1,U_2,...,U_m$ then
\begin{eqnarray}
\tilde\gamma^{(a)}_{\alpha}= \max_{j=1...n,k=1...m}\left[\left(\frac{g_{jk}(\alpha)}{M_{jk}}\right)^{\sigma_a}\right], \label{function_ya2}\\
g_{jk}(\alpha)=(f_d(\alpha,\alpha_j)+f_d(\alpha,U_k))^{-1} \quad \forall j=1...n,k=1...m,\label{function_gj2}\\
M_{jk}=\max_{\alpha \in \R}g_{jk}(\alpha)  \quad \forall j=1...n,k=1...m.\label{function_Mj2}
\end{eqnarray}
Then we construct $p^{a}_{\alpha,\beta}$ following  \eqref{function_ga} and \eqref{pab}.
Once again,  $p^{(a)}_{\alpha,\beta}$ assumes its greatest values  if $\alpha$ and $\beta$ are along each
minimal paths going from the given cell $\alpha_j$ to some exit $U_k$,
while $p^{(a)}_{\alpha,\beta}$ decreases to zero when the direction from cell $\alpha$ to cell $\beta$ is not along a minimal path.

\subsubsection{Without interaction (Setting $S_3^{ld}$)}
\begin{figure}
\begin{center}
\includegraphics[width=8cm]{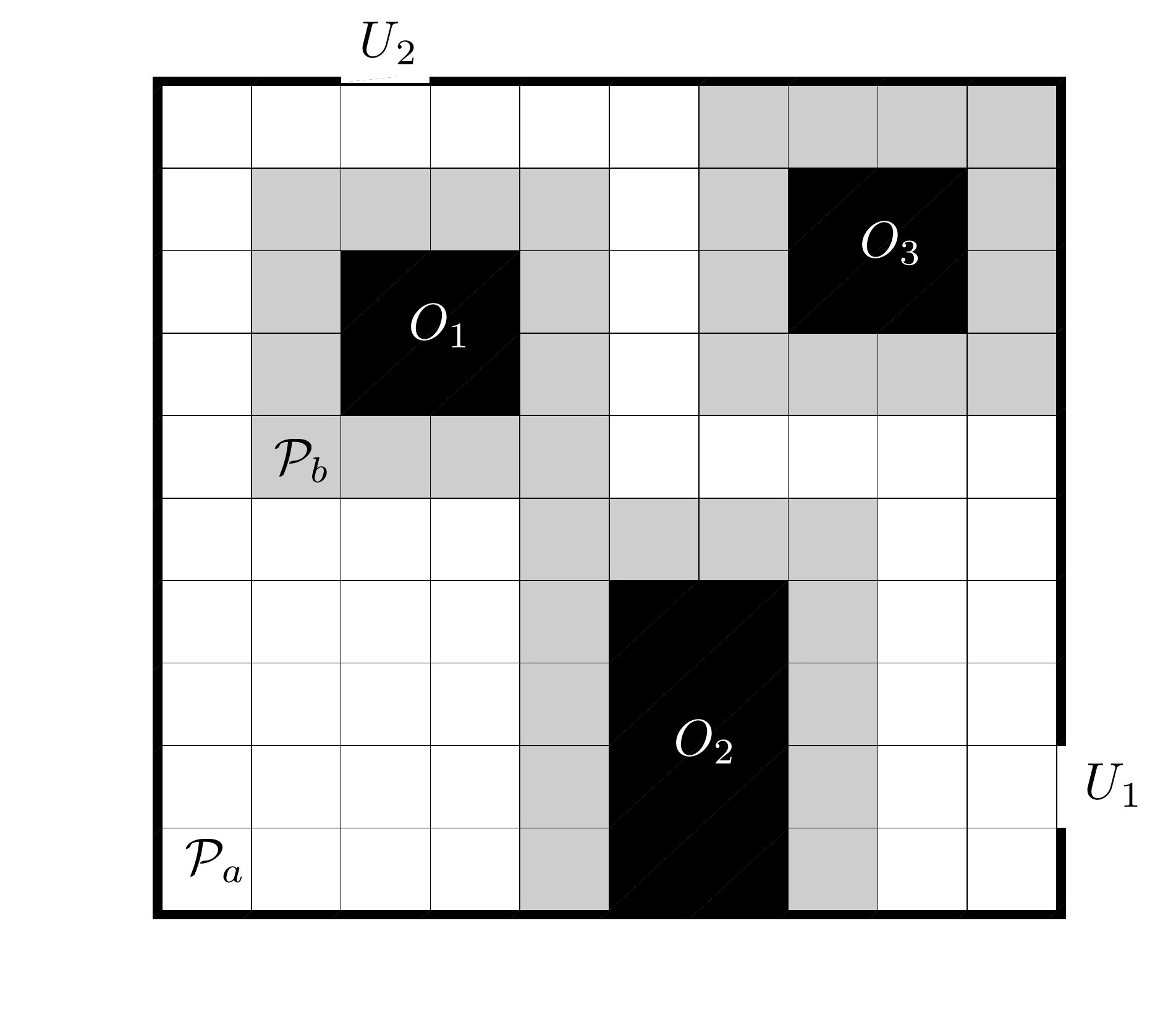}
\vspace*{-0.25cm}\caption{\textbf{Setting $S_3^{ld}$}: at $t=0$ the populations $\Pc_a$ and $\Pc_b$
are located in the cells $(1,1)$ and $(2,6)$ respectively. As usually the black cells represents the obstacles.
The exit cells $U_1, U_2$ are located at (11,2) and (3,11).}
\label{setup_p2_u2_po}
\end{center}
\end{figure}

\begin{figure}
\vspace*{-0.25cm}
\begin{center}
\includegraphics[width=8cm]{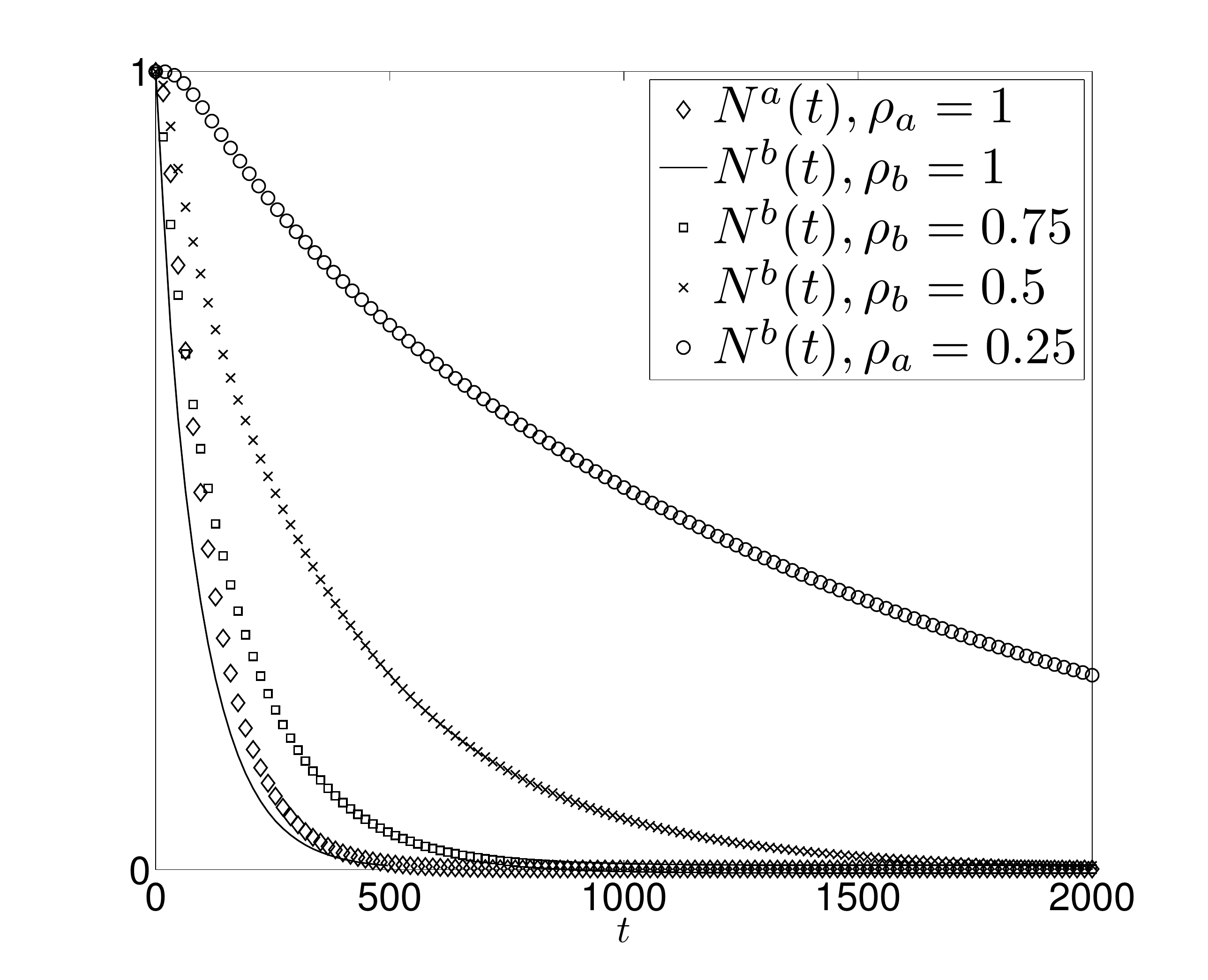}
\caption{\textbf{Setting $S_3^{ld}$}: the total densities $N^a(t),N^b(t)$ are shown for $\Delta T=0.08, N^a_{trh}=N^b_{trh}=10^{-5},
\omega^a_{\alpha}=\omega^b_{\alpha}=1\quad \forall \alpha, \rho_a=1$, and different values of the parameter $\rho_b$.
The population $\Pc_b$ is globally closer to the exit cells than $\Pc_a$ and therefore $N^b(t)$ decays faster than $N^a(t)$
if the populations have the same mobility. If we decrease $\rho_b$ we tune the speed of the population $\Pc_b$ making it slower,
and therefore $N^b(t)$ decays slower than $N^a(t)$. }
\label{2p_2u_td_vargamma}
\end{center}
\end{figure}

\begin{figure}
\vspace*{-1.0cm}
\begin{center}
\includegraphics[width=8cm]{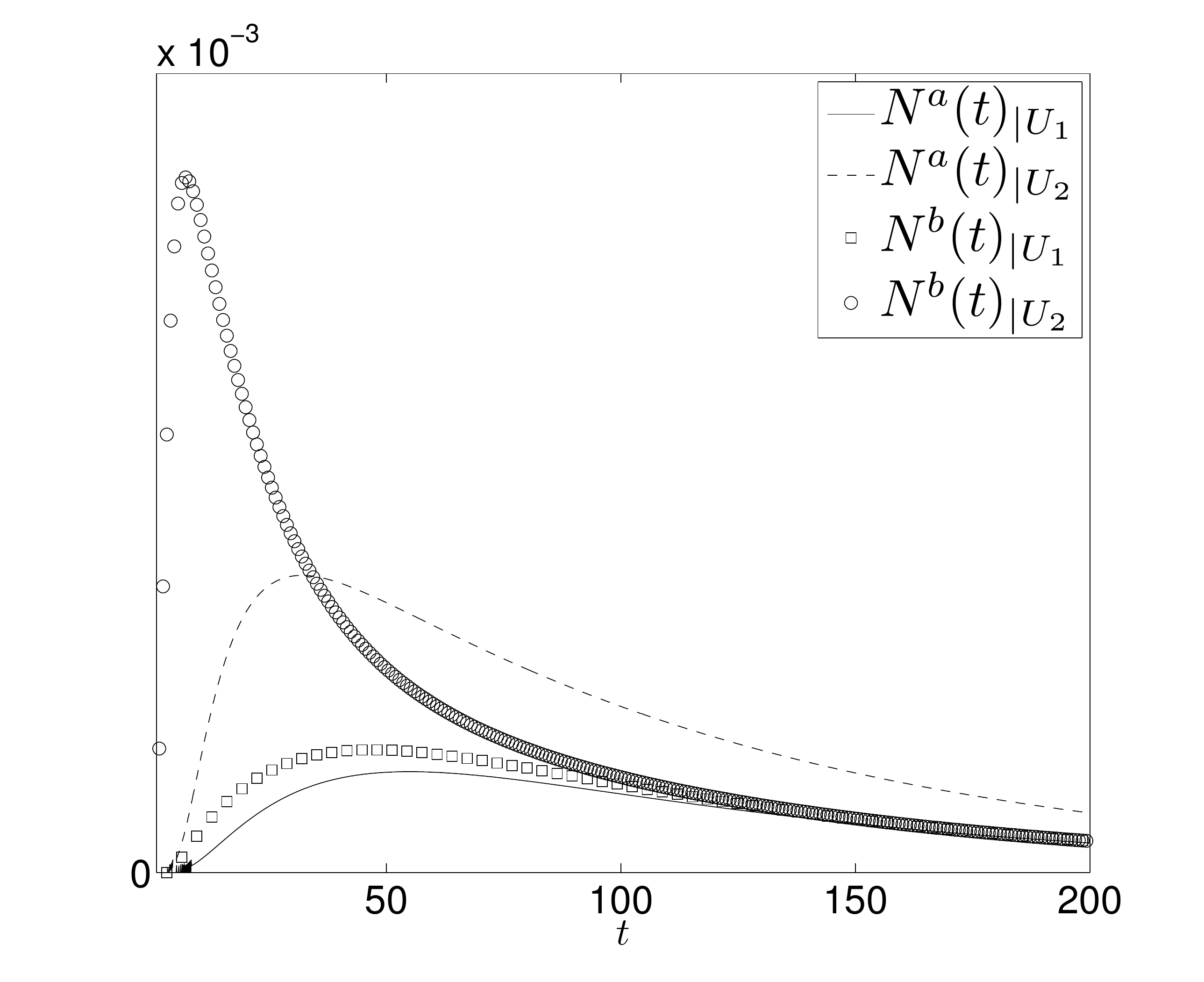}
\caption{\textbf{Setting $S_3^{ld}$}: the densities $N^a(t),N^b(t)$  in the exit cells $U_1$ and $U_2$ are shown for $\Delta T=0.08, N^a_{trh}=N^b_{trh}=10^{-5},
\omega^a_{\alpha}=\omega^b_{\alpha}=1\quad \forall \alpha , \rho_a=\rho_b=1$. The population $\Pc_b$ is globally closer to the exit cells than $\Pc_a$ and therefore it take
 less time to arrive in the exit cells. The exit cell $U_2$ is the more accessible for both populations, and in fact
both populations accumulate faster in $U_2$ than $U_1$. }
\label{2p_2u_td_U}
\end{center}
\end{figure}

The two populations $\Pc_a$ and $\Pc_b$ are originally
located as shown in Figure \ref{setup_p2_u2_po} and they have the same initial density, $N^a(0)=N^b(0)=1$. As in Section \ref{sect_s2ld}
for the Setting  $S_2^{ld}$,
we neglect for the moment the effects of any possible interaction between the populations due to their low initial densities. Therefore, we take $\lambda_{\alpha}=0$ for all $\alpha$.
As before, motivated by our previous analysis, we have taken $\Delta T=0.08, N^a_{trh}=N^b_{trh}=10^{-5}$, $\omega^a_{\alpha}=\omega^b_{\alpha}=1$, $\rho_a=\rho_b=1$.
In this configuration the population $\Pc_b$ is globally closer to the exit cells $U_1$ and $U_2$, as the sum of all the minimal paths going from the cell initially occupied to $U_1$ and $U_2$
is 14, while for $\Pc_a$ this $distance$ is 20: we therefore expect that if the populations have the same mobility \footnote{Some snapshots of the densities $N_{\alpha}^a (t)$, $N_{\alpha}^b(t)$ in $\R$ are shown in the Appendix, Figures \ref{m22a}-\ref{m22b}}, \emph{i.e.}, $\rho_a=\rho_b$, then $N^b(t)$ decays faster than $N^a(t)$
as it is actually shown by Figure \ref{2p_2u_td_vargamma}, where $\rho_a=\rho_b=1$.
{ In the same Figure, $N^b(t)$ is also shown for low values of $\rho_b$, and as observed previously for the Setting $S_1$ and $S_2^{ld}$ (see Figures \ref{p1_u1_vargamma} and \ref{2p_1u_td_vargammab}), if we decrease $\rho_b$ then $\Pc_b$ slows down and $N^b(t)$ decays slower than $N^a(t)$.}
We also show in Figure \ref{2p_2u_td_U} the densities (removed if greater than $N^a_{trh}$ or $N^b_{trh}$) in the exit cells $U_1,U_2$ when the  populations have the same mobility
($\rho_a=\rho_b=1$):
for both populations the cell $U_2$ is more accessible. This can be easily understood  for $\Pc_b$
which is very close to $U_2$ at the initial time. Regarding $\Pc_a$, although the lengths of the minimal paths going from the cell initially occupied by $\Pc_a$ to the exit cells
$U_1$ and $U_2$ are both 10, the cell $U_2$ turns out to be more accessible since there exist more minimal
 paths going to $U_2$ than paths going to $U_1$.
Furthermore, a realistic psychological suggestion is given by the position of the obstacles,
which gives the perception that, for $\Pc_a$, $U_2$ is a more natural way out, since $U_1$ is not even visible from their original cell.

\begin{figure}
\subfigure[$N^a(t)$]{\hspace*{-1.75cm}\includegraphics[width=8cm]{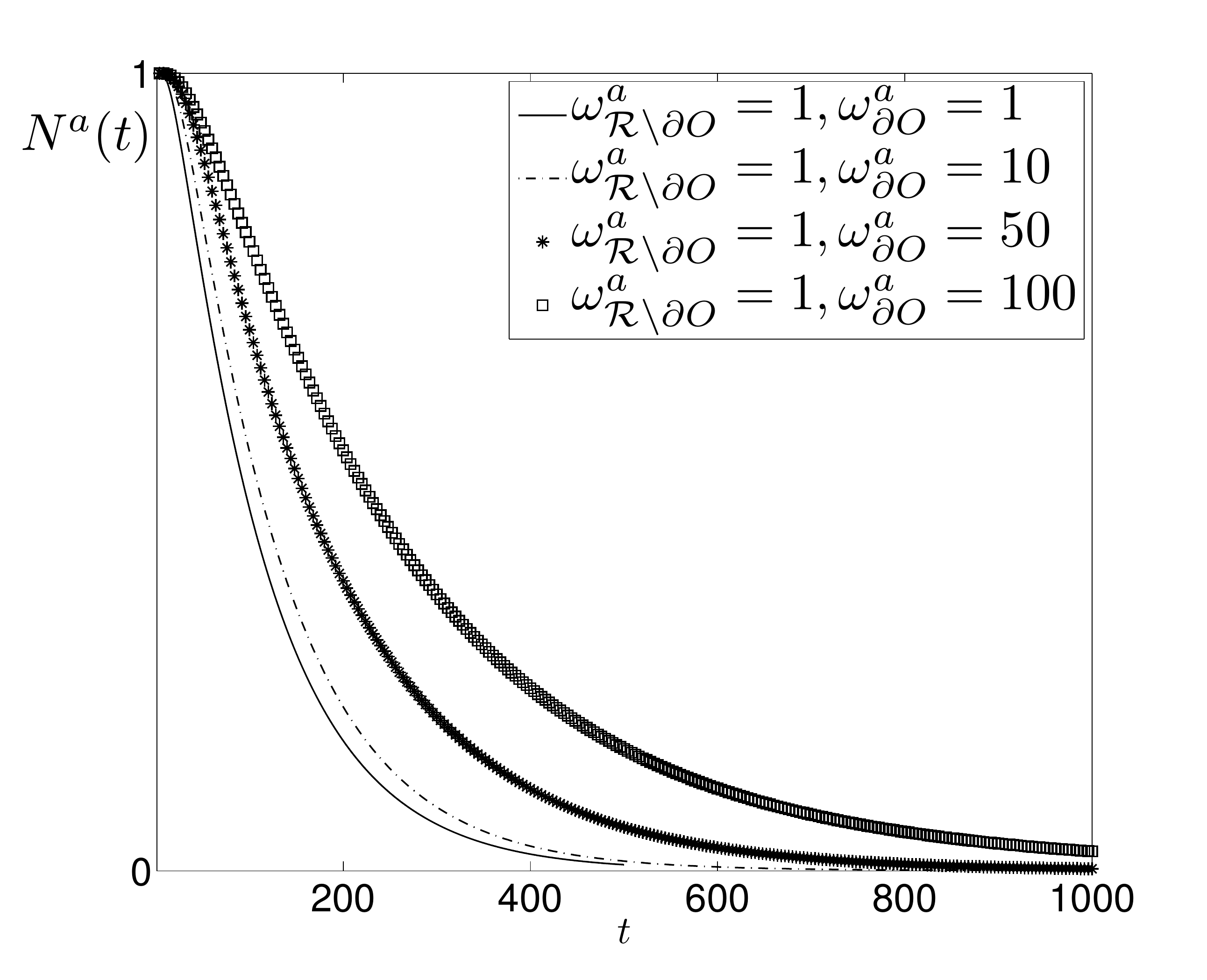}}
\subfigure[$N^b(t)$]{\hspace*{-0.5cm}\includegraphics[width=8cm]{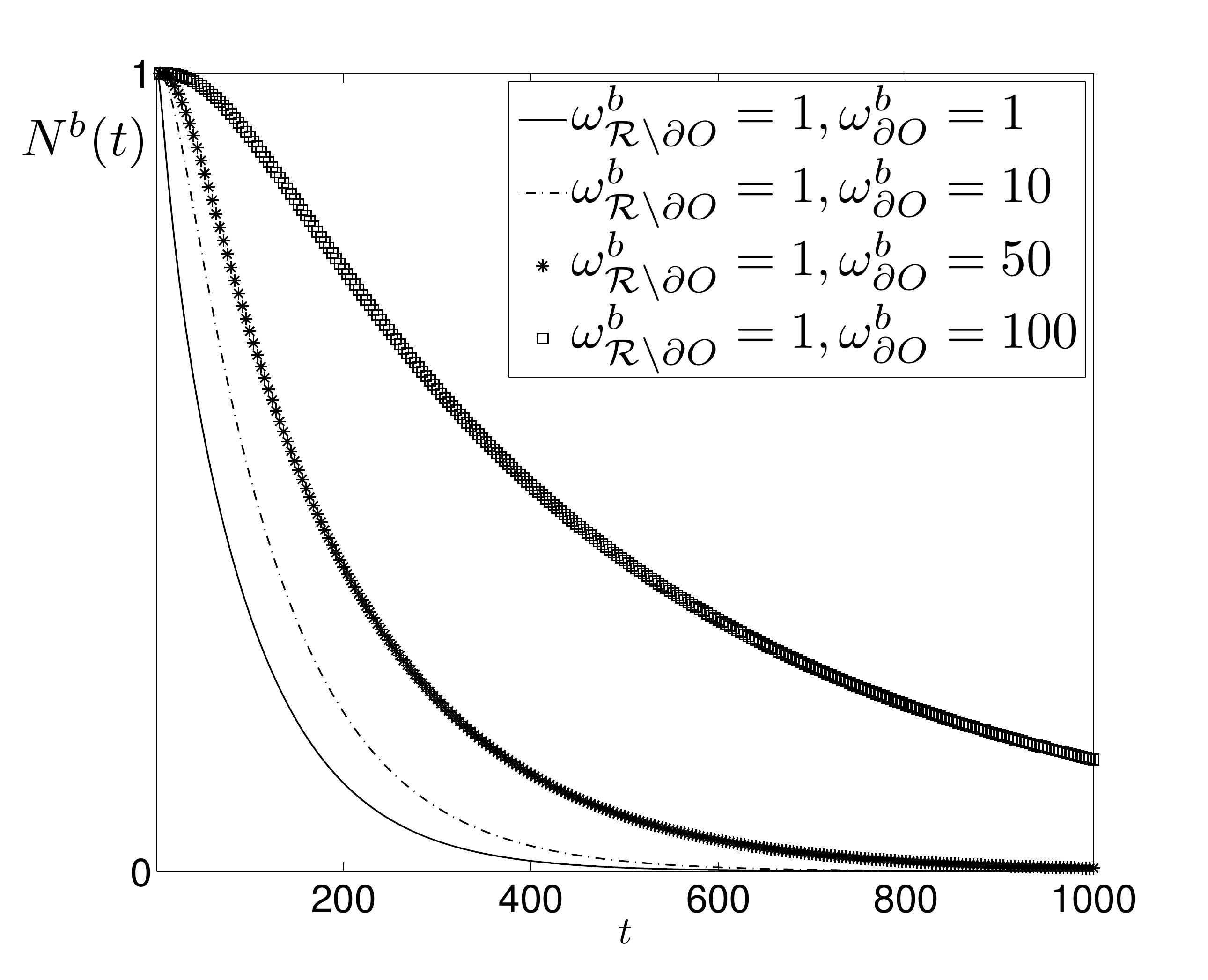}}
\vspace*{-0.5cm}\caption{\textbf{Setting $S_3^{ld}$}: the total densities $N^a(t),N^b(t)$ are shown for $\Delta T=0.08, N^a_{trh}=N^b_{trh}=10^{-5},
 \rho_a=\rho_b=1$ and different values of $\omega^a,\omega^b$. A strong inhomogeneity within $\R$ has the effect to slow down both populations,and in particular increasing values
of $\omega^a,\omega^b$ in $\partial O$ means more staticity in $\partial O$ for both the populations.  }
\label{2p_2u_na_varomega}
\end{figure}

 { To confirm the meaning of the parameters $\omega^a_{\alpha},\omega^b_{\alpha}$, that were previously interpreted as the \textit{inertia}
of the populations, we show in Figures \ref{2p_2u_na_varomega}(a)-\ref{2p_2u_na_varomega}(b) the densities of the populations for different values of $\omega^a_{\alpha},\omega^b_{\alpha}$ within $\R$ : analogously to what we have seen
in Figure \ref{p1_u1_varomega} and \ref{2p_1u_na_varomega} for the Setting $S_1$ and $S_2^{ld}$,
when $\omega^a,\omega^b$ inside $\partial O$ are much bigger than outside, then the two populations are quite slower because they become \emph{more
static} in $\partial O$.}


\subsubsection{With interaction (Setting $S_3^{hd}$)}
\begin{figure}
\vspace*{-0.5cm}
\begin{center}
\includegraphics[width=8cm]{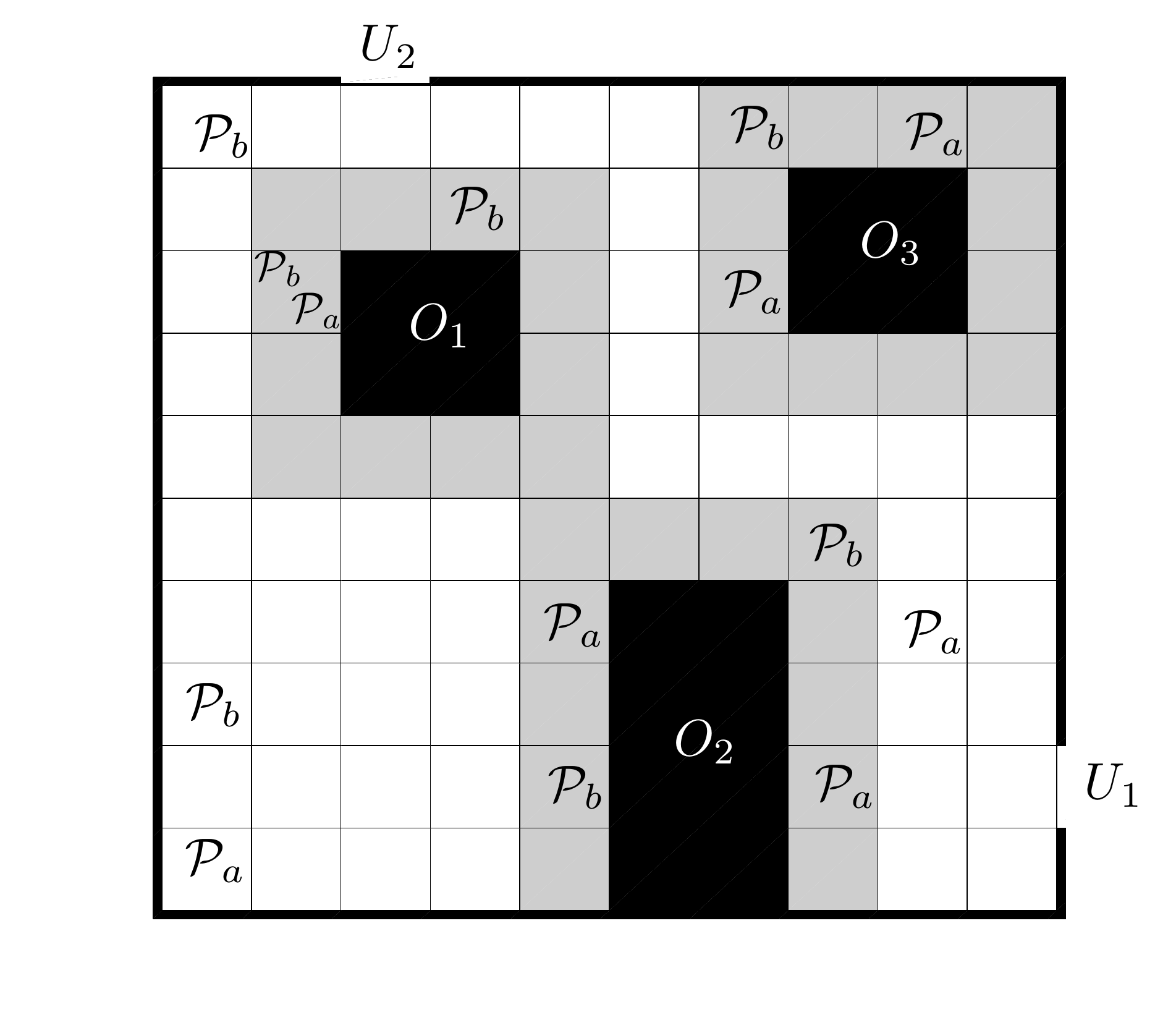}
\caption{\textbf{Setting $S_3^{hd}$}: at $t=0$ the population $\Pc_a$ is located in the cells  $(1,1),(8,2),(5,4),(9,4),(2,8),(7,8),(9,10)$, while
$\Pc_b$ is located in the cells $(5,2),(1,3),(8,5),(2,7),(2,8),(4,9), (1,10),(7,10)$. The
 black cells $O_1,O_2,O_3$ represents the obstacles, and the exit cells $U_1,U_2$ are located at (11,2) and (3,11).}
\label{setup_p2_u2_po_hd}
\end{center}
\end{figure}

\begin{figure}
\vspace*{-1cm}
\begin{center}
\includegraphics[width=8cm]{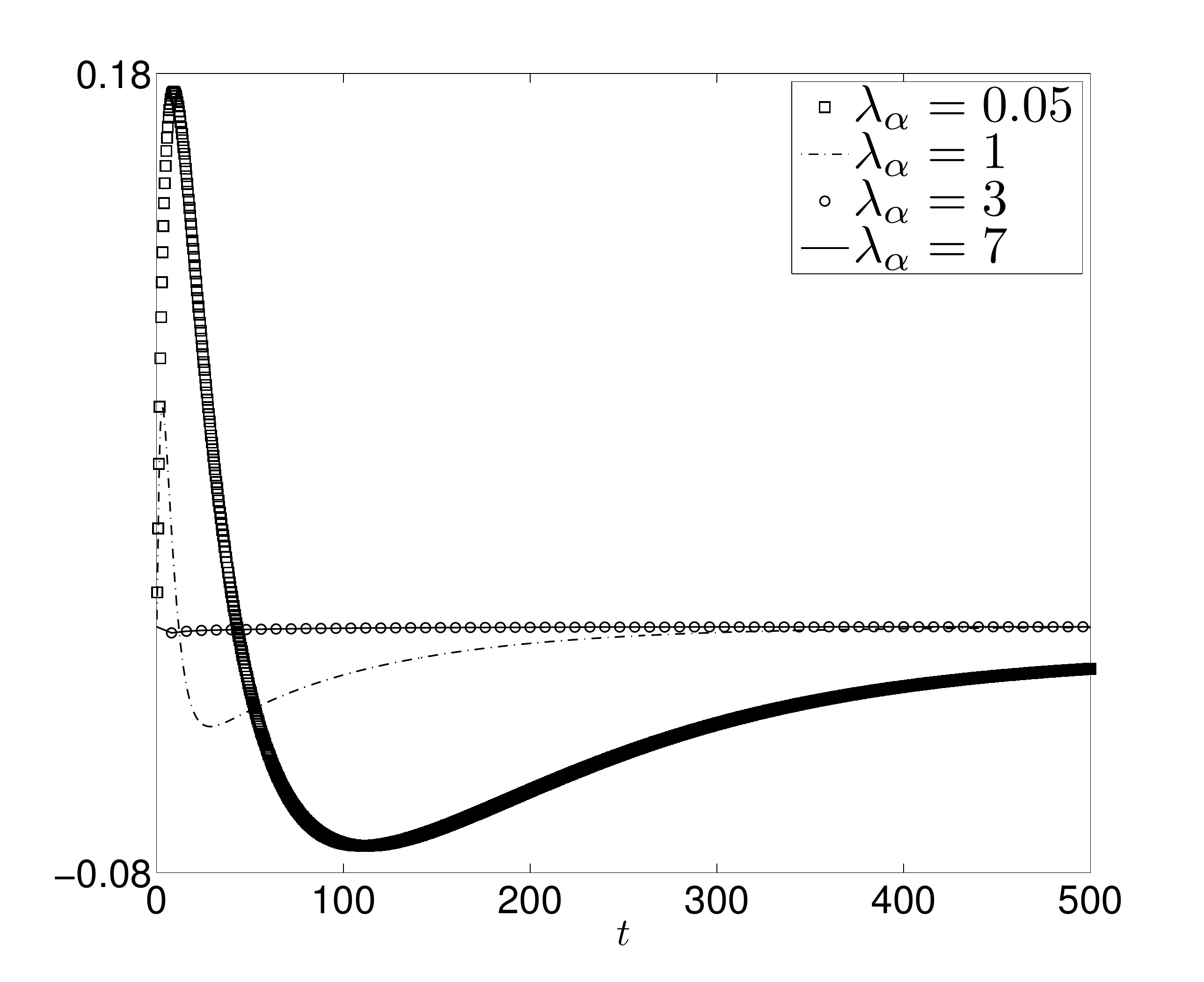}
\caption{\textbf{Setting $S_3^{hd}$}: the difference $N^a(t)-N^b(t)$ of the densities $N^a(t),N^b(t)$  for different values of $\lambda_{\alpha}$.
As
$\lambda_{\alpha}$ increases,   $N^a(t)-N^b(t)$ decreases, because the interaction parameter $\lambda_{\alpha}$ acts like an equalizer for $N^a(t)$ and $N^b(t)$.}
\label{2p_2u_diff_hd}
\end{center}
\end{figure}

\begin{figure}
\vspace*{-0.5cm}
\begin{center}
\includegraphics[width=8cm]{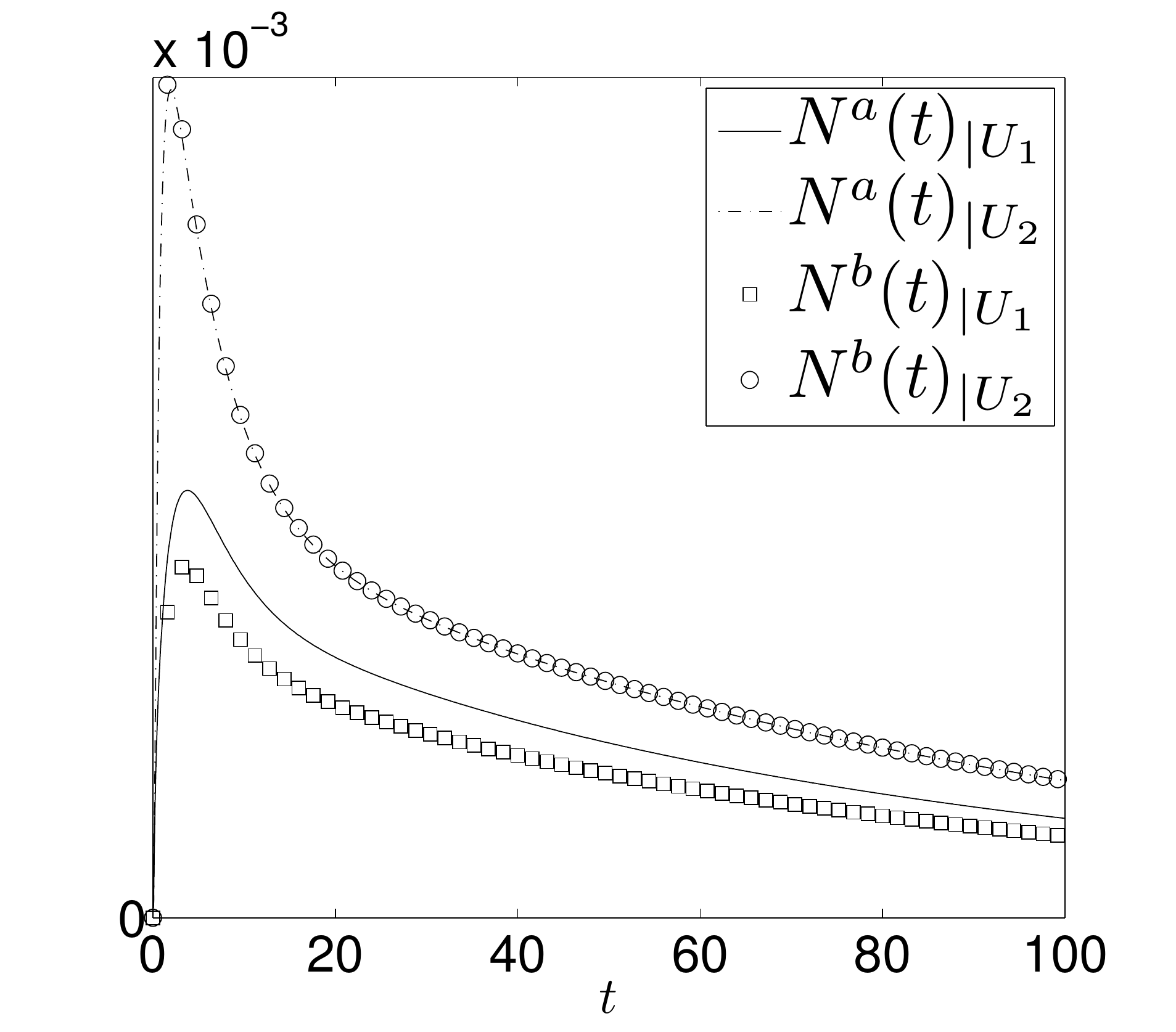}
\caption{\textbf{Setting $S_3^{hd}$}: the densities $N^a(t),N^b(t)$  in the exit cells $U_1$ and $U_2$ are shown for $\Delta T=0.08, N^a_{trh}=N^b_{trh}=10^{-5},
\omega^a_{\alpha}=\omega^b_{\alpha}=1,\lambda_{\alpha}=1 \quad \forall \alpha , \rho_a=1,\rho_b=1$. The exit cell $U_2$ is the more accessible for both populations, and in fact
both populations accumulate faster in $U_2$ than $U_1$. }
\label{2p_2u_td_U_hd}
\end{center}
\end{figure}

Suppose now that the populations $\Pc_a$ and $\Pc_b$ are originally
located as shown in Figure \ref{setup_p2_u2_po_hd} and they have same initial density, $N^a(0)=N^b(0)=7$.
As done before in Setting $S_2^{hd}$, we consider the physical effect of the interaction between the populations due to their sufficiently high initial densities.
When compared to the Setting $S_2^{hd}$, in which only one exit was present, here the second exit cell $U_2$ creates an easy way out for the population $\Pc_b$, and we observe a kind of equilibrium between the two populations: in fact,
the sum of all the lengths of the minimal paths going from the initial cells occupied by the populations to the exit cells is 90 for $\Pc_a$ and 89 for $\Pc_b$, and therefore we should expect that $N^a(t)$
and $N^b(t)$ decay in a similar way, at least if they have a similar mobility.
In Figure \ref{2p_2u_diff_hd}  the difference $N^a(t)-N^b(t)$ is shown for  $\lambda_{\alpha}=0.05,1,3,7$: as we increase
$\lambda_{\alpha}$,   $N^a(t)-N^b(t)$ decreases, and therefore also in this case the interaction parameter $\lambda_{\alpha}$
equalizes $N^a(t)$ and $N^b(t)$ as observed for the Setting $S_2^{hd}$ (for $\lambda_{\alpha}=3,7$ we have in practice $N^a(t)\approx N^b(t)$).
 We can also distinguish in Figure \ref{2p_2u_diff_hd} a first time range (depending on $\lambda_{\alpha}$) in which $N^a(t)>N^b(t)$,
due to the fact that $\Pc_b$ is initially  closer to $U_2$ than $\Pc_a$ and therefore $N^b(t)$ decays initially faster than $N^a(t)$.
At a subsequent time we get  $N^a(t)<N^b(t)$, due to the fact that the amount of $\Pc_b$ which goes to $U_1$ takes a longer time to exit
than $\Pc_a$. Figure \ref{2p_2u_td_U_hd} clearly shows this behavior.

Changing the value of $\rho_b$, or the values of $\omega_\alpha^{a,b}$ inside $\partial O$, produces exactly the same phenomena we have already described in the previous settings, and the conclusions are exactly the same: to speed up the escape procedure, it is better to keep $\R$ homogeneous.

\section{Conclusions}
\label{sec:conclusions}
In this paper we have consider a system consisting of one or two populations staying in a certain room $\R$, with one or two exits and with some obstacles all around, and we have discussed what happens in case of danger, when the populations have to leave the room as fast as they can. We have adopted an operatorial approach, in which the dynamics is deduced by an operator, the hamiltonian of the system. The two populations differ because of their different initial dispersion in $\R$, and because they have, in general, different mobilities. Apart from quite natural conclusions, as the fact that the path toward the exit(s) should be clearly identified and that a larger mobility means less time needed to go out of the room, we have also deduced that another improvement in the escape procedure is obtained when that part of $\R$ which can be occupied by the populations is \emph{homogeneous}, meaning with that that putting in $\R$ some \emph{points of interest} can slow down the escape procedure. We have also seen that the effect of the interaction between the two populations is to slow down the fast population and to speed up the slow one,
 so that the speed of the two populations become essentially equal, at least for moderate-high values of the interaction parameters. We have also analyzed the role of the parameter $\sigma_a$ in (\ref{function_ya}), deducing that not many differences arise when modifying its value. This could be understood as follows: when $\sigma_a$ increases the path is narrow but direct to the exit. On the other hand, when $\sigma_a$ decreases, the path is large, not so direct, but a larger amount of populations can use that path simultaneously.

It may be worth to observe that, despite the apparent difficulty of the model,  we have been able to recover analytically the densities of the populations in each cell in a very simple way (see Eq. \eqref{36}), with obvious advantages in term
of computational effort. Other fluid-dynamic models working on a macroscopic scale, for example, can work well in certain environments, but the presence
of high non--linearities in the model equation usually produce serious difficulties in any numerical approach (see Refs.~\cite{xiaoping,zheng} and the references therein).
Moreover, some restrictions of other methodologies used to describe crowd evacuation can be overcome in our approach.
For example to incorporate the typical high-pressures
phenomena, not properly simulated without an appropriate force-model, we can simply create a correlation in each cell between the density of a population
and the mobility parameters $\rho_{a,b}$ (high density-low values for $\rho_{a,b}$, and vice versa),
even if in this way we are obliged to recompute the coefficients $f_{\alpha,\beta}(t)$ in \eqref{36} at each time step.
We can also consider the effect of a chaotic escape
by including some randomness in the mobility and inertia parameters $\rho_{a,b}$ and $\omega^{a,b}_{\alpha}$  in each cell, or in the definition of the $p^{(a,b)}_{\alpha,\beta}$.
These changes can lead to a more complete and satisfactory model describing the the escape of two populations from a room, along with the
inclusion of some nonlinearities in the model due to a different form of the hamiltonian, and this is just part of our future works.

{ We conclude observing that, not unexpectedly, these aspects are somehow related: nonlinearities
in the differential equations arise from a non quadratic hamiltonian, which naturally replaces
the one used here if we want to modify the mobility parameters $\rho_{a,b}$ in order to take into account 
the role of the density in the speed of movement of $\Pc_a$ and $\Pc_b$. In fact, in this case, we expect 
$\rho_{a,b}$ to be functions of $\hat n_\alpha^{(a)}$ and $\hat n_\alpha^{(b)}$.}

\section{Appendix}

In the following figures we show the snapshots of densities $N_{\alpha}^a(t)$ and $N_{\alpha}^b(t)$ in each cell of $\R$ at various time for the setting $S_2^{ld},S_3^{ld}$. In these figures the behavior of the populations during the escape from $\R$ seems to be quite reasonable, as the populations are always directed toward the exit cells and the densities globally diminish in time within $\R$.
 \begin{figure}
\subfigure[$t=0.1$]{\hspace*{-1cm}\includegraphics[width=7.5cm]{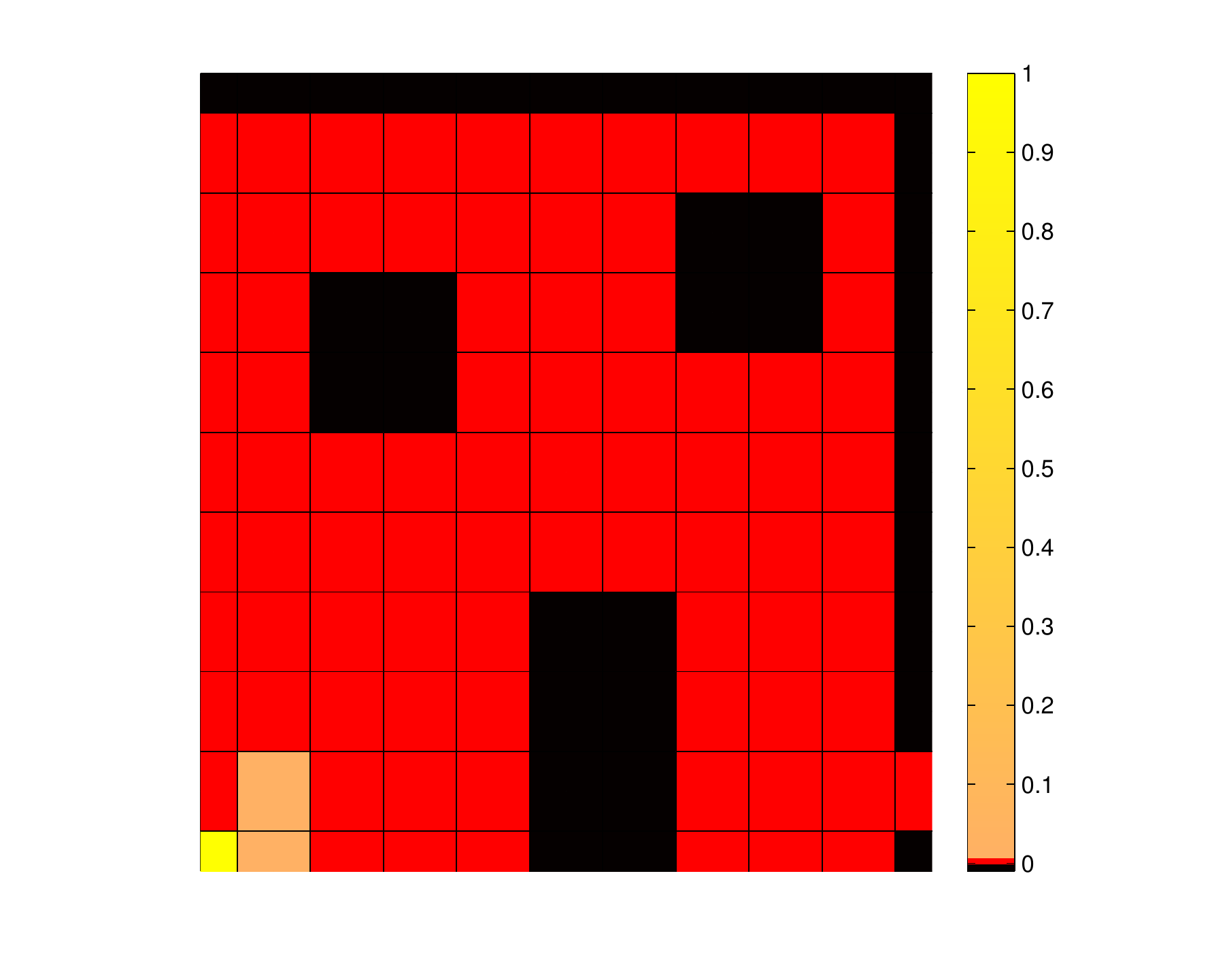} }
\subfigure[$t=4$]{\includegraphics[width=7.5cm]{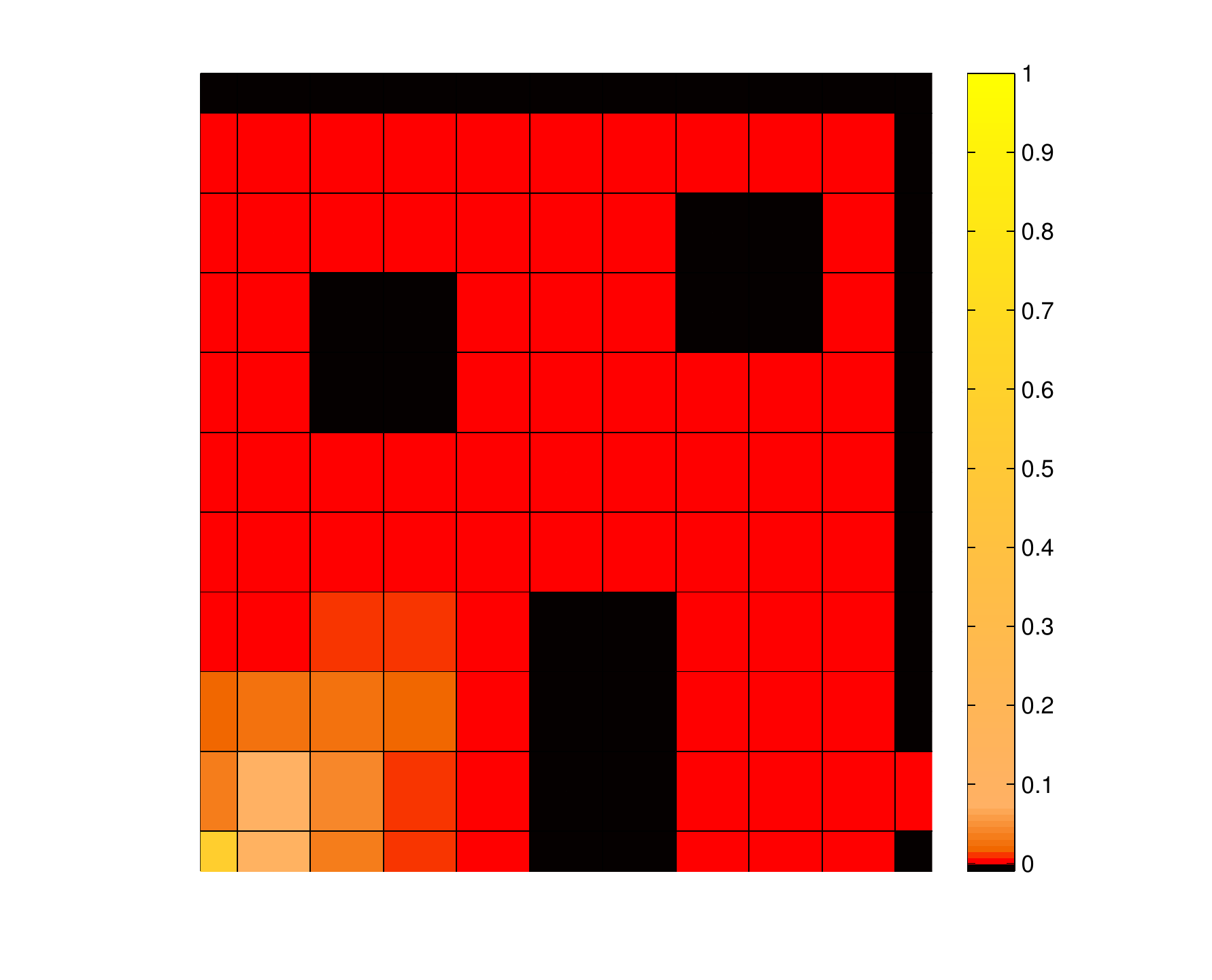}}
\subfigure[$t=80$]{\vspace*{-0.5cm}\hspace*{-1cm}\includegraphics[width=7.5cm]{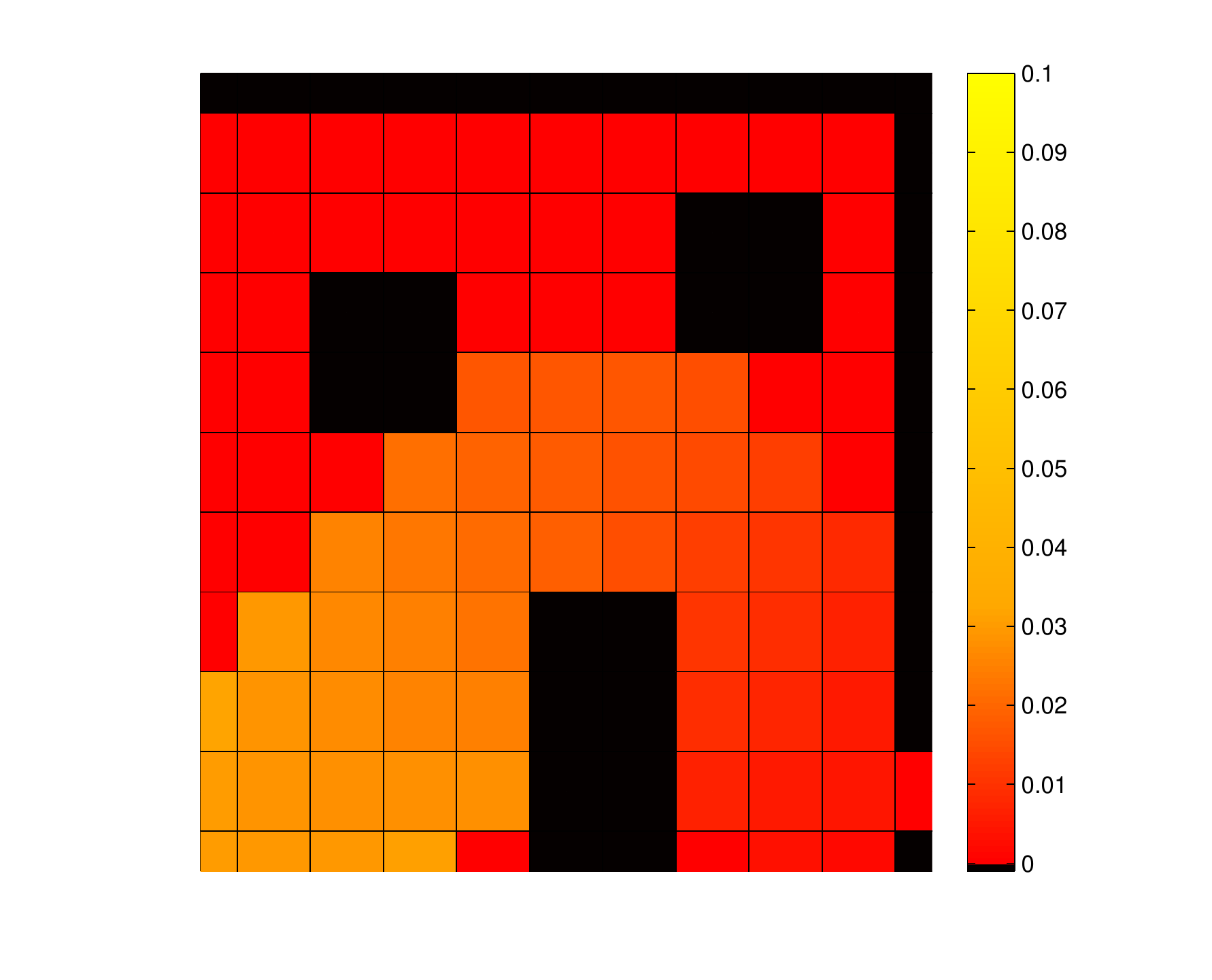} }
\subfigure[$t=480$]{\includegraphics[width=7.5cm]{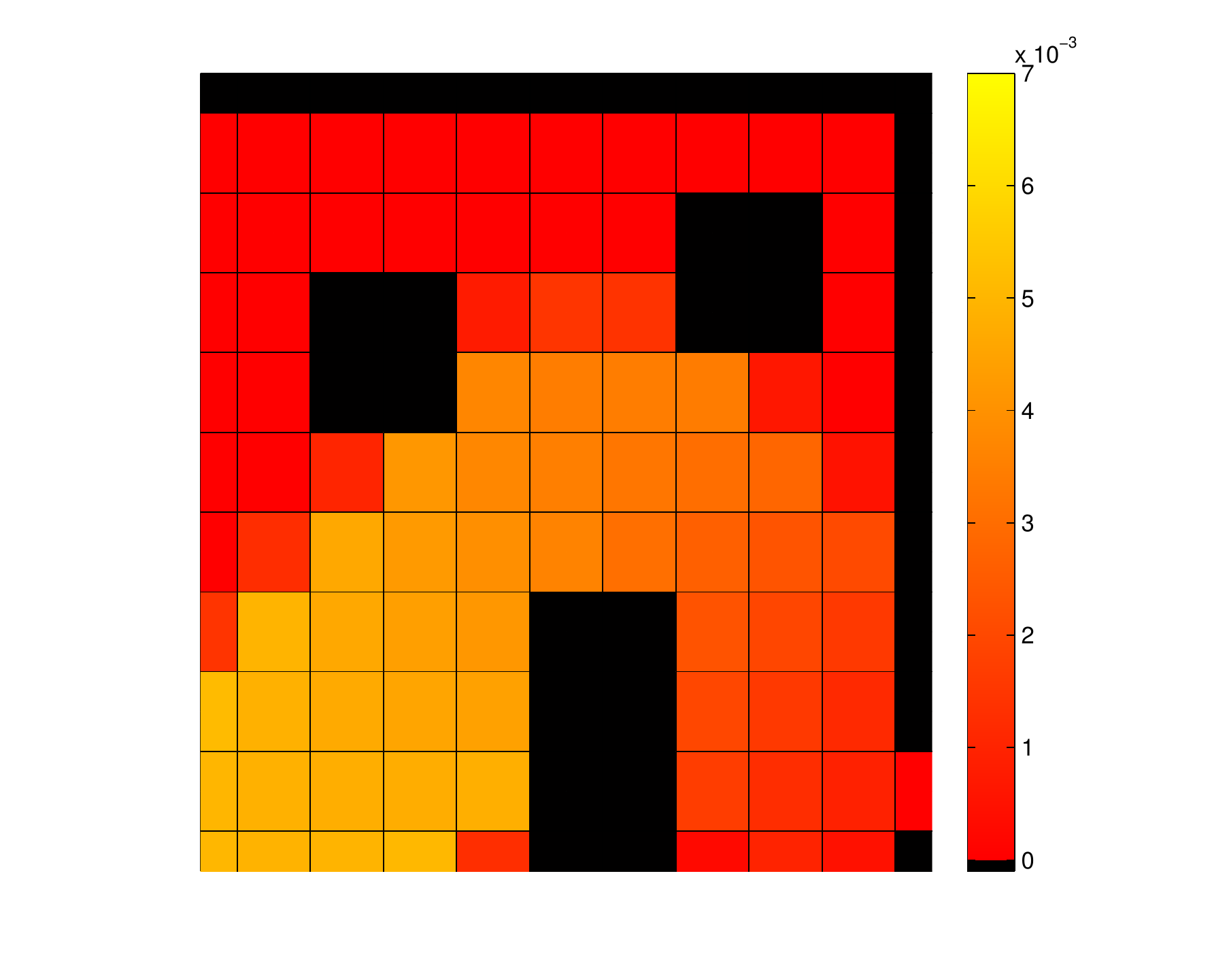} }
\vspace*{-0.5cm}\caption{\textbf{Setting $S_2^{ld}$}: the density $N_{\alpha}^a$ of the population $\Pc_a$ in the various cells of $\R$ at various time. Yellow/light colors mean high density, red/dark colors mean low density. The obstacles are the black cells.  The parameters are $\Delta T=0.08, N^a_{trh}=N^b_{trh}=10^{-5},
\omega^a_{\alpha}=\omega^b_{\alpha}=1\, \rho_a=\rho_b=1,\lambda_{\alpha}=0$.  }
\label{m21a}
\end{figure}
 \begin{figure}
\subfigure[$t=0.1$]{\hspace*{-1cm}\includegraphics[width=7.5cm]{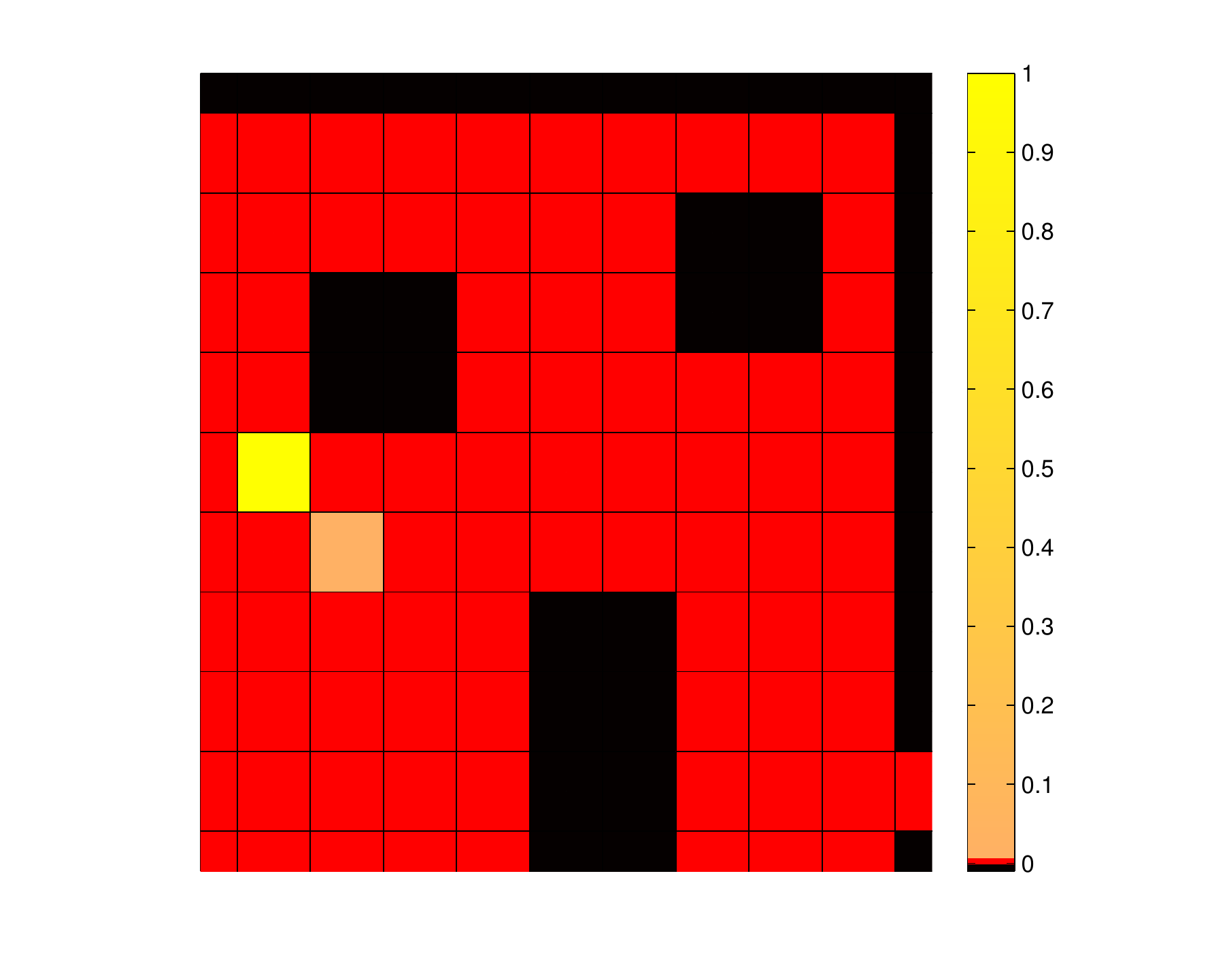} }
\subfigure[$t=4$]{\includegraphics[width=7.5cm]{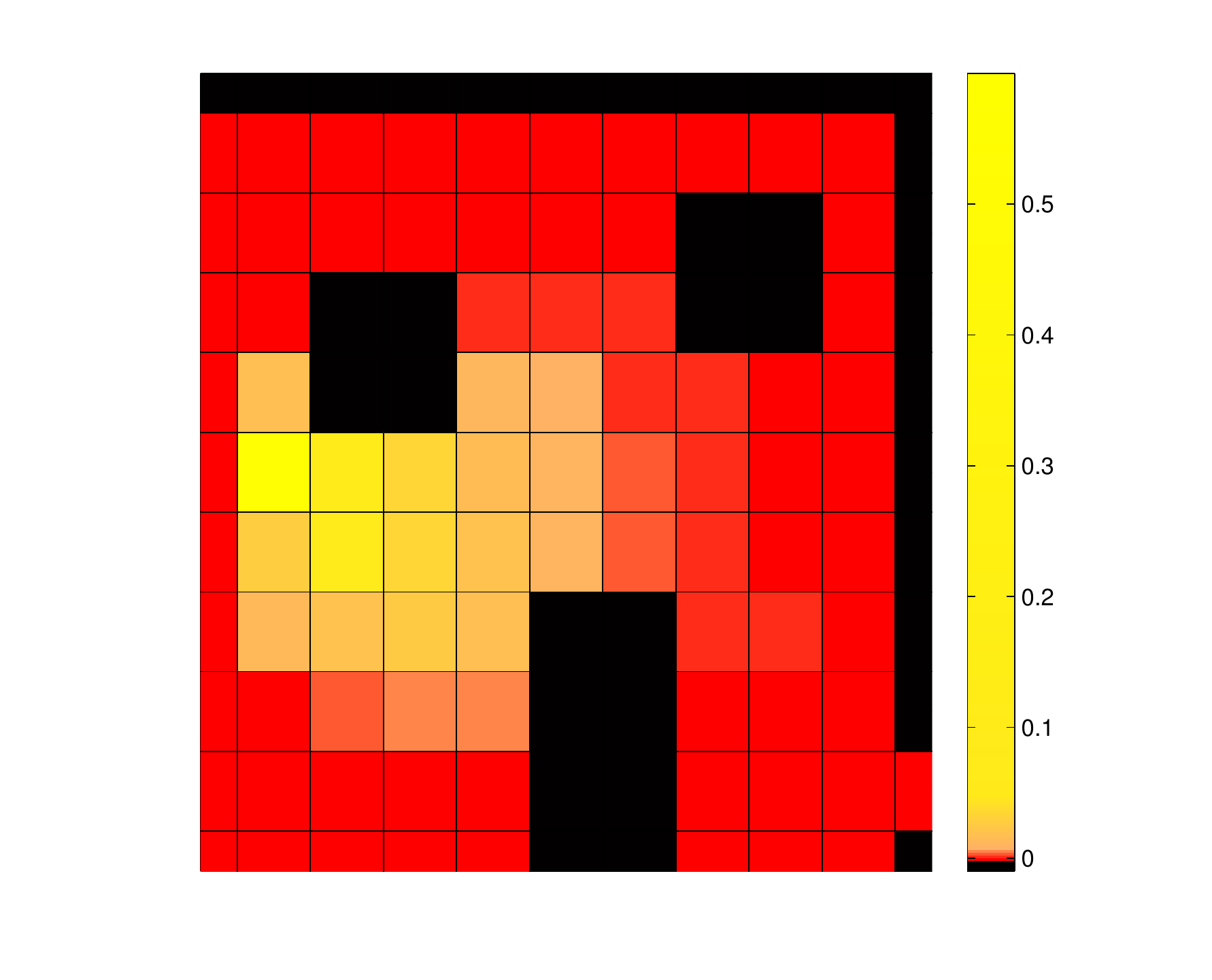}}
\subfigure[$t=80$]{\vspace*{-0.5cm}\hspace*{-1cm}\includegraphics[width=7.5cm]{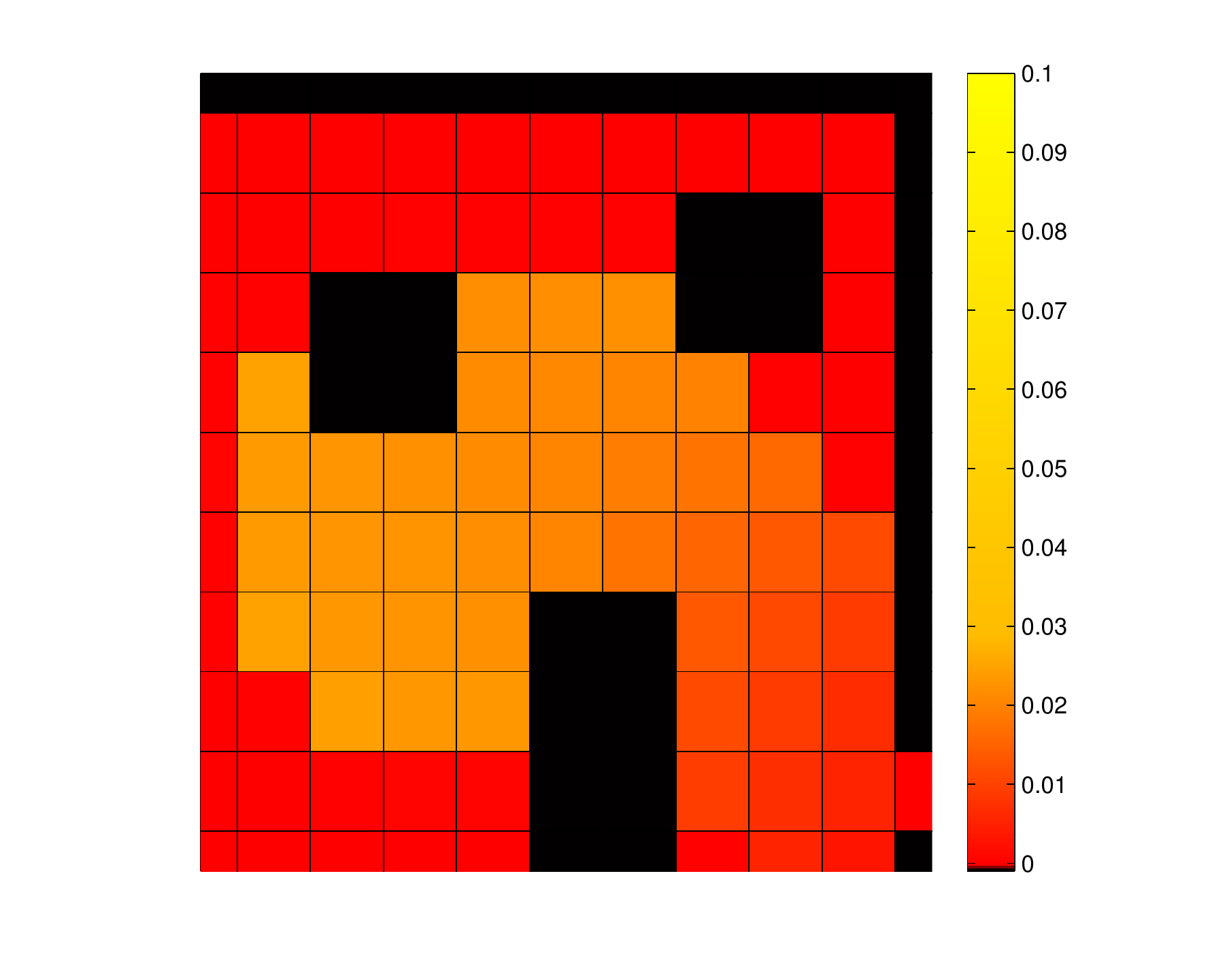} }
\subfigure[$t=480$]{\includegraphics[width=7.5cm]{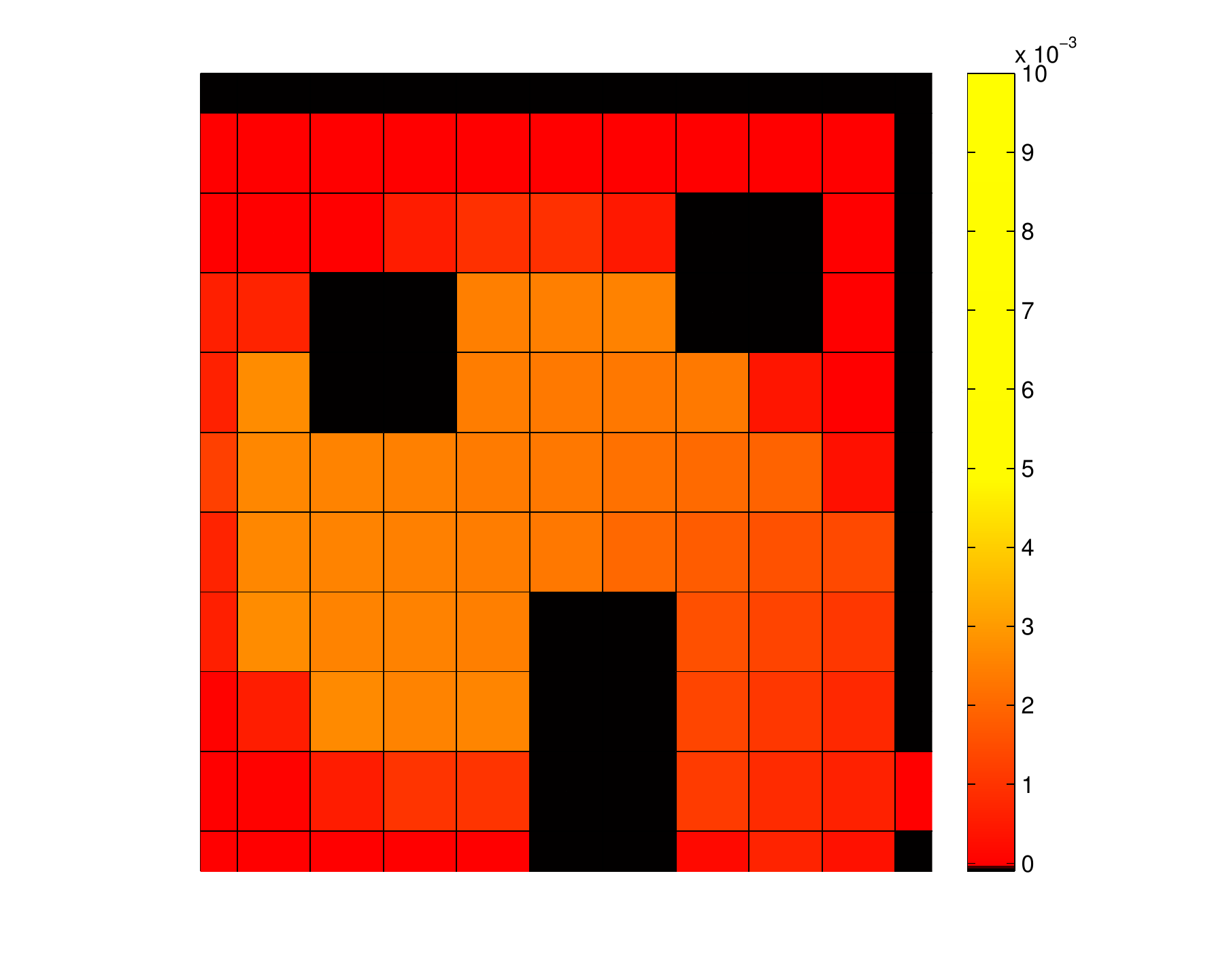} }
\vspace*{-0.5cm}\caption{\textbf{Setting $S_2^{ld}$}: the density $N_{\alpha}^b$ of the population $\Pc_b$ in the various cells of $\R$ at various time. Yellow/light colors mean high density, red/dark colors mean low density. The obstacles are the black cells. The parameters are $\Delta T=0.08, N^a_{trh}=N^b_{trh}=10^{-5},
\omega^a_{\alpha}=\omega^b_{\alpha}=1\, \rho_a=\rho_b=1,\lambda_{\alpha}=0$.  }
\label{m21b}
\end{figure}

 \begin{figure}
\subfigure[$t=0.1$]{\hspace*{-1cm}\includegraphics[width=7.5cm]{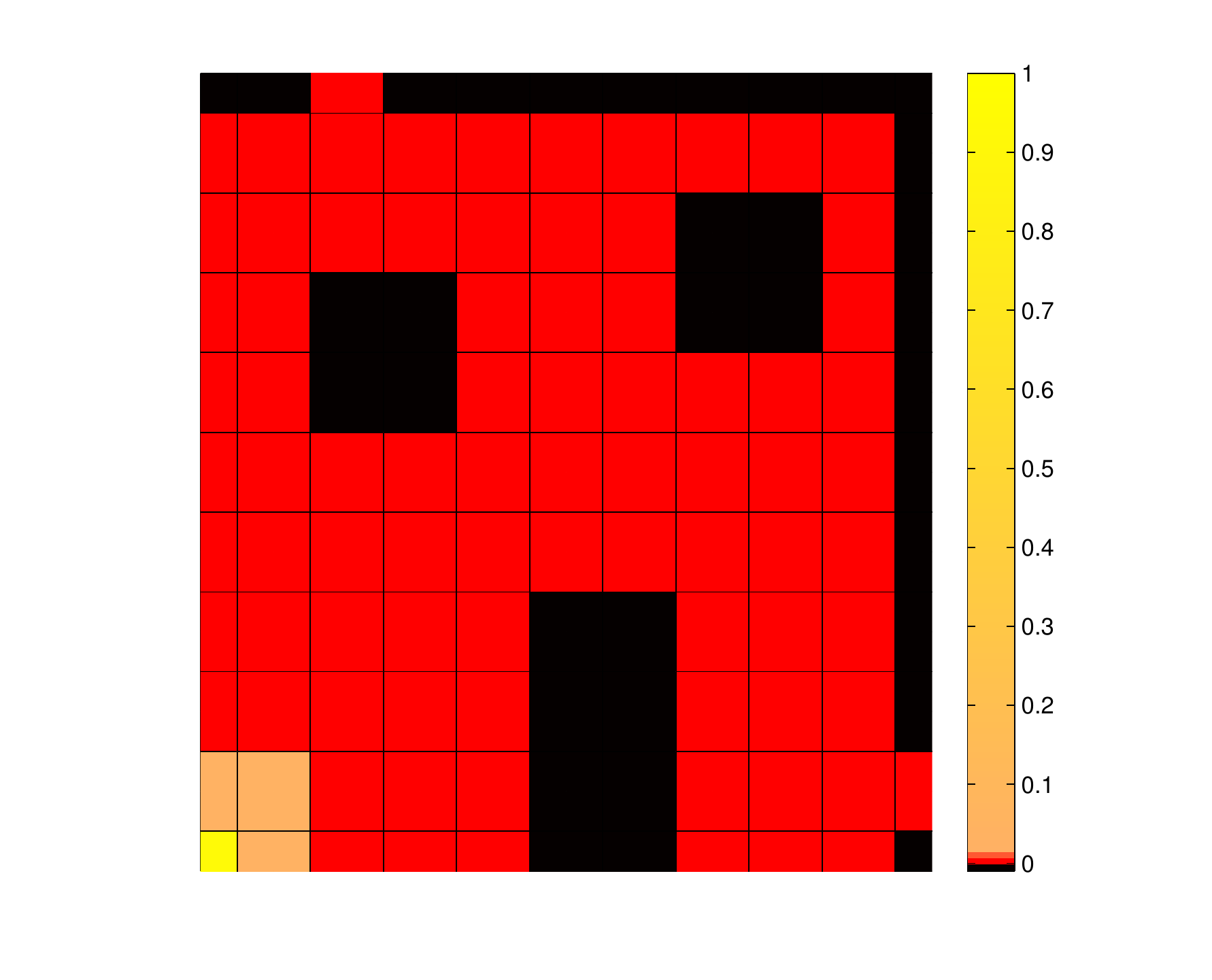} }
\subfigure[$t=4$]{\includegraphics[width=7.5cm]{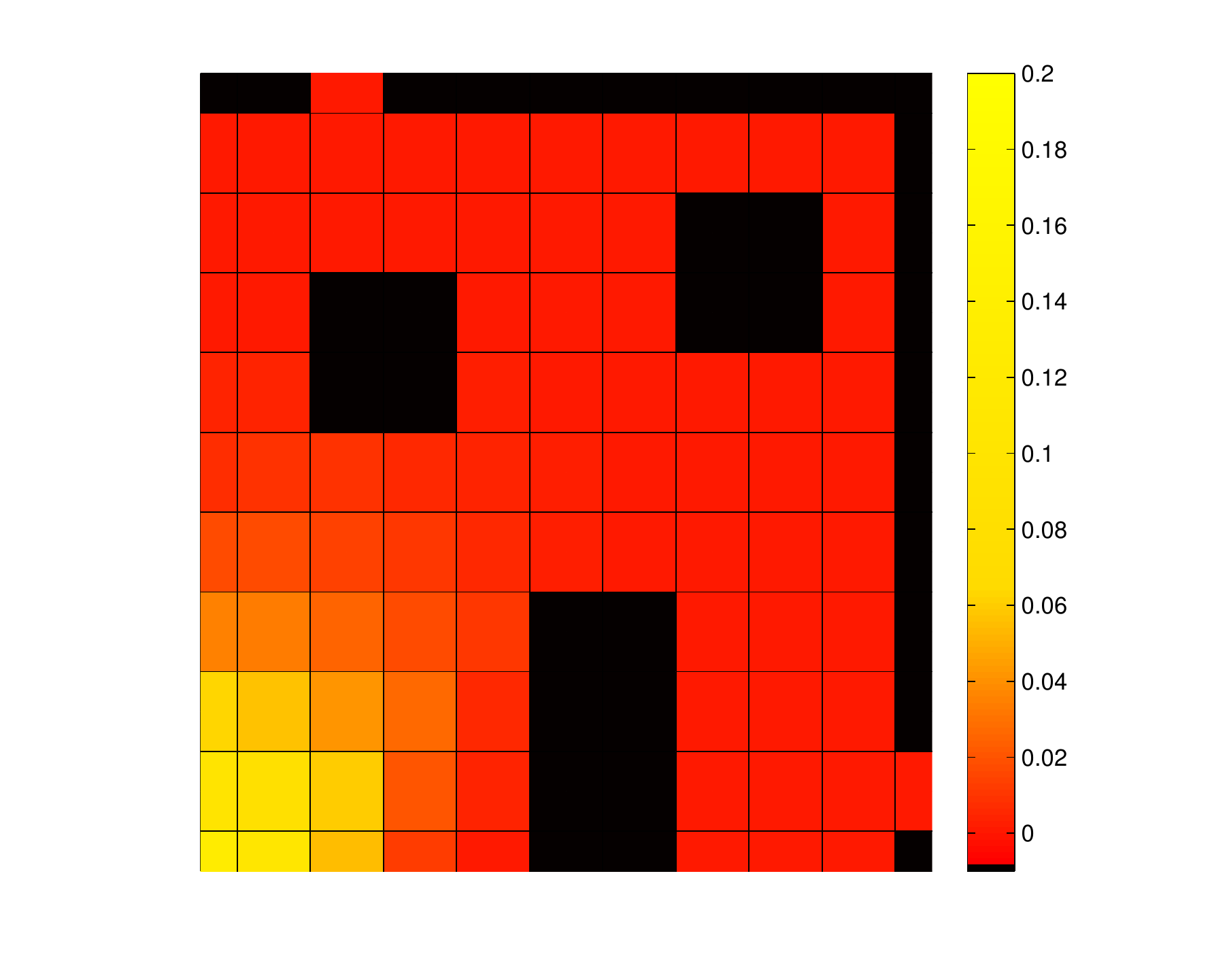}}
\subfigure[$t=80$]{\vspace*{-0.5cm}\hspace*{-1cm}\includegraphics[width=7.5cm]{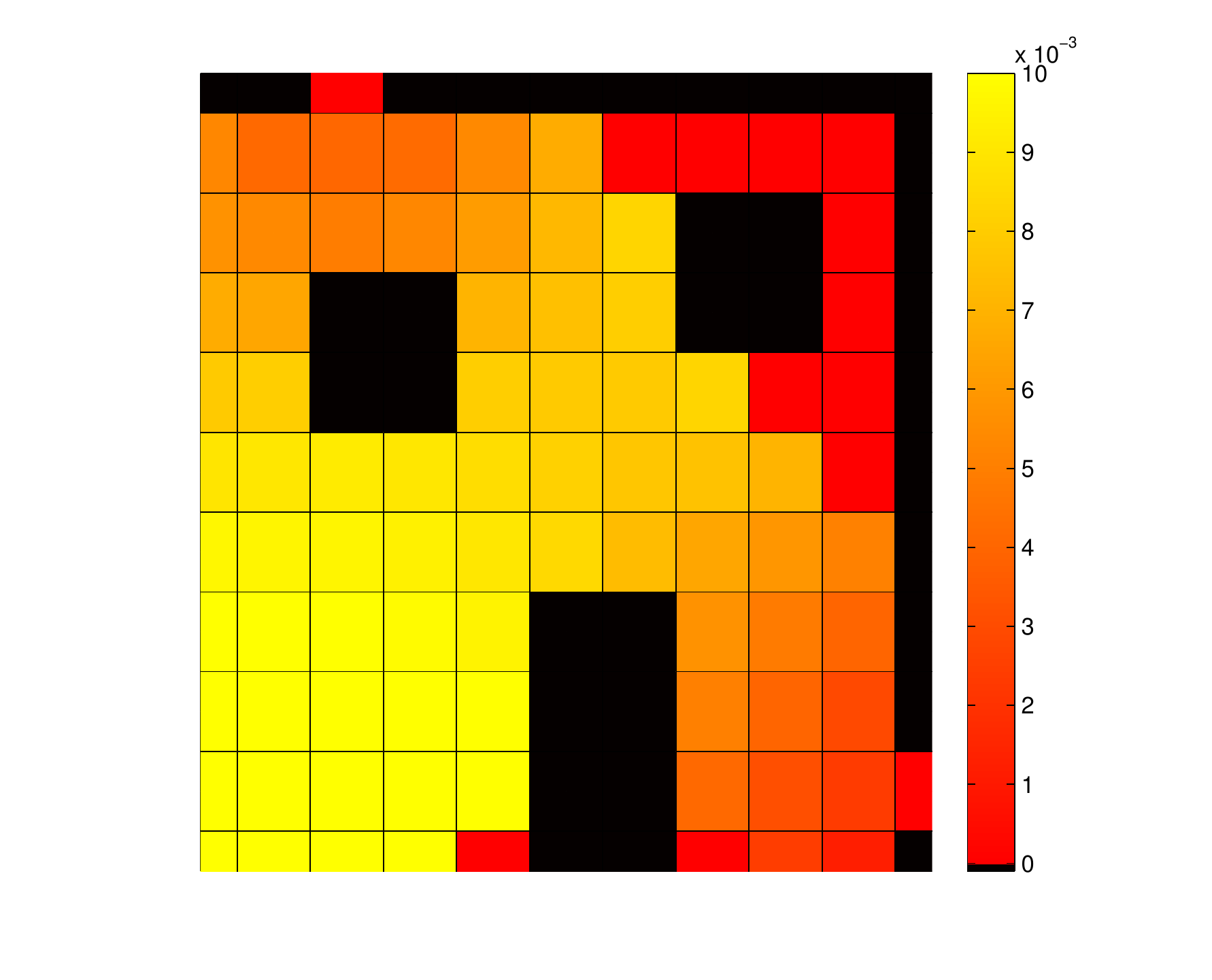}}
\subfigure[$t=145$]{\includegraphics[width=7.5cm]{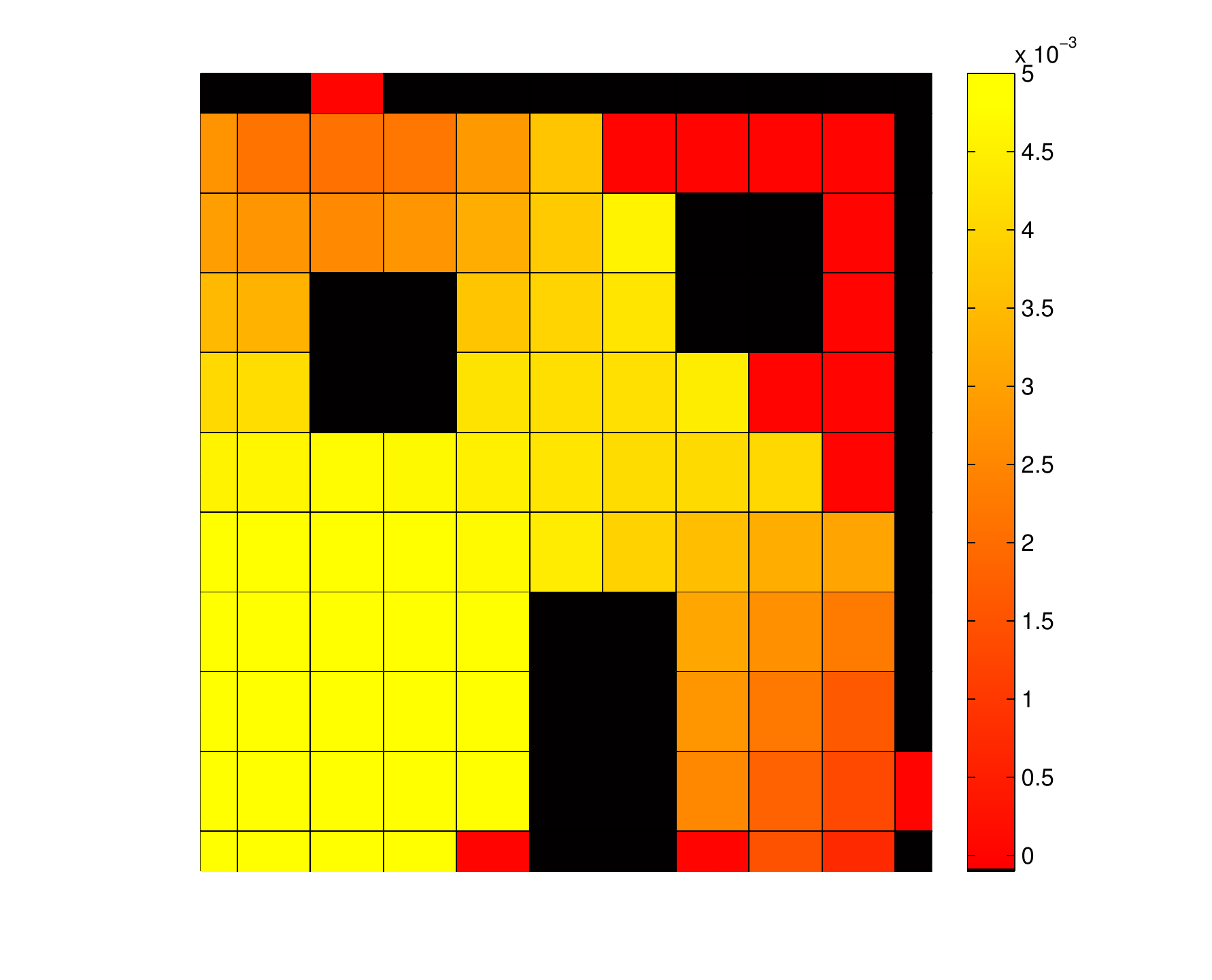} }
\vspace*{-0.5cm}\caption{\textbf{Setting $S_3^{ld}$}: the density $N_{\alpha}^a$ of the population $\Pc_a$ in the various cells of $\R$ at various time. Yellow/light colors mean high density, red/dark colors mean low density. The obstacles are the black cells.  The parameters are $\Delta T=0.08, N^a_{trh}=N^b_{trh}=10^{-5},
\omega^a_{\alpha}=\omega^b_{\alpha}=1\, \rho_a=\rho_b=1,\lambda_{\alpha}=0$.  }
\label{m22a}
\end{figure}
 \begin{figure}
\subfigure[$t=0.1$]{\hspace*{-1cm}\includegraphics[width=7.5cm]{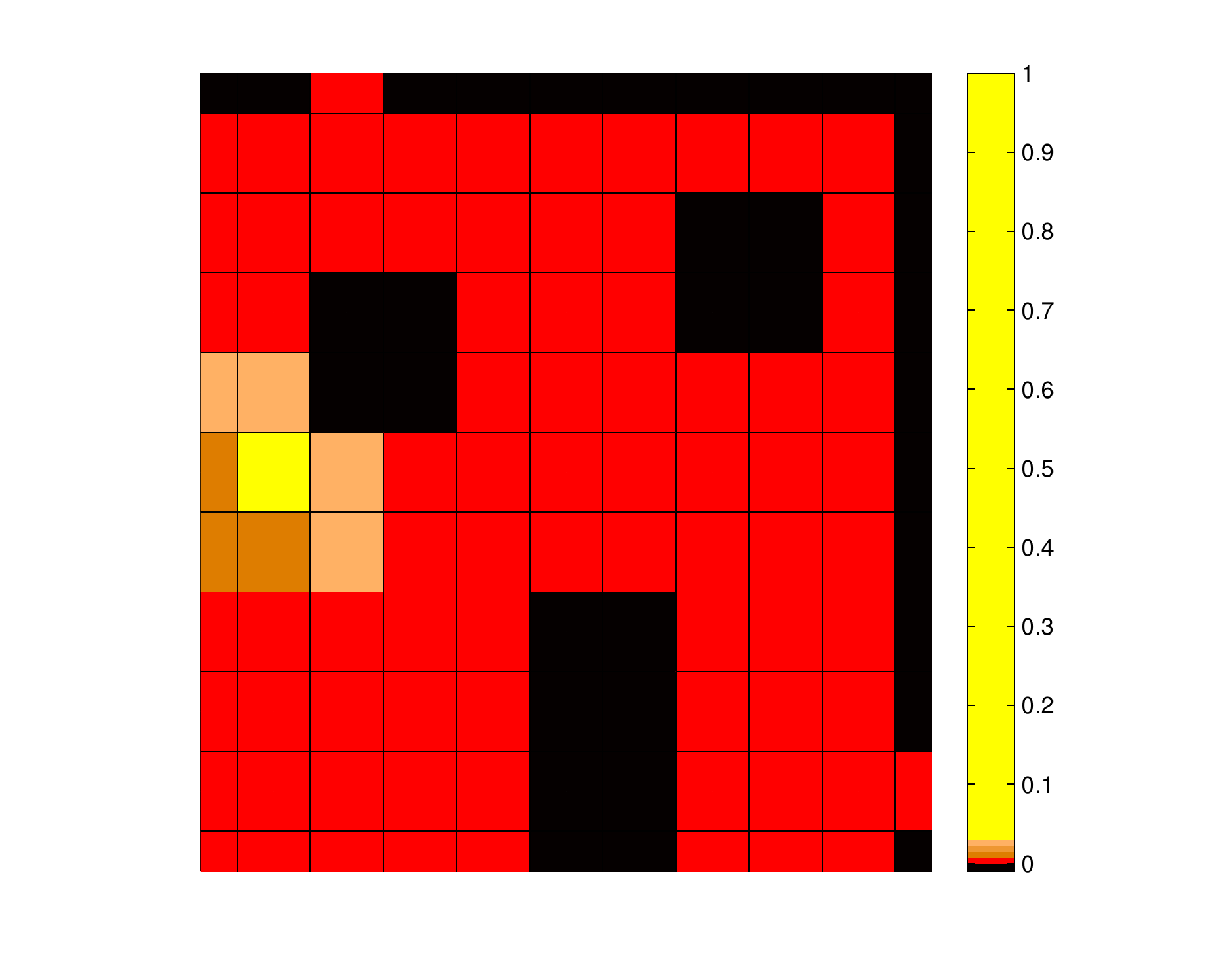} }
\subfigure[$t=4$]{\includegraphics[width=7.5cm]{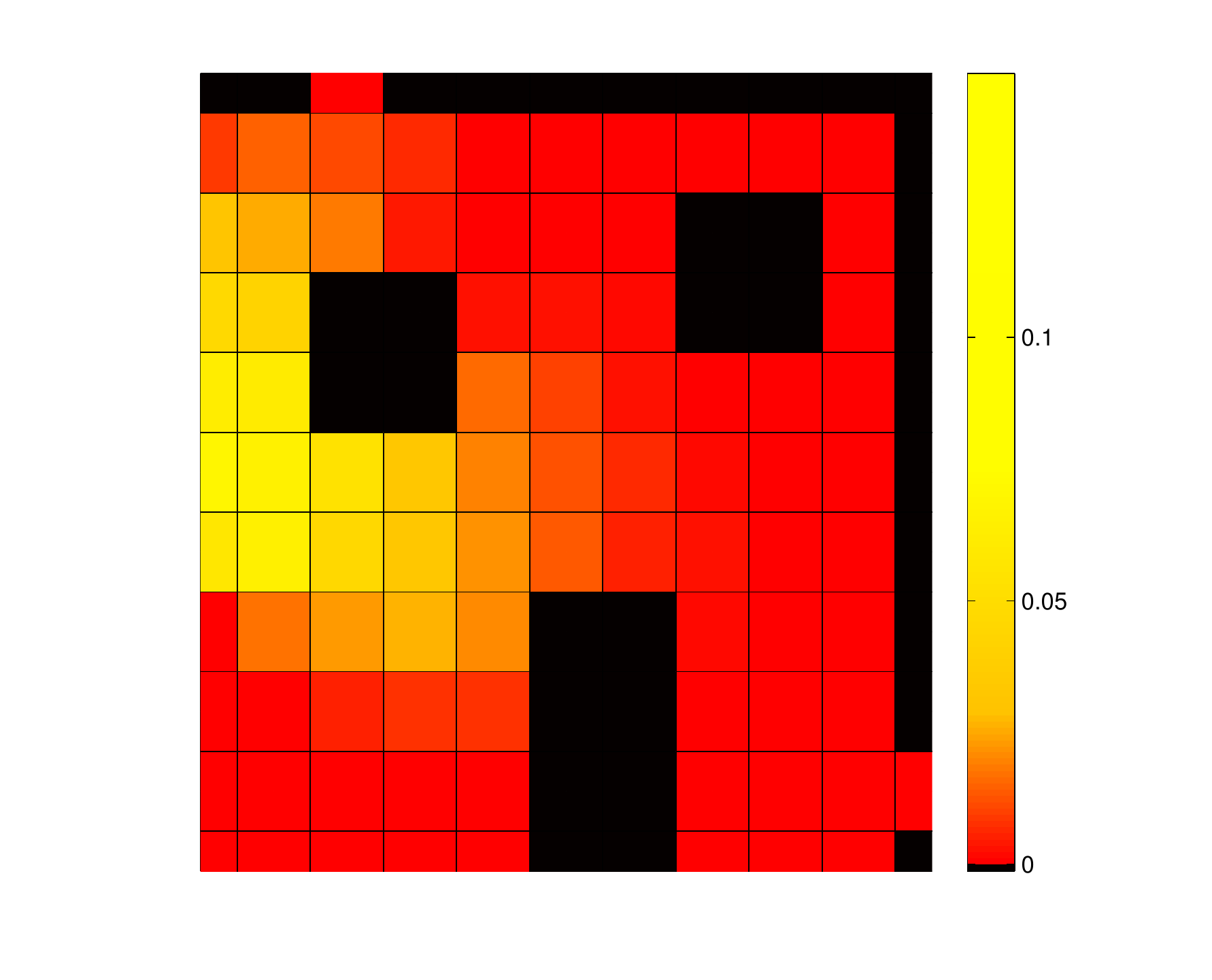}}
\subfigure[$t=80$]{\vspace*{-0.5cm}\hspace*{-1cm}\includegraphics[width=7.5cm]{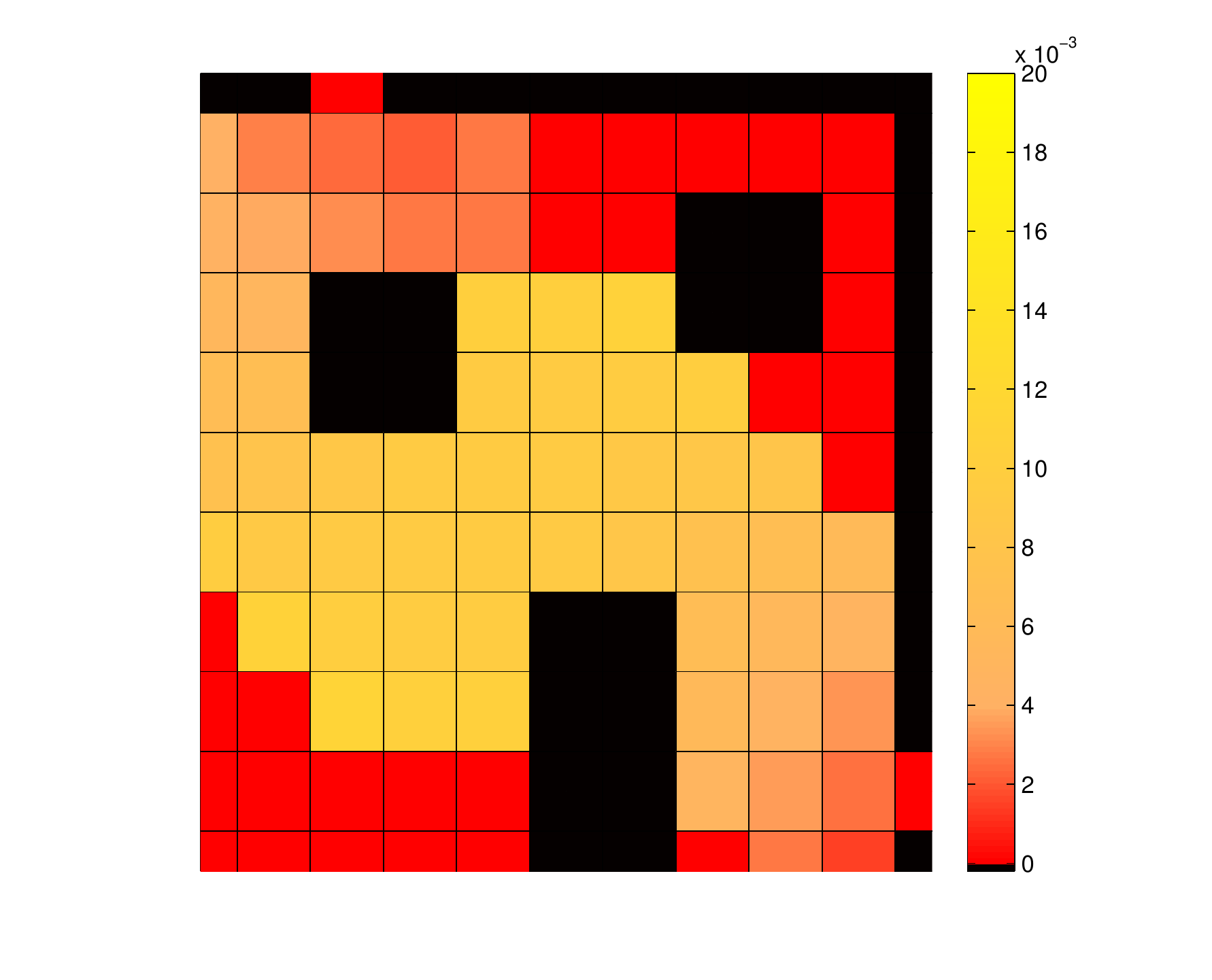} }
\subfigure[$t=145$]{\includegraphics[width=7.5cm]{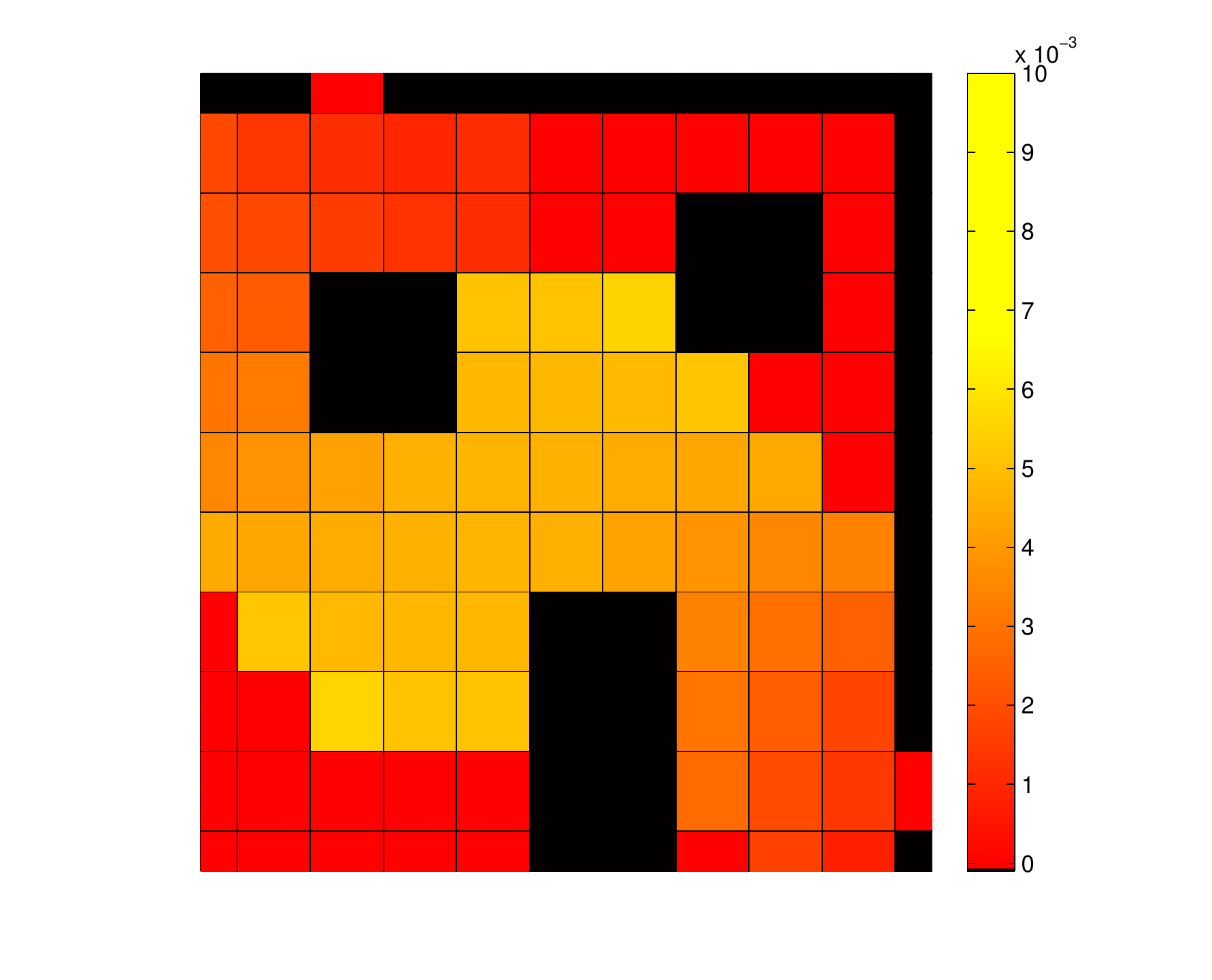} }
\vspace*{-0.5cm}\caption{\textbf{Setting $S_3^{ld}$}: the density $N_{\alpha}^b$ of the population $\Pc_b$ in the various cells of $\R$ at various time. Yellow/light colors mean high density, red/dark colors mean low density. The obstacles are the black cells. The parameters are $\Delta T=0.08, N^a_{trh}=N^b_{trh}=10^{-5},
\omega^a_{\alpha}=\omega^b_{\alpha}=1\, \rho_a=\rho_b=1,\lambda_{\alpha}=0$.  }
\label{m22b}
\end{figure}

\end{document}